\documentclass[longauth]{aa}  

\usepackage{graphicx}
\usepackage[varg]{txfonts}
\usepackage{float}
\usepackage{color}
\usepackage{natbib,twoopt}
\usepackage{wasysym}
\usepackage{multicol}
\usepackage{lscape}
\usepackage{pdflscape}
\usepackage{longtable}
\usepackage{bigints}
\usepackage[breaklinks=true]{hyperref} 

\definecolor{darkblue}{rgb}{0.0,0.0,0.4}
\hypersetup{colorlinks,breaklinks, linkcolor=darkblue,urlcolor=darkblue, anchorcolor=darkblue,citecolor=darkblue}

\def\kms{km.s$^{-1}$}         
\def\ms{\hbox{m.s$^{-1}$}}         
\def\cmss{\hbox{cm.s$^{-2}$}}       
\def\kms{\hbox{km.s$^{-1}$}}       
\def\vsini{\hbox{$\upsilon \sin i_{\star}$}}      
\def\Msun{\hbox{$\mathrm{M}_{\astrosun}$}}             
\def\Rsun{\hbox{$\mathrm{R}_{\astrosun}$}}
\def\Mjup{\hbox{$\mathrm{M}_{\jupiter}$}}
\def\Rjup{\hbox{$\mathrm{R}_{\jupiter}$}}
\def\Mearth{\hbox{$\mathrm{M}_{\earth}$}}

\def\degr{\hbox{$^\circ$}}
\def\teff{T$_{\rm eff}$}
\def\logg{log {\it g}}
\def\met{[Fe/H]}

\def\vmicro{$\upsilon_\mathrm{micro}$}
\def\vmacro{$\upsilon_\mathrm{macro}$}

\def\figw{\columnwidth}

\begin{document}

   \title{SOPHIE velocimetry of \textit{Kepler} transit candidates}

   \subtitle{XII. KOI-1257~b: a highly eccentric three-month period transiting exoplanet\thanks{Based on observations made with SOPHIE on the 1.93 m telescope at Observatoire de Haute-Provence (CNRS), France, and with the Italian Telescopio Nazionale Galileo (TNG) operated on the island of La Palma by the Fundaci\'on Galileo Galilei of the INAF (Istituto Nazionale di Astrofisica) at the Spanish Observatorio del Roque de los Muchachos of the Instituto de Astrofisica de Canarias. Part of the observations were made with the IAC80 operated on the Spanish Observatorio del Teide of the Instituto de Astrof\'isica de Canarias.}}

   \author{A.~Santerne\inst{\ref{caup}}\fnmsep\inst{\ref{IA}}\fnmsep\inst{\ref{lam}}\fnmsep\thanks{\email{alexandre.santerne@astro.up.pt}}
          \and G.~H\'ebrard \inst{\ref{iap}}\fnmsep\inst{\ref{ohp}}
          \and M.~Deleuil \inst{\ref{lam}}
          \and M.~Havel \inst{\ref{oca}}\fnmsep\inst{\ref{caup}}
          \and A.~C.~M.~Correia \inst{\ref{aveiro}}\fnmsep\inst{\ref{imcce}}
          \and J.-M.~Almenara\inst{\ref{lam}}
          \and R.~Alonso \inst{\ref{iac}}\fnmsep\inst{\ref{laguna}}
          \and L.~Arnold \inst{\ref{ohp}}
          \and S.~C.~C.~Barros\inst{\ref{lam}}
          \and R.~Behrend \inst{\ref{geneva}}\fnmsep\inst{\ref{CdRCdL}}
          \and L.~Bernasconi \inst{\ref{engarouines}}
          \and I.~Boisse \inst{\ref{caup}}\fnmsep\inst{\ref{lam}}
          \and A.~S.~Bonomo \inst{\ref{torino}}
          \and F.~Bouchy \inst{\ref{lam}}\fnmsep\inst{\ref{geneva}}
          \and G.~Bruno \inst{\ref{lam}}
          \and C.~Damiani \inst{\ref{lam}}
          \and R.~F.~D\'iaz \inst{\ref{lam}}\fnmsep\inst{\ref{geneva}}
          \and D.~Gravallon \inst{\ref{ohp}}
          \and T.~Guillot \inst{\ref{oca}}
          \and O.~Labrevoir \inst{\ref{labrevoir}}
          \and G.~Montagnier \inst{\ref{iap}}\fnmsep\inst{\ref{ohp}}
          \and C.~Moutou \inst{\ref{lam}}\fnmsep\inst{\ref{cfht}}
         \and C.~Rinner \inst{\ref{moos}}
          \and N.~C.~Santos \inst{\ref{caup}}\fnmsep\inst{\ref{porto}}
          \and L.~Abe \inst{\ref{oca}}
          \and M.~Audejean \inst{\ref{chinon}}
          \and P.~Bendjoya \inst{\ref{oca}}
          \and C.~Gillier \inst{\ref{cala}}
          \and J.~Gregorio \inst{\ref{atalaia}}
          \and P.~Martinez \inst{\ref{adagio}}
          \and J.~Michelet \inst{\ref{cala}}
          \and R.~Montaigut \inst{\ref{cala}}
	 \and R.~Poncy \inst{\ref{poncy}}
         \and J.-P.~Rivet \inst{\ref{oca}}
          \and G.~Rousseau \inst{\ref{CdRCdL}}
          \and R.~Roy \inst{\ref{blauvac}}
          \and O.~Suarez \inst{\ref{oca}}
          \and M.~Vanhuysse \inst{\ref{oversky}}
          \and D.~Verilhac \inst{\ref{verilhac}}
          }

   \institute{
Centro de Astrof\'isica, Universidade do Porto, Rua das Estrelas, 4150-762 Porto, Portugal\label{caup}
\and Instituto de Astrof\'isica e Ci\^{e}ncias do Espa\c co, Universidade do Porto, CAUP, Rua das Estrelas, PT4150-762 Porto, Portugal\label{IA}
\and Aix Marseille Universit\'e, CNRS, LAM (Laboratoire d'Astrophysique de Marseille) UMR 7326, 13388, Marseille, France\label{lam}
\and Institut d'Astrophysique de Paris, UMR7095 CNRS, Universit\'e Pierre \& Marie Curie, 98bis boulevard Arago, 75014 Paris, France\label{iap}
\and Observatoire de Haute-Provence, Universit\'e d'Aix-Marseille \& CNRS, 04870 Saint Michel l'Observatoire, France\label{ohp}
\and Laboratoire Lagrange, UMR7239, Universit\'e de Nice Sophia-Antipolis, CNRS, Observatoire de la Cote d'Azur, F-06300 Nice, France \label{oca}
\and Departamento de F\'isica, I3N, Universidade de Aveiro, Campus de Santiago, 3810-193 Aveiro, Portugal\label{aveiro}
\and Astronomie et Syst\`emes Dynamiques, IMCCE-CNRS UMR8028, 77 Av. Denfert-Rochereau, 75014 Paris, France \label{imcce}
\and Instituto de Astrof\'isica de Canarias (IAC), E-38200 La Laguna, Tenerife, Spain\label{iac}
\and Universidad de La Laguna, Dept. Astrof\'isica, E-38206 La Laguna, Tenerife, Spain\label{laguna}
\and Observatoire Astronomique de l'Universit\'e de Gen\`eve, 51 chemin des Maillettes, 1290 Versoix, Switzerland\label{geneva}
\and CdR \& CdL Group: Lightcurves of Minor Planets and Variable Stars, Switzerland\label{CdRCdL}
\and Observatoire des Engarouines, 1606 chemin de Rigoy, 84570 Mallemort-du-Comtat, France\label{engarouines}
\and INAF -- Osservatorio Astrofisico di Torino, via Osservatorio 20, 10025 Pino Torinese, Italy\label{torino}
\and Centre d'Astronomie, Plateau du Moulin \`a Vent, 04870, St-Michel-l'Observatoire, France\label{labrevoir}
\and CNRS, Canada-France-Hawaii Telescope Corporation, 65-1238 Mamalahoa Hwy., Kamuela, HI 96743, USA\label{cfht}
\and Observatoire Oukaimeden 40273 Oukaimeden, Morocco\label{moos}
\and Departamento de F\'isica e Astronomia, Faculdade de Ci\^encias, Universidade do Porto, Portugal\label{porto}
\and Observatoire de Chinon B92, Mairie de Chinon, 37500 Chinon, France\label{chinon}
\and Club d'Astronomie Lyon Amp\`ere, b\^{a}t. Plan\'etarium, Place de la Nation, 69120 Vaulx-en-Velin, France\label{cala}
\and Atalaia Group -- Crow Observatory, Portalegre, Portugal \label{atalaia} 
\and Observatoire ADAGIO, 31540 B\'elesta, France\label{adagio}
\and 2 rue des \'Ecoles, 34920 Le Cres, France\label{poncy}
\and Observatoire de Blauvac, 293 Ch. de St Guillaume 84570 Blauvac, France \label{blauvac}
\and OverSky, 47 all\'ee des Palanques, BP 12, 33127, Saint-Jean d'Illac, France\label{oversky}
\and Montreurs de Grande Ourse, les Mairies, 26190 St Laurent en Royans, France\label{verilhac}
}

   \date{Received 8 May 2014; accepted 23 June 2014}

 
  \abstract
   {
   In this paper we report a new transiting warm giant planet: KOI-1257~b. It was first detected in photometry as a planet-candidate by the \textit{Kepler} space telescope and then validated thanks to a radial velocity follow-up with the SOPHIE spectrograph. It orbits its host star with a period of 86.647661 d $\pm$ 3 s and a high eccentricity of 0.772 $\pm$ 0.045. 
   The planet transits the main star of a metal-rich, relatively old binary system with stars of mass of 0.99 $\pm$ 0.05 \Msun\, and 0.70 $ \pm $ 0.07 \Msun\, for the primary and secondary, respectively. This binary system is constrained thanks to a self-consistent modelling of the \textit{Kepler} transit light curve, the SOPHIE radial velocities, line bisector and full-width half maximum (FWHM) variations, and the spectral energy distribution. However, future observations are needed to confirm it. 
   The \texttt{PASTIS} fully-Bayesian software was used to validate the nature of the planet and to determine which star of the binary system is the transit host. 
   By accounting for the dilution from the binary both in photometry and in radial velocity, we find that the planet has a mass of 1.45 $ \pm $ 0.35 \Mjup, and a radius of 0.94 $ \pm $ 0.12 \Rjup, and thus a bulk density of 2.1 $ \pm $ 1.2 g.cm$^{-3}$. The planet has an equilibrium temperature of 511 $\pm$ 50 K, making it one of the few known members of the warm-jupiter population. 
   The HARPS-N spectrograph was also used to observe a transit of KOI-1257~b, simultaneously with a joint amateur and professional photometric follow-up, with the aim of constraining the orbital obliquity of the planet. However, the Rossiter-McLaughlin effect was not clearly detected, resulting in poor constraints on the orbital obliquity of the planet.
   }


\maketitle
\section{Introduction}

Transiting giant exoplanets still have many secrets to reveal. Most of the known giant planets with both a measured mass and radius orbit their host stars with periods of a few days (the so-called hot jupiters). This population of planets has been deeply explored thanks to large ground-based photometric surveys such as Super-WASP \citep{2007MNRAS.375..951C} and HAT-Net \citep{2007ApJ...656..552B}. They revealed a large diversity in terms of bulk density and internal structure, from dense giant planets such as, HAT-P-20~b \citep[$\rho_{p} \approx 13.78$ g.cm$^{-3}$ ;][]{2011ApJ...742..116B}, WASP-18~b \citep[$\rho_{p} \approx 10.3$ g.cm$^{-3}$ ;][]{2009Natur.460.1098H}, and HAT-P-2~b \citep[$\rho_{p} \approx 7.29$ g.cm$^{-3}$ ;][]{2010MNRAS.401.2665P}, to low-density, highly-inflated giant planets such as, WASP-17~b \citep[$\rho_{p} \approx 0.19$ g.cm$^{-3}$ ;][]{2010ApJ...709..159A} and HAT-P-32~b \citep[$\rho_{p} \approx 0.14$ g.cm$^{-3}$ ;][]{2011ApJ...742...59H}. \\

Different physical processes might explain this diversity. They were reviewed in \citet{2014arXiv1401.4738B}, and references therein. They are mostly driven by the fact that the planet is very close to its host star. Therefore, the planet receives a strong irradiation and/or an efficient tidal heating (as in the case of Io with Jupiter). However, to further probe and understand those physical processes, it is important to compare the properties of hot jupiters with a population of warm and cool giants. \citet{2011ApJS..197...12D} found a lack of inflated radii for moderate-irradiated giant planet-candidates detected by \textit{Kepler} \citep{2009Sci...325..709B}. However, giant \textit{Kepler} Objects of Interest (KOIs) are known to be biased by a significant rate of false positives \citep{2012A&A...545A..76S}. To constrain the physics of planets, it is important to consider only well-established planets for which the physical parameters have been accurately determined. \\

Giant planets that orbit at larger separation than the hot jupiters are also of great interest to understand the migration processes of planets \citep[e.g.][]{2009A&A...501.1139M, 2009A&A...501.1161M}. Clearly, the hot-jupiter population (with orbital separation of less than $\sim$ 0.1 AU) should have a different or more efficient migration process than the population of cold giants (with orbital separation of more than $\sim$ 1 AU, like Jupiter and Saturn). Between these two populations resides the so-called period-valley planets \citep{2003A&A...407..369U} which might have another process of formation, migration, and dynamical evolution. \\

Radial velocity planet detections have already provided some constraints, as discussed in, e.g. \citet{2013A&A...560A..51A} and \citet{2013ApJ...767L..24D}. However, characterising longer-period giant \textit{transiting} planets would permit to constrain even further those formation and migration processes, by measuring their orbital obliquity through the Rossiter-McLaughlin effect \citep{1924ApJ....60...15R, 1924ApJ....60...22M}. Indeed, it was proposed that measuring the orbital obliquity of transiting planets provides constraints on the migration and dynamical history of those exoplanets \citep[e.g.][]{2005ApJ...631.1215W, 2010A&A...524A..25T, 2014arXiv1403.4095B}.\\

While more than 150 transiting giant planets were already discovered and characterised with orbital periods of less than one month, the number of accurately characterised (with both measured mass and radius) transiting giant planets with orbital periods greater than one month remain rare\footnote{This selection is based on the NASA Exoplanet Archive for planets with a secured mass and radius, a period longer than 30 days, and a constrained mass greater than 0.3\Mjup. We removed from this selection the circumbinary planets.}: HD80606~b \citep[111-day period,][]{2001A&A...375L..27N, 2009A&A...498L...5M, 2010A&A...516A..95H} ; CoRoT-9~b \citep[95-day period,][]{2010Natur.464..384D} ; Kepler-30~c \citep[60-day period,][]{2012ApJ...750..114F} and Kepler-87~b \citep[115-day period,][]{2014A&A...561A.103O}.\\

The characterisation of new longer-period planets is therefore important to further constrain the theories of planet formation, migration, and evolution. However, because of their long orbital periods, these planets are unlikely be detected by a ground-based photometric survey. This requires a high duty-cycle mission, with timespan observations of several months or years, such as \textit{CoRoT} \citep{2006cosp...36.3749B} or \textit{Kepler} \citep{2009Sci...325..709B}. Future space missions like \textit{TESS} \citep{2010AAS...21545006R} and \textit{CHEOPS} \citep{2013EPJWC..4703005B} will target bright stars, but will focus mainly on short-period planets. Only the future space mission \textit{PLATO} \citep{2013arXiv1310.0696R}, expected to be launched in 2024, with observing runs of 2-3 years will be able to detect new warm giant planets orbiting bright stars. Therefore, the population of warm and cool giant planet candidates detected by \textit{Kepler} is the only one known until the next decade. \\

Since only a few giant transiting planets are secured with a period longer than a month, in 2012 we decided to start an independent spectroscopic follow-up of \textit{Kepler} candidates with the SOPHIE spectrograph \citep{2008SPIE.7014E..17P, 2009A&A...505..853B}. This follow-up aims at increasing the statistics of such population of planets by establishing and characterising new longer-period giant transiting planets. We observed all candidates reported by \citet{2013ApJS..204...24B} that respect (1) an orbital period between 25 and $\sim$ 400 days, (2) a transit depth between 0.4\% and 3\%, and (3) a host star brighter than K$_{p}$ = 14.7. These candidates have an estimated equilibrium temperature that range between 250 K and 750 K. This sample of warm giant planet candidates completes the sample of short-period candidates transiting the brightest \textit{Kepler} targets presented in \citet{2012A&A...545A..76S}. The result of this new sample will be presented in a forthcoming paper (Santerne et al., in prep.). \\

In this paper, we report the fifth transiting giant exoplanet with an orbital period longer than one month, whose planetary nature has been established and whose mass has been characterised. The transiting planet candidate KOI-1257.01 has an orbital period of almost 3 months. It was first detected in photometry by the \textit{Kepler} space telescope in the first four months of the mission \citep{2011ApJ...736...19B}. We used the SOPHIE spectrograph of the 1.93 m telescope at Observatoire de Haute-Provence (France) to establish its planetary nature and to measure its mass and eccentricity. We also used the new HARPS-N spectrograph mounted on the 3.6 m Telescopio Nazionale Galileo (TNG) at La Palma (Spain) in an attempt to detect its Rossiter-McLaughlin effect \citep{1924ApJ....60...15R, 1924ApJ....60...22M}. This spectroscopic transit was simultaneously observed in photometry by several professional and amateur telescopes to secure the time of the transit.\\

In Sect. \ref{Obs&Data} we present the observations of KOI-1257 and their data reduction that we analyse in section \ref{Analysis}. In section \ref{Blend}, we perform a blend analysis and statistically validate the planet. In section \ref{SystemProp}, we present the physical properties of the KOI-1257 system and perform a study of the internal structure and dynamical evolution of the planetary system in section \ref{Dynamic}. In section \ref{Discut}, we discuss this system and finally we draw the conclusion of the paper in section \ref{Conclu}.

\section{Observations and data reduction}
\label{Obs&Data}

\subsection{\textit{Kepler} observations}
\label{KepObs}

\begin{figure*}[]
\begin{center}
\includegraphics[width=\textwidth]{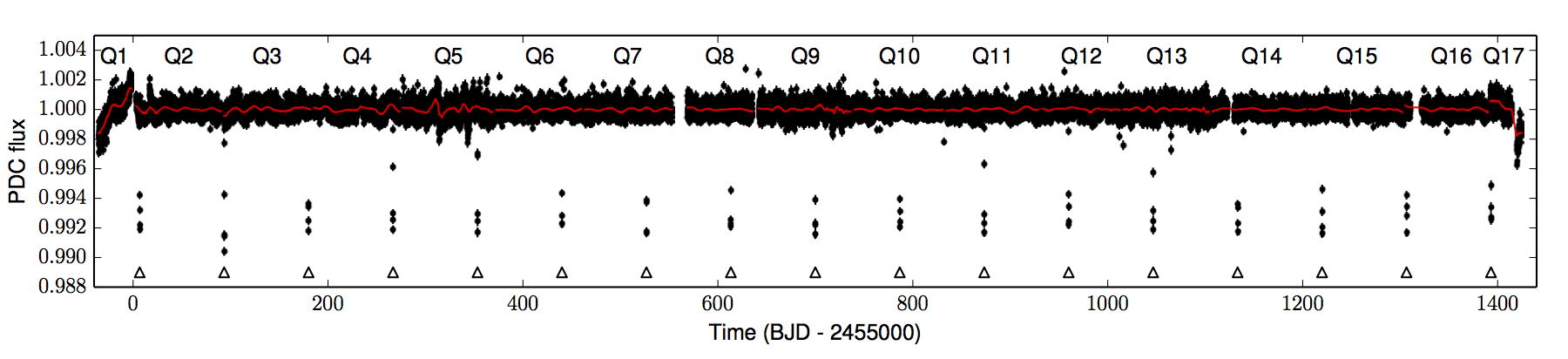}
\caption{The seventeen quarters of the \textit{Kepler} light curve of KOI-1257. The transit times of KOI-1257.01 are highlighted with the white triangles. The red line is a 7-day sliding median of the data.}
\label{KepLC}
\end{center}
\end{figure*}

The target KOI-1257 was observed during the four years of the \textit{Kepler} mission, from 2009 May 13 to 2013 May 11 in long-cadence mode only (29.6 minutes) with a typical photometric precision of $\sim$ 250 ppm per cadence. The various identifiers, coordinates, and magnitudes of KOI-1257 are listed in Table \ref{1257IDs}. Figure \ref{KepLC} displays the Presearch Data Conditioning (PDC) \textit{Kepler} light curve \citep{2010ApJ...713L..87J} as available from the MAST archive\footnote{\url{http://archive.stsci.edu/kepler/data_search/search.php?action=Search&ktc_kepler_id=8751933}}. The photometric data present a periodic transit with a depth of nearly 1\% which is characteristic of a transiting giant planet candidate. Only one transiting candidate has been found so far in this light curve. The transit analysis performed by \citet{2013ApJS..204...24B} shows that the candidate KOI-1257.01 has an orbital period of $\sim$ 86.65 days, a transit depth of 0.7\%, and a transit duration of 4.25 hours. This transit duration is relatively short for this orbital period. Indeed, if the candidate is transiting a solar twin in a circular orbit and a central transit, we might expect a transit duration of about 8 hours \citep[according to Eq. 14 of][]{2010arXiv1001.2010W}. This short transit duration can reveal either a small host star, a grazing transit, and/or an eccentric orbit.\\

\begin{table}[h!]
\caption{KOI-1257 identifiers, coordinates, and magnitudes}
\begin{minipage}[t]{\figw} 
\begin{center}
\begin{tabular*}{8.2cm}{lc}
\hline
\hline
\textit{Kepler} Input Catalog (KIC) & 8751933\\
\textit{Kepler} Object of Interest (KOI) & 1257\\
\textit{Kepler} exoplanet catalogue (Kepler) & 420\\
\hline
Right Ascension (J2000) & 19:24:54.039 \\
Declinaison (J2000) & 44:55:38.57  \\
\hline
\textit{Kepler} magnitude $K_{p}$ & 14.65 ~\tablefootmark{(a)}\\
Johnson B  & 15.830 $\pm$ 0.098 ~\tablefootmark{(b)}\\
Johnson V & 14.867 $\pm$ 0.062 ~\tablefootmark{(b)}\\
Sloan g$'$ & 15.285 $\pm$ 0.079 ~\tablefootmark{(b)}\\
Sloan r$'$ & 14.605 $\pm$ 0.080 ~\tablefootmark{(b)}\\
Sloan i$'$ & 14.344 $\pm$ 0.068 ~\tablefootmark{(b)}\\
2MASS J & 13.222 $\pm$ 0.024 ~\tablefootmark{(c)}\\
2MASS H & 12.827 $\pm$ 0.023 ~\tablefootmark{(c)}\\
2MASS Ks  &  12.726 $\pm$ 0.023 ~\tablefootmark{(c)}\\
WISE W1 & 12.557 $\pm$ 0.024 ~\tablefootmark{(d)}\\
WISE W2 & 12.644 $\pm$ 0.025 ~\tablefootmark{(d)}\\
WISE W3 & 12.034 $\pm$ 0.157 ~\tablefootmark{(d)}\\
\hline
\hline
\end{tabular*}
\end{center}
\vspace{-0.3cm}
\tablefoot{Magnitudes from: \tablefoottext{a}{the \textit{Kepler} Input Catalog \citep{2011AJ....142..112B}, }
\tablefoottext{b}{the APASS catalogue \citep{2009AAS...21440702H}, }
\tablefoottext{c}{the 2MASS catalogue \citep{2006AJ....131.1163S}, }
\tablefoottext{d}{the WISE All-Sky catalogue \citep{2010AJ....140.1868W, 2012yCat.2311....0C}.}}
\end{minipage}
\label{1257IDs}
\end{table}%

To further analyse this system, we corrected the seventeen transits of KOI-1257.01 observed by \textit{Kepler} (see Fig. \ref{KepLC}) by fitting a parabola to the out-of-transit long-cadence raw data (SAP\_FLUX). We removed from the analyses the 12$^\mathrm{th}$ transit (at BJD $\sim$ 2455960) which occurs a few hours after one of the data downlinks of the quarter 12 and is affected by thermal changes of the telescope. The remaining sixteen transits are displayed in Fig. \ref{1257data}, phase-folded, together with the best models as described in section \ref{Analysis}. We did not correct the light curves from the background stellar contamination as this will be taken into account in our analysis (see section \ref{Analysis}). Since we did not detect significant transit time variations (see section \ref{TTVs}) we did not correct the times of the transits. We note that the star is photometrically quiet, no spot modulation is seen in the raw \textit{Kepler} light curve.

\subsection{SOPHIE spectroscopic follow-up}
\label{SOPHIEObs}

We performed a spectroscopic follow-up of the KOI-1257 system with the SOPHIE spectrograph \citep{2008SPIE.7014E..17P, 2009A&A...505..853B} mounted on the 1.93m telescope at Observatoire de Haute-Provence (France). SOPHIE is a fibre-fed high-resolution stable spectrograph dedicated to high-precision radial velocity (RV) measurement. Thanks to an upgrade of the fibre paths in 2011 June, the SOPHIE spectrograph has an improved radial velocity stability over long timespan \citep{2011SPIE.8151E..37P, 2013A&A...549A..49B}. This allows us to explore more reliably the regime of lower mass and a longer period planets. This improved capability motivated us to follow up giant KOIs with much longer orbital periods (between one month and one year) than previously observed \citep{2012A&A...545A..76S}. This led to the characterisation of a few new systems, such as KOI-1257, as well as several false positives (Santerne et al., in prep.).\\

We obtained 28 spectra of KOI-1257 with SOPHIE\footnote{OHP programme IDs: 12A.PNP.MOUT, 12B.PNP.MOUT, 13A.PNP.MOUT and 13B.PNP.HEBR} between 2012 August 17 and 2013 December 01. These spectra were observed with an exposure time ranging from 800s to 3600s (see Table \ref{1257RVtable} in the online material) resulting in a signal-to-noise ratio (S/N) per pixel at 550nm between 5 and 22. All observations were conducted using the high efficiency mode (HE) of SOPHIE, which has a resolution of $\sim$ 39000 at 550nm, and the slow read-out mode of the charge-coupled device (CCD). We reduced the spectra using the online pipeline and derived the RVs by computing the weighted cross-correlation function (hereafter CCF) of the observed spectrum with a numerical mask of a G2V star as described by \citet{1996A&AS..119..373B} and \citet{2002A&A...388..632P}. This mask corresponds to the same spectral type as the target. Radial velocity uncertainties were estimated as described in \citet{2001A&A...374..733B}. From the observed CCF, we also measured the line asymmetry diagnosis V$_\mathrm{span}$ as proposed by \citet{2011A&A...528A...4B} and the full width half maximum (FWHM), for which we estimated the uncertainties as two and four times the uncertainties of the radial velocities, respectively.\\

Observations of faint stars such as KOI-1257 in the presence of the Moon might result in a systematic shift in the measure of the radial velocity. This effect might be strong when the spectrum of the star and that of the Sun reflected by the Moon and scattered by the atmosphere are blended \citep{1996A&AS..119..373B}. To correct this effect, we followed the procedure described in \citet{1996A&AS..119..373B} and \citet{2010A&A...520A..65B}, using the second fibre of the HE mode. This fibre observes the sky scattered light simultaneously with the target star. The observations corrected from the Moon background light are flagged and their correction is given in the online table \ref{1257RVtable}.\\

The CCD of the SOPHIE spectrograph suffers from charge transfer inefficiency (CTI) that affects radial velocities measured on spectra with different S/N ratios \citep{2009EAS....37..247B}. To correct this systematic effect, we followed the procedure described in \citet{2012A&A...545A..76S}. The amplitude of the CTI correction ranges between $\sim$ 30 \ms\, and 140 \ms.\\

Even if the SOPHIE upgrade improves the scrambling of the light within the fibres resulting in a better radial velocity stability, the HE mode of the spectrograph still suffers from long-term instrumental variations at the level of $\sim$ 10\ms that are not well understood . To account for any instrumental instability of the spectrograph, our strategy was to observe each night the star HD185144 ($\sigma$ Dra) which was measured to be stable at the level of a few \ms\, by \citet{2010Sci...330..653H} and \citet{2013A&A...549A..49B}. This constant star is extremely bright (V$\sim$4.7) and close to the \textit{Kepler} field of view. We corrected each radial velocity measurement of KOI-1257 by interpolating the RV variations of HD185144. Thanks to the high stability of HD185144 and the accuracy of our measurements (our mean uncertainty is below 1\ms\, for this target) this correction does not affect significantly the uncertainties of KOI-1257, which have a median of 22\ms. More details about this correction are provided in Appendix \ref{constcorr}. \\

Among the 28 observed spectra of KOI-1257 with SOPHIE, two of them have a low S/N ($\sim$ 6). The first was observed with an insufficient exposure time of $\sim$ 800s and the second under poor weather conditions. The measurement of precise radial velocities on such spectra is difficult since several systematics might affect the data without being accounted for. For example, the accuracy of the correction of the CTI proposed by \citet{2012A&A...545A..76S} is unknown below a S/N (at 550 nm) of 10. We therefore decided to remove these two points from the analysis performed in section \ref{Analysis}. We removed another measurement which also has a relatively low S/N (10.7) and was observed in the presence of the Moon. The real S/N from the target spectrum only is therefore lower. The impact of the Moon scattered light, estimated by computing the RV difference between the corrected and the non-corrected measurement, is of 130 \ms\, for this observation, which is larger than the amplitude of the planet we want to detect. The remaining radial velocities we used in the analysis are displayed in Fig. \ref{1257data} together with the best model described in section \ref{Analysis}. They are listed in the online table \ref{1257RVtable}. The radial velocity data clearly shows a long-term drift revealing an outer companion in the system, in addition to the transiting candidate.

\subsection{HARPS-N spectroscopic follow-up}
\label{HARPSNObs}

We observed the spectroscopic transit of KOI-1257 with the new HARPS-N spectrograph mounted at the 3.6 m Telescopio Nazionale Galileo (TNG) located on the island of La Palma (Spain). HARPS-N \citep{2012SPIE.8446E..1VC} is an improved copy of the HARPS spectrograph on the ESO 3.6 m telescope at La Silla Observatory (Chile). HARPS-N was commissioned in early 2012 with one of its main goals being to follow up \textit{Kepler} objects of interest. HARPS-N has already characterised two transiting giant exoplanets in synergy with SOPHIE \citep[KOI-200~b and KOI-889~b:][]{2013A&A...554A.114H}, as well as the transiting rocky planet Kepler-78~b \citep{2013ApJ...774...54S, 2013Natur.503..377P}.\\

\begin{figure*}[t!]
\begin{center}
\begin{tabular}{cc}
\includegraphics[width=\figw]{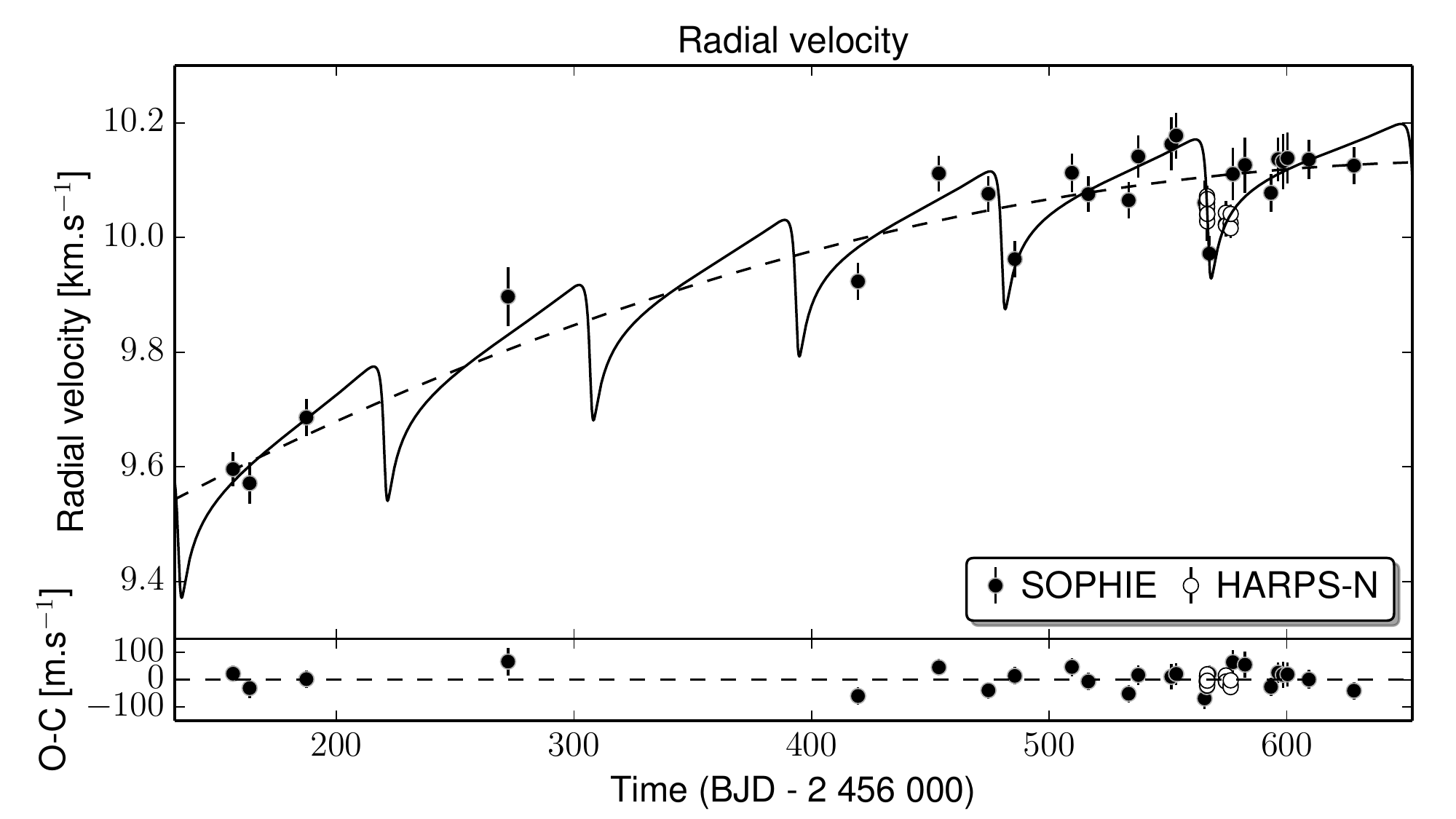} & \includegraphics[width=\figw]{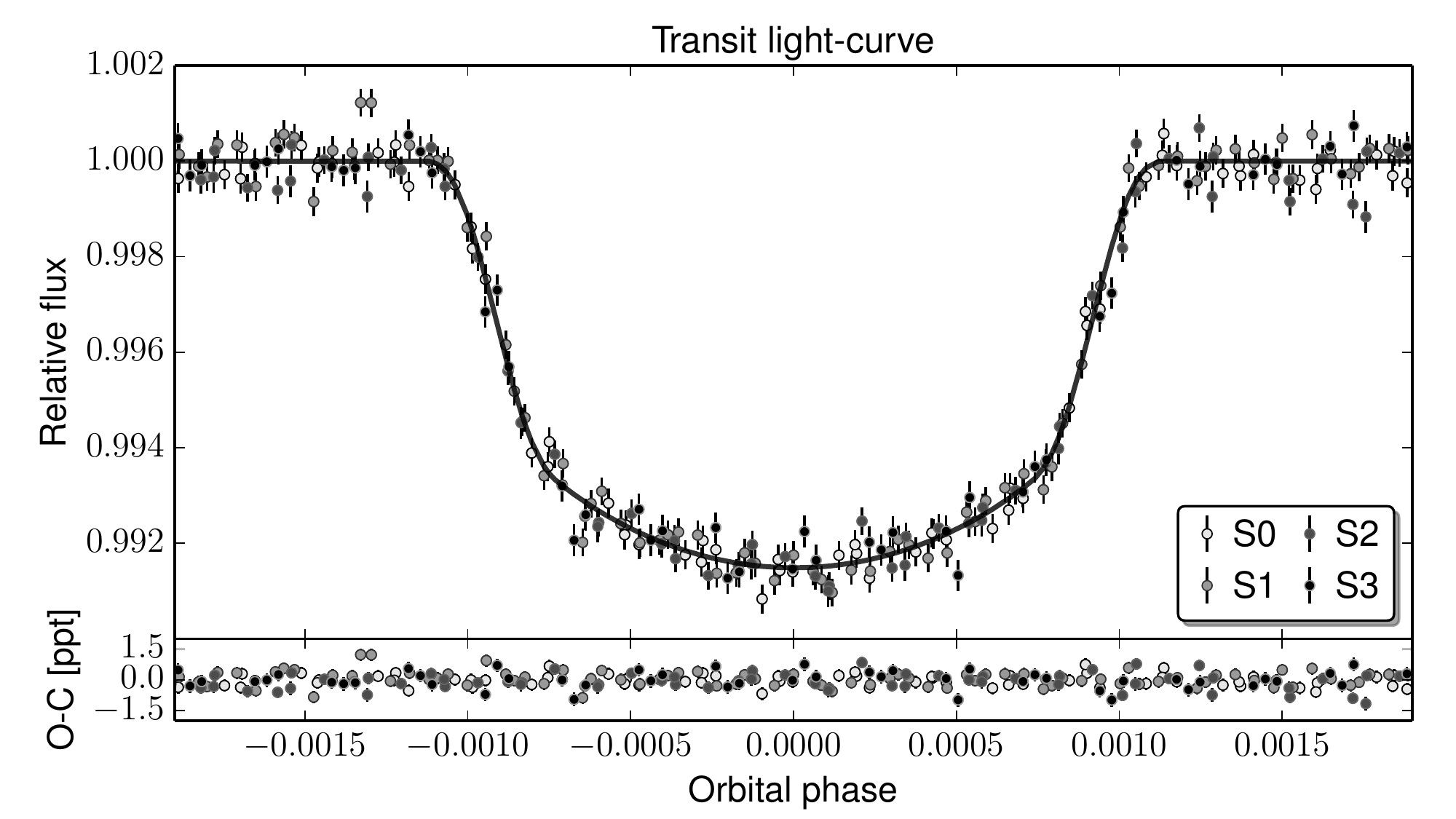}\\
\includegraphics[width=\figw]{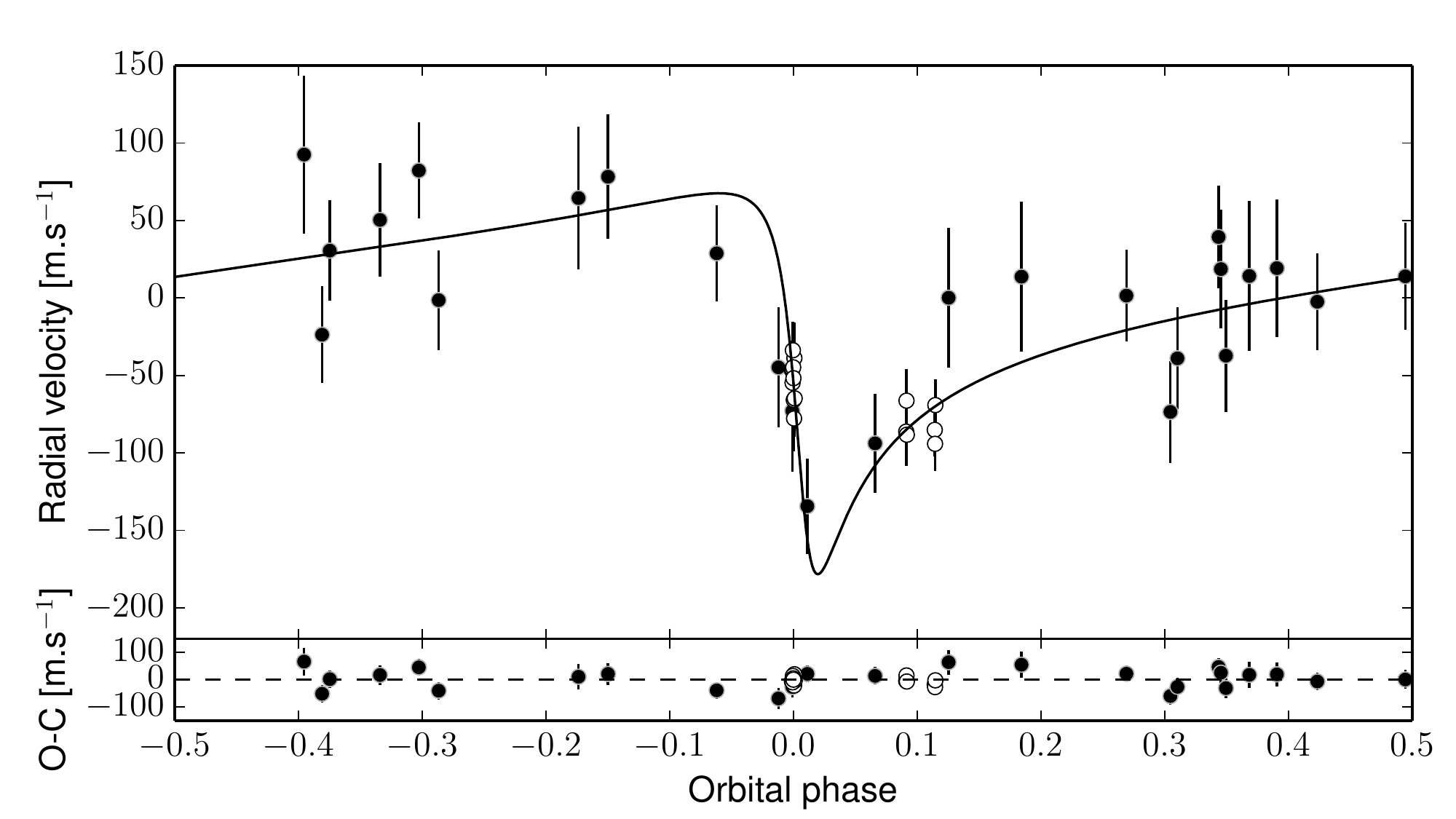} & \includegraphics[width=\figw]{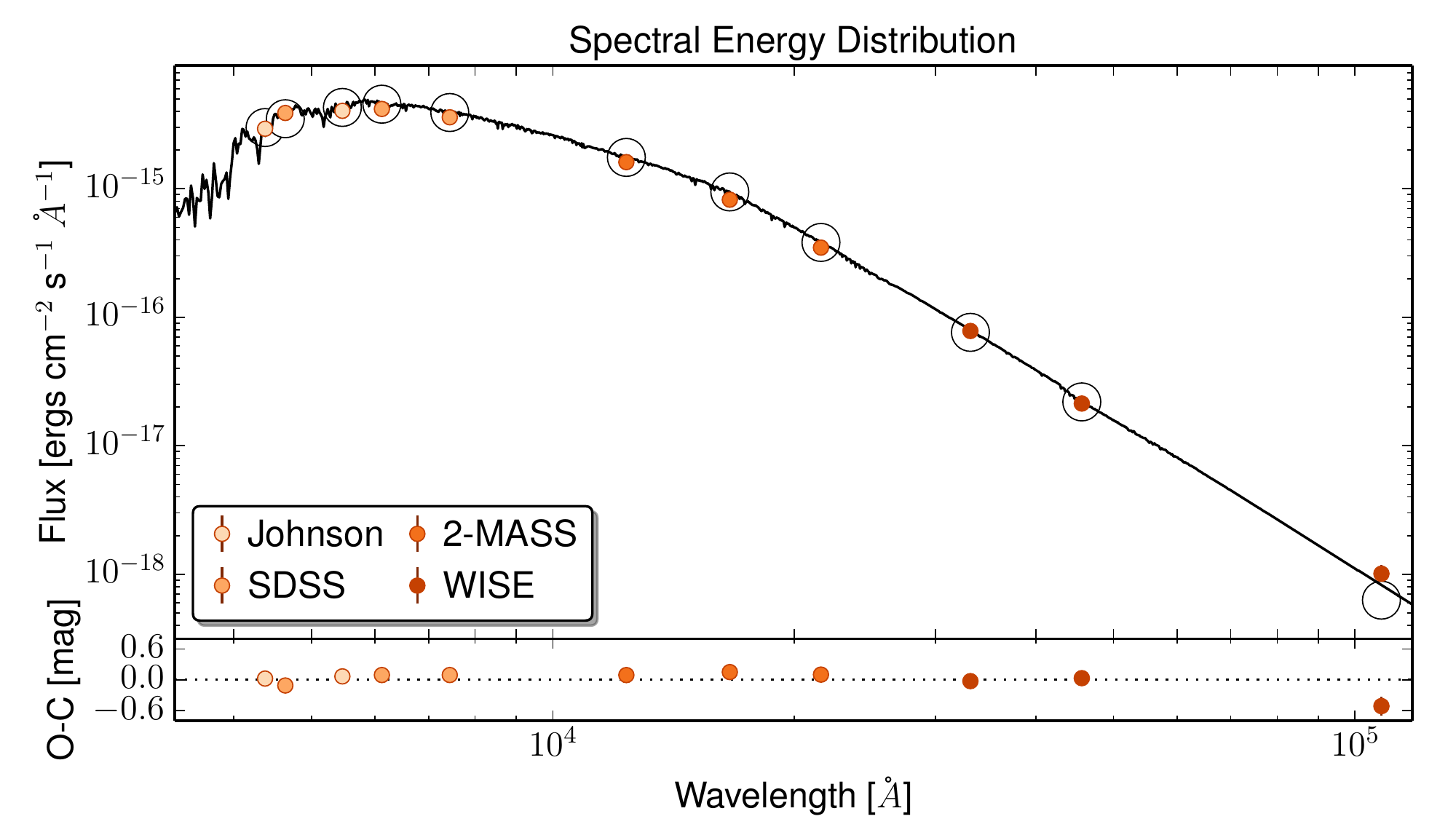}
\end{tabular}
\caption{\textit{Top-left panel:} SOPHIE and HARPS-N time series superimposed with the best-fit model of the Keplerian orbit of KOI-1257.01 and the quadratic drift of the outer companion (dashed line). The best-fit residuals are shown. \textit{Bottom-left panel:} Phase-folded radial velocities from SOPHIE and HARPS-N after removing the quadratic drift. The best Keplerian model and its residuals are also displayed. \textit{Top-right panel:} Phase-folded transit light curve of KOI-1257.01 as observed during the four years of the \textit{Kepler} mission. The best-fit model is superimposed to the data and the residuals from the best-fit model (in parts per thousand -- ppt) are also displayed. The four different seasons of the \textit{Kepler} spacecraft have different marks (see legend). \textit{Bottom-right panel:} Spectral energy distribution of KOI-1257 as shown in Table \ref{1257IDs}. The best stellar atmosphere model from the BT-SETTL library is also shown here together with the residuals. The best displayed models are those from Model C described in section \ref{ModelC}. The error bars displayed in these plots were increased quadratically by the jitter value fitted in the analyses.}
\label{1257data}
\end{center}
\end{figure*}

We obtained ten spectra of KOI-1257 with HARPS-N\footnote{TNG programme ID: A28DDT2} on the transit night 2013 September 30 (transit epoch of BJD=2456566.45273), and three additional spectra on each night of 2013 October 8 and 2013 October 10. We used HARPS-N in its normal mode with a fast read-out mode of the CCD. Exposure times were set to 1800s which result in S/N ranging from 8.2 to 10.6 at 550nm. HARPS-N radial velocities are also affected by the CTI effect, but with an amplitude that is at least five times smaller than for SOPHIE (Lovis, private comm.). Given the range of S/N of our HARPS-N observations and our photon noise level, this effect is negligible. Observations were reduced with the online pipeline using a G2V mask in the same way as for the SOPHIE observations. The corresponding radial velocity measurements present uncertainties ranging from 13.6 \ms\, to 23.5 \ms. They are listed in the online table \ref{1257RVtable} and are displayed in Fig. \ref{1257data}.\\

\subsection{Joint professional and amateur photometric follow-up}
\label{ProAmObs}

The analysis of the Rossiter-McLaughlin effect requires accurate times on the transit ephemeris, especially for low-S/N observations \citep{2011A&A...533A.113M, 2013arXiv1305.3647M}. The \textit{Kepler} spacecraft was not in operation during the transit night of 2013 September 30 because of the reaction-wheel failure that happened in 2013 May. Therefore, we called for amateur photometric observations to detect the transit of KOI-1257.01 on 2013 September 30. Nearly 20 amateurs replied to our call, with observatories located in France, Belgium, Portugal and Morocco. We also obtained observing time on the professional OHP-T120 telescope \citep{2009A&A...498L...5M, 2011A&A...533A.113M} in the Observatoire de Haute-Provence (France) and on the IAC80 telescope \citep{2009A&A...506..343D} in the Observatorio del Teide (Spain). However, because of the bad weather conditions in the South-West of Europe during the night the transit occurred, only five observatories were able to detect the relatively small transit of KOI-1257.01 (about 7mmag), in spite of the faintness of the target (V$\sim$14.9). The list and characteristics of the professional and amateur observatories that detected the transit of KOI-1257.01 on the night 2013 September 30 are listed in the online table \ref{ListPhot}. \\

\begin{table}[h!]
\caption{List of ground-based observatories that observed and detected the transit of KOI-1257.01 on the night of 2013 September 30.}
\begin{center}
\setlength{\tabcolsep}{4pt}
\begin{tabular}{lcccc}
\hline
\hline
Observatory & IAU code & Aperture & Focal ratio & Filter \\
\hline
OHP-T120 & 511 & 1.2m & f/5 & r$'$\\
IAC80 & 954 & 80cm & f/11.3 & clear\\
ROTAT & 511 & 60cm & f/3.2 & R\\
MOOS & J43 & 50cm & f/3 & clear\\
Engarouines & A14 & 50cm & f/3 & R\\
\hline
\hline
\end{tabular}
\end{center}
\label{ListPhot}
\end{table}%

The photometry was extracted by performing an aperture photometry relatively to the neighbour stars. The time of each exposure was then converted in Barycentric Dynamical Time (BJD$_\mathrm{TDB}$), so that this set of data could be analysed simultaneously with the \textit{Kepler} data. For this correction, we used the online tool\footnote{\url{http://astroutils.astronomy.ohio-state.edu/time/utc2bjd.html}} provided by \citet{2010PASP..122..935E}. The data are shown in the online table \ref{PhotData} and are displayed in Fig. \ref{1257ProAm}.

\section{Data analysis}
\label{Analysis}

We performed different analyses of all the available data that we present below. First, we present in section \ref{SpA} an analysis of the observed HARPS-N spectra to derive the stellar atmosphere parameters. Then, we performed several analyses of the \textit{Kepler} transit light curves, the SOPHIE and HARPS-N radial velocities, and the spectral energy distribution that we named models A to E as follows:
\begin{itemize}
\item Model A: analysis of the \textit{Kepler} transit light curve alone, using the constraints from the spectral analysis (Section \ref{ModelA}).
\item Model B: analysis of the SOPHIE and HARPS-N radial velocities alone, using the constraints from the transit ephemeris (Section \ref{ModelB}).
\item Model C: combined analysis of the SOPHIE, and HARPS-N radial velocities (without the Rossiter-McLaughlin effect), the \textit{Kepler} transit light curve and the spectral energy distribution, \textit{using} the constraints from the spectral analysis (Section \ref{ModelC}).
\item Model D: combined analysis of the SOPHIE, and HARPS-N radial velocities (without the Rossiter-McLaughlin effect) and the \textit{Kepler} transit light curve, \textit{without} the constraints from the spectral analysis (Section \ref{ModelC}).
\item Model E: combined analysis of the SOPHIE, and HARPS-N radial velocities (including the Rossiter-McLaughlin effect), the \textit{Kepler} transit light curve, the ground-based photometry, and the spectral energy distribution, using the constraints from the spectral analysis (Section \ref{ModelE}).
\end{itemize}
Models B to E include both a Keplerian orbit and a quadratic drift to fit the SOPHIE and HARPS-N radial velocities.\\

The constraints on the eccentricity independently derived by Models A and B are then compared in section \ref{photecc}. Model C is a stellar-model-dependent combined analysis of the photometric and spectroscopic data while Model D is an analysis without use of stellar models which aims at independently confirming the results obtained in Model C. Model E is the analysis that accounts for all the data to constrain the orbital obliquity of the planet.\\

Finally, we performed another two analyses: the first one to provide some constraints on the outer companion that imprints a radial velocity drift in the SOPHIE data (Section \ref{OuterCompanion}), and the second to search for transit time variations (Section \ref{TTVs}) that might be caused by this outer companion.

\subsection{\texttt{PASTIS} Bayesian analyses}
\label{PASTIS}
For the analyses we present in Sects. \ref{ModelA} -- \ref{OuterCompanion} and \ref{Scenario0} -- \ref{Scenario3}, we used the \texttt{PASTIS} fully-Bayesian software, which is described in detail in \citet{2014arXiv1403.6725D} and references therein. It uses a Markov Chain Monte Carlo (MCMC) method to sample the posterior distribution of the parameters. The posterior distribution is described as

\begin{equation}
\mathcal{P}\left(\theta\, \Big\vert\, \mathcal{D}, \mathcal{M}, \mathcal{I}\right) = \frac{\pi\left(\theta\, \Big\vert\,\mathcal{M}, \mathcal{I}\right)\cdot\mathcal{P}\left(\mathcal{D}\, \Big\vert\,\theta,\mathcal{M},\mathcal{I}\right)}{\mathcal{P}\left(\mathcal{D}\, \Big\vert\,\mathcal{M}, \mathcal{I}\right)}\, ,
\end{equation}
where $\mathcal{P}\left(\theta\, \Big\vert\, \mathcal{D}, \mathcal{M}, \mathcal{I}\right)$ represents the probability of the parameters $\theta$ given the available data $\mathcal{D}$, the assumed model $\mathcal{M}$, and available information $\mathcal{I}$. The symbol $\pi$ represents the a priori probability. The term $\mathcal{P}\left(\mathcal{D}\, |\,\mathcal{M}, \mathcal{I}\right)$ is the marginalised likelihood (also called the evidence) which does not depend on $\theta$. Thus this term is just a normalisation factor here. \\ 

The complete lists of priors used for these Bayesian analyses are provided in the online tables \ref{PriorModel} and \ref{ScenarioPriors} for analyses of Sects. \ref{Analysis} and \ref{Blend}, respectively. We assumed that the distribution of the errors follows a Normal distribution and we used a likelihood function $\mathcal{L}$ with the form
\begin{equation}
\mathcal{P}\left(\mathcal{D}\, \Big\vert\,\theta,\mathcal{M},\mathcal{I}\right) = \mathcal{L} = \prod_{i=0}^{n}\frac{1}{\sqrt{2\pi}\sigma_{i}}\exp\left[-\frac{1}{2}\cdot\frac{\left(x_{i} - \mu_{i}\left(\theta\right)\right)^{2}}{\sigma_{i}^{2}}\right]\, ,
\end{equation}
where $x_{i}$ and $\sigma_{i}$ represent the data point and its 63.8\% uncertainty, respectively, as $\mathcal{D}$:$\left\{[x_{0}, \sigma_{0}], [x_{1}, \sigma_{1}], \dots, [x_{n},\sigma_{n}]\right\}$, and $\mu_{i}$ is the corresponding model value. By doing this, we assume here that all the measurements are independent from each other, within each dataset and between the various datasets. For most of the analyses, we ran between 20 and 40 independent chains of 10$^{6}$ iterations each, randomly started from the joint prior distribution. For the parameters that were already constrained in previous analyses, we decided to start those parameters from their median value in order to speed up the convergence burn-in of the chain. This is especially true for models C, D, and E.\\

We analysed the resulting chains by first rejecting all the chains that did not converge significantly to the same maximum of posterior and removed the burn-in phase of each converged chain. Then, we computed the autocorrelation function for each parameter of each chain. We evaluate the correlation length of each parameter and each chain when the value of autocorrelation function drops below 1/e. We use this correlation length to thin each chain with the maximum of correlation length among all the parameters \citep[e.g.][]{2004PhRvD..69j3501T}. Each thinned chain is then expected to be composed of independent samples of the posterior distribution. Finally, we merged all the thinned chains together and we made sure that we had a minimum of 1000 independent samples to derive the median values of the parameters and their 68.3\% confidence interval. We give these values in the online tables \ref{ModelResult} and \ref{ScenarioResult} for the analyses performed in sections \ref{Analysis} and \ref{Blend}, respectively. If this threshold of 1000 independent samples was not reached, we re-ran new chains until it was reached.

\subsection{Spectral analysis}
\label{SpA}

To perform the spectroscopic analysis (SpA) of the host star, the HARPS-N spectra were co-added after correcting from the velocity variation of the target and the barycentric radial velocity of the Earth. It results in a co-added spectrum with a S/N measured per element of resolution of 270 at 560nm in the continuum. We chose to carry out the spectral analysis on this spectrum because of its good quality and the much higher spectral resolution of HARPS-N (R $\simeq$ 110 000) compared to SOPHIE (R $\simeq$ 39 000). The spectroscopic analysis was performed using the semi-automated package VWA  \citep[e.g.][]{2010MNRAS.405.1907B} on this co-added spectrum, normalised and with all the orders concatenated in a single master spectrum. The method, described in detail by \citet{2010A&A...519A..51B}, consists in minimising the correlations of the \ion{Fe}{i} abundance with both equivalent width and excitation potential to derive the atmospheric parameters : the effective temperature \teff, the surface gravity \logg, and the microturbulence velocity.  We checked that the surface gravity derived from the \ion{Fe}{i}  and \ion{Fe}{ii} agrees with the estimate obtained from the pressure-sensitive lines: the \ion{Mg}{1}b and \ion{Ca}{i} lines at 6122\AA\ and 6262\AA, respectively.\\

We obtained an effective temperature \teff\, of 5540 $\pm$ 90 K, a surface gravity \logg\, of 4.30 $\pm$ 0.15 \cmss, a iron abundance \met\, of 0.26 $\pm$ 0.10 dex, and a sky-projected rotational velocity \vsini\, of 4 $\pm$ 2 \kms. We also derived a microturbulence velocity \vmicro\, of 0.80 \kms\, and a macroturbulence velocity \vmacro\, of 1.7 \kms. We did not detect emission features in the \ion{Ca}{II} H and K lines that would indicate chromospheric activity, which is compatible with the absence of photometric modulation. We also did not detect the \ion{Li}{i} doublet at 6707\AA\, in both the HARPS-N and SOPHIE spectra, which is consistent with an age of several Gyr and the quite slow rotation of the star.\\

We also derived the atmospheric properties of the observed star using the procedure described in \citet{2013A&A...556A.150S} and references therein. We measured the equivalent width on the co-added HARPS-N spectrum of 193 and 22 \ion{Fe}{i} and \ion{Fe}{ii} weak lines, respectively \citep{2013A&A...555A.150T}, by imposing excitation and ionisation equilibrium assuming local thermal equilibrium. We found \teff\, = 5528 $\pm$ 54 K, \logg\, = 4.10 $\pm$ 0.11 \cmss, \met\, = 0.22 $\pm$ 0.04 dex and \vmicro\, = 1.01 $\pm$ 0.07 \kms. These results are fully compatible with those derived by VWA. We adopted the parameters derived by the VWA analysis for homogeneity with our previous results \citep[e.g.][]{2011A&A...528A..63S, 2011A&A...536A..70S, 2013A&A...554A.114H, 2014A&A...561L...1B, 2014arXiv1401.6811D}.

\subsection{Model A: light-curve analysis}
\label{ModelA}

We analysed the \textit{Kepler} transit light curve of the sixteen transits that were reduced as described in section \ref{KepObs}. We modelled the light curve using the \texttt{EBOP} code \citep{1972ApJ...174..617N, 1981psbs.conf..111E, 1981AJ.....86..102P} extracted from the \texttt{JKTEBOP} package \citep{2008MNRAS.386.1644S}. The model is described with 23 free parameters: the orbital period $P$, the epoch of transit $T_{0}$, the orbital inclination $i_{p}$, the orbital eccentricity $e_{p}$, the argument of periastron $\omega_{p}$, the radius ratio $r_{p}/R_{\star}$, the linear and quadratic limb darkening coefficients $u_{a}$ and $u_{b}$ (respectively), the stellar effective temperature \teff, the iron abundance \met, and the stellar density $\rho_\star$, as well as the contamination, the flux out-of-transit, and an extra source of white noise (jitter) for each of the four seasons of the \textit{Kepler} data as already done in \citet{2013A&A...554A.114H}. We used the Dartmouth evolutionary tracks \citep{2008ApJS..178...89D} to estimate the stellar parameters. We used non-informative priors (uniform and Jeffreys distributions) for the parameters, except for $P$ and $T_{0}$ for which we used the ephemeris of \citet{2013ApJS..204...24B} as Normal distribution with a width increased by 2 orders of magnitude to avoid biasing the results. For the parameters \teff, \met, and $\rho_{\star}$, we used as prior the results of the spectral analysis described in section \ref{SpA} after converting the \logg\, into $\rho_{\star}$ using the same tracks, which gives $\rho_{\star}$ = 0.33$^{_{+0.36}}_{^{-0.10}}~\rho_{\odot}$. The exhaustive list of free parameters and their a priori distribution is provided in the online table \ref{PriorModel}. To account for the long cadence of the \textit{Kepler} data, we oversampled the light-curve model by a factor of 10 before binning them and computing the likelihood, as recommended by \citet{2010MNRAS.408.1758K}.\\

We ran 40 MCMC chains and analysed them as described in section \ref{PASTIS}. We derived the median values and the 68.3\% confidence intervals which are given in the online table \ref{ModelResult}. We find that the transit ephemeris and the transit depth are fully compatible with the ones derived by \citet{2013ApJS..204...24B}. The fitted values for the contamination are also in agreement with the expected ones provided in the MAST archive for the four seasons. We note that the analysis converged toward a high-eccentric orbit ($e = 0.77 \pm 0.08$) in order to explain the relatively short transit duration in spite of the long-orbital period of the planet-candidate and the low density of the star.

\subsection{Model B: radial velocity analysis}
\label{ModelB}

We analysed the SOPHIE and HARPS-N data by modelling a Keplerian orbit with a quadratic drift component (see Fig. \ref{1257data}) using the \texttt{PASTIS} software. We also tested a linear drift, but this model is not able to fit the data. The model is described by two fixed parameters ($P$, $T_{0}$ which are set to the median values found in the Model A) and nine free parameters: $e_{p}$, $\omega_{p}$, the systemic radial velocity $\gamma$, the radial velocity semi-amplitude $K$, the linear and quadratic terms $d_{1}$ and $d_{2}$, respectively, as well as an extra source of white noise (jitter) for both SOPHIE and HARPS-N and a RV offset between the two instruments. The time reference for the drift was set to the day of 2013 September 30 (BJD = 2456566.0) where the Rossiter-McLaughlin effect was observed. All the free parameters used in the analysis and their a priori distribution are provided in the online table \ref{PriorModel}. In this model, we did not model the Rossiter-McLaughlin effect since this will be done in Model E (see Section \ref{ModelE}). However, we used the HARPS-N observations obtained during the transit. Since the Rossiter-McLaughlin effect is not clearly detected (see Section \ref{ModelE}), we do not expect these data to significantly bias the results.\\

We ran another 40 MCMC chains and analysed them as presented in section \ref{PASTIS}. The median values of the fitted parameters and their 68.3\% confidence intervals are given in online table \ref{ModelResult}. We find in this analysis that the orbit of the transiting candidate is highly eccentric ($0.72^{_{+0.05}}_{^{-0.10}}$). However, low-eccentricity values and even circular orbits cannot be rejected within 99.7\% of confidence based on this analysis. Because of this uncertainty on the eccentricity, the amplitude of the radial velocity is detected with a significance of only 98.6\% in this model (K = 86$^{_{+20}}_{^{-35}}$ \ms).

\subsection{The photoeccentric effect}
\label{photecc}

\begin{figure}[h!]
\begin{center}
\includegraphics[width=\figw]{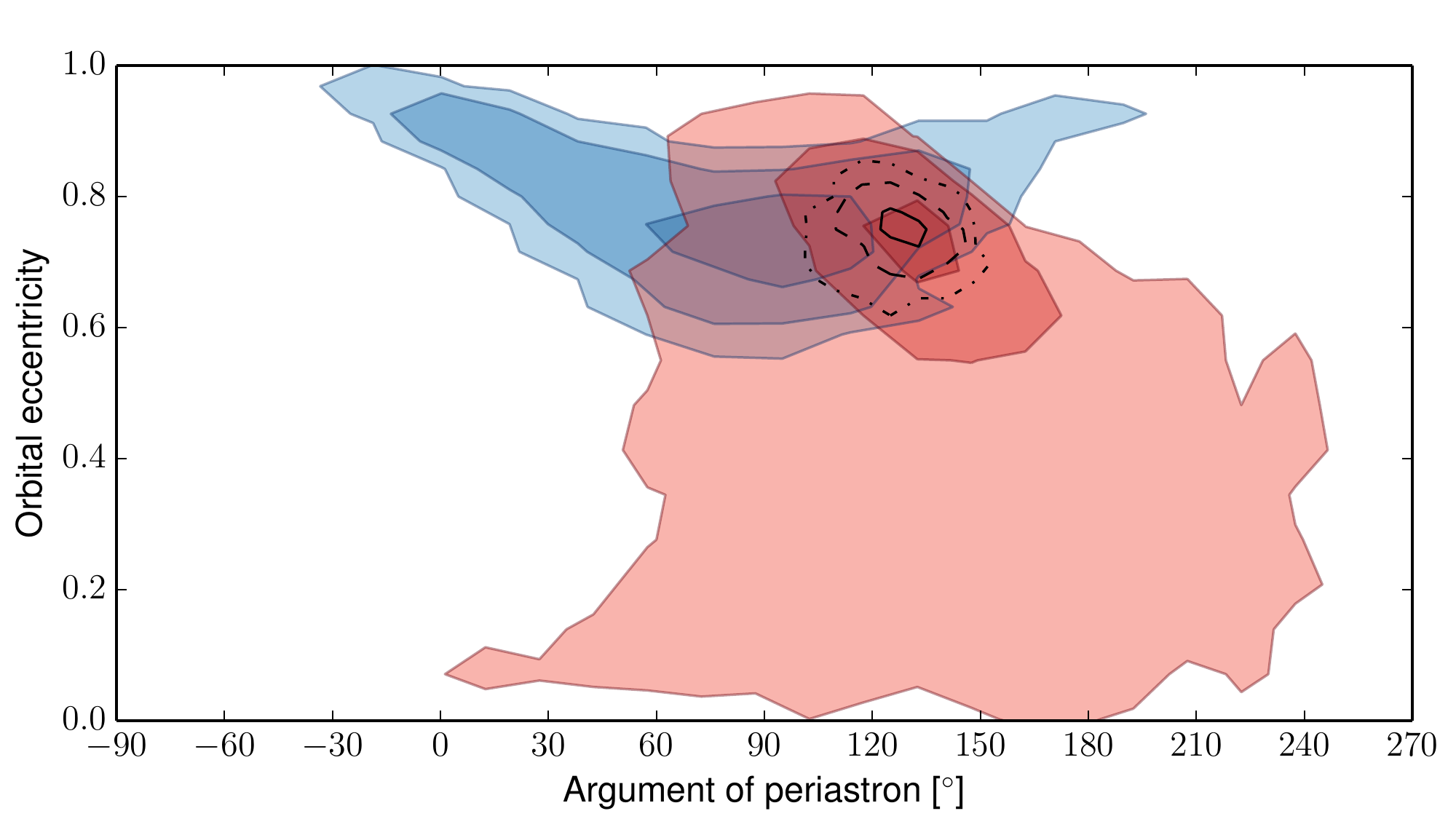}
\caption{Posterior distribution of the orbital eccentricity versus the argument of periastron as constrained independently by the \textit{Kepler} transit light curve (model A, in blue), the radial velocities (model B, in red), and the combined analysis (model C, with the black contours). The regions represent the 68.3\%, 95.5\%, and 99.7\% confidence intervals.} 
\label{PhotEccEffect}
\end{center}
\end{figure}

The analysis described in Sects. \ref{ModelA} and \ref{ModelB} model independently the \textit{Kepler} light curve on the one hand and the SOPHIE and HARPS-N radial velocities on the other hand. They both provide a constraint on the orbital eccentricity. While the eccentricity is naturally constrained by the Keplerian solution in the radial velocities (model B), the eccentricity is constrained in model A thanks to the prior on the stellar density ($\rho_{\star}$ = 0.33$^{_{+0.36}}_{^{-0.10}}~\rho_{\odot}$), derived by the spectral analysis, through the transit duration assuming the third law of Kepler and the orbital period. The low density of the host star cannot explain the relatively short transit duration of KOI-1257.01 already noticed in section \ref{KepObs}. Since the impact parameter found is small ($b = 0.11 \pm 0.11$), a grazing transit cannot explain the short transit duration either. Therefore, the orbit needs to have a high eccentricity in order to explain the observed transit. This effect was presented by \citet{2008ApJ...678.1407F}. It was named ``the photoeccentric effect'' and was illustrated in \citet{2012ApJ...756..122D} and \citet{2012ApJ...761..163D}. In Fig. \ref{PhotEccEffect}, we superimposed the posterior distribution of the orbital eccentricity as a function of the argument of periastron derived by both the analysis of the light curve only (model A) and of the radial velocity data only (model B). The two distributions are fully compatible with each other and both confirm that the orbit of KOI-1257.01 is highly eccentric. We also show in Fig. \ref{PhotEccEffect} the posterior distribution from the combined analysis (model C, see section \ref{ModelC}) for comparison.

\subsection{Model C and D: combined analysis of the light curve and radial velocities}
\label{ModelC}

In this section we analyse simultaneously the \textit{Kepler} transit light curve as in Model A, the SOPHIE and HARPS-N radial velocities as in Model B, and the spectral energy distribution (SED). The bandpasses, magnitudes and their uncertainties are given in Table \ref{1257IDs}. We modelled the SED using the PHOENIX/BT-SETTL synthetic spectral library \citep{2012IAUS..282..235A}. The model uses the 30 free parameters already described in sections \ref{ModelA} and \ref{ModelB} and three new free parameters relative to the fit of the SED: the distance of the system from the Earth $D$, the interstellar extinction E(B$-$V), and an extra source of white noise (jitter) for the observed magnitudes. We chose an non-informative prior for those new parameters (see Table \ref{PriorModel}, online, for the list of free parameters and their prior distribution). As for model B, we did not model the Rossiter-McLaughlin effect in this analysis, but we used the HARPS-N observations obtained during the transit.\\

We ran 20 new MCMC chains and analysed them following the description presented in section \ref{PASTIS}. We show in Fig. \ref{1257data} the phase-folded \textit{Kepler} transit light curve, the time series and phase-folded SOPHIE and HARPS-N data, and the SED of KOI-1257 together with the best-fit model and its residuals. We derived the median and 63.8\% confidence intervals for the parameters that we present in the online table \ref{ModelResult}. All parameters are compatible within 63.8\% with the parameters derived independently in models A and B. The derived uncertainties on the parameters are all equivalent to or smaller than the ones for the independent analysis. We note that the fitted value of the interstellar extinction E(B$-$V) = 0.16 $\pm$ 0.04 is fully compatible with the expected value of 0.175 for the coordinates and the distance of KOI-1257 from the Galactic three-dimensional extinction model of \citet{2005AJ....130..659A}. The fitted values of the limb-darkening coefficients ($u_{a}$ = 0.57 $\pm$ 0.07, $u_{b}$ = 0.00 $\pm$ 0.17) are compatible with those expected for the KOI-1257 host star and the \textit{Kepler} bandpass from \citet{2011A&A...529A..75C}: $u_{a_{t}}$ = 0.463 $\pm$ 0.022, $u_{b_{t}}$ = 0.230 $\pm$ 0.013.\\

As shown in Section \ref{photecc}, low-eccentric orbits are rejected by the transit light curve while this is not the case from the analysis of the radial velocity alone (see Section \ref{ModelB}). By combining the radial velocities and the transit light curve, the degeneracy between the eccentricity and the radial velocity amplitude is lifted. In this combined analysis, the radial velocity amplitude (K = 94 $\pm$ 21 \ms) is detected with a confidence of 99.9992\%.\\

To test the dependence of our results on the stellar evolution tracks, we performed a combined analysis of both the \textit{Kepler} transit light curves and the SOPHIE and HARPS-N velocimetry. However, in comparison with Model C, in this analysis (hereafter Model D), we did not model the SED and we used the system scale parameter $a/R_\star$ to model the transit instead of the stellar density $\rho_{\star}$. We chose an non-informative prior for $a/R_\star$ (see Table \ref{PriorModel}, online) and ran another 20 chains randomly started from the uniform distribution for $a/R_\star$ and from the median values found in Model C for the other parameters. We analysed the chains as already described in section \ref{PASTIS} and we derived the median values and their 63.8\% confidence intervals that we give in the online table \ref{ModelResult}. All the parameters derived in Model D are compatible within 63.8\% with the ones derived in Model C. We estimated the stellar density using the third law of Kepler, the fitted system scale $a/R_{\star}$ and the fitted orbital period and find a value of $\rho_{\star} / \rho_{\odot} = 0.70 \pm 0.37$ which fully agrees within 68.3\% with the one derived by the spectral analysis and the models A and C.\\

\subsection{Model E: analysis of the Rossiter-McLaughlin effect}
\label{ModelE}

\begin{figure}[h!]
\begin{center}
\includegraphics[width=\figw]{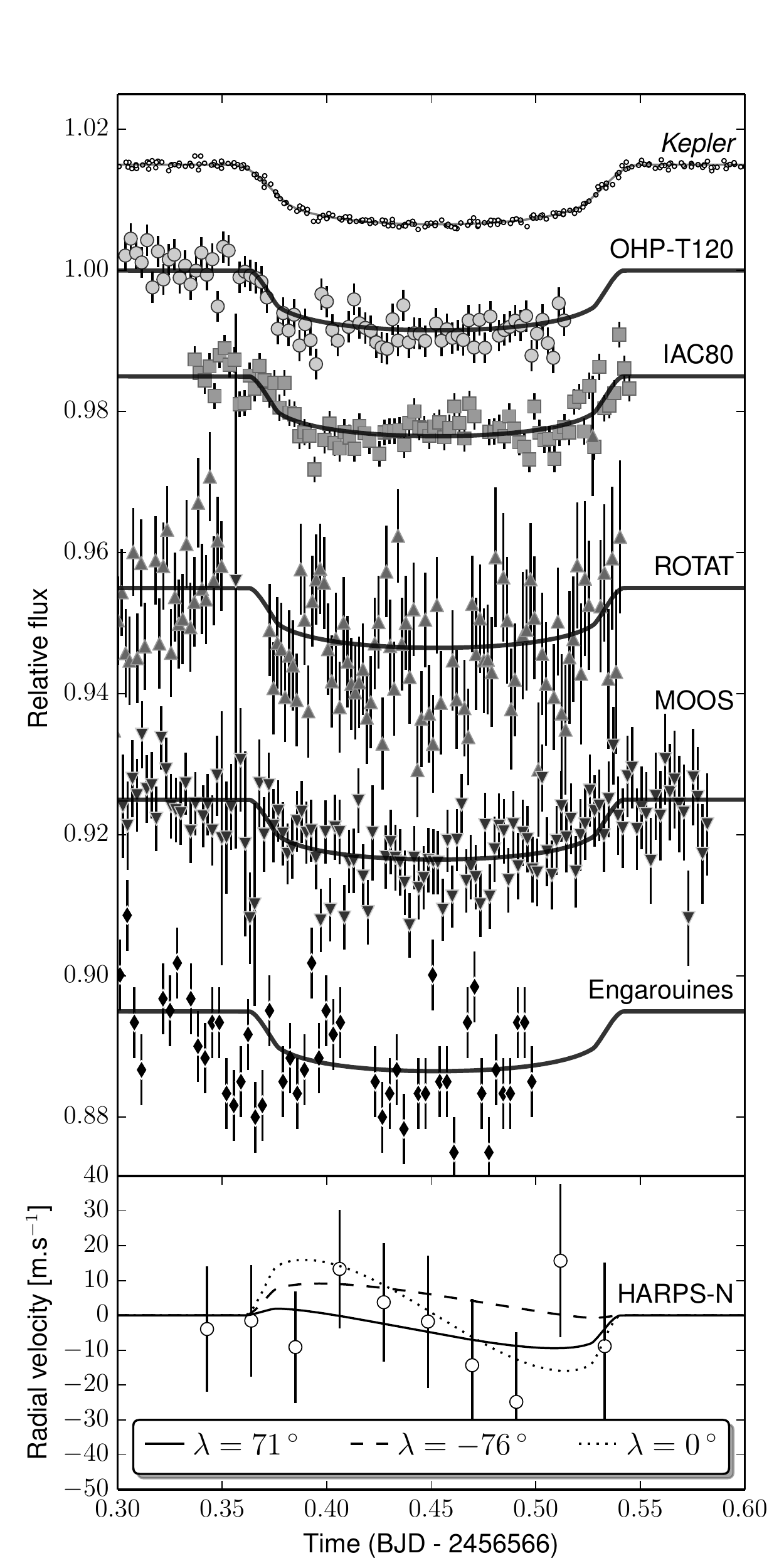}
\caption{\textit{Upper panel:} Transit light curves of KOI-1257.01 obtained during the transit night of 2013 September 30. The phase-folded \textit{Kepler} light curve was shifted in time for comparison. From top to bottom: \textit{Kepler}, OHP-T120, IAC80, ROTAT, MOOS, and Engarouines. More information about the instrumental configurations of the ground-based observatories can be found in Table \ref{ListPhot}. The best model E that fits all the data is shown with the black curve. Each light curve was arbitrarily shifted in flux. \textit{Lower panel:} HARPS-N radial velocities obtained on the night 2013 September 30 during the transit of KOI-1257.01. The data were corrected using the best Keplerian orbit. The two median-fit models of the Rossiter-McLaughlin effect (from Model E) are displayed with a solid line (for $\lambda$ = -76 \degr) and a dashed line (for $\lambda$ = 71 \degr). The aligned model (with $\lambda$ = 0 \degr) is displayed with the dotted line for comparison.}
\label{1257ProAm}
\end{center}
\end{figure}

We repeated the analysis performed in Model C, this time including the model of the Rossiter-McLaughlin effect \citep{1924ApJ....60...15R, 1924ApJ....60...22M} and the ground-based photometric observations obtained simultaneously with the HARPS-N data (see section \ref{ProAmObs}). We modelled the Rossiter-McLaughlin effect using the \texttt{Arome} code \citep{2013A&A...550A..53B} as implemented into the \texttt{PASTIS} software\footnote{We correct the typo that occurs in Eq. 42 of \cite{2013A&A...550A..53B}. The equation is as follows: $H_{xy}^{(n)} = -\alpha x^{n-1}yI_{\alpha-2}(x,y) + \alpha(\alpha -2)x^{n+1}y^{2}I_{\alpha-4}(x,y)$ (Bou\'e, private communication). This typo is however not present in the public code of \texttt{Arome} available at \url{http://www.astro.up.pt/resources/arome/}.}. Compared with Model C, here we added two free parameters: (1) the sky-projected spin-orbit angle $\lambda$ for which we used an non-informative prior and (2) the sky-projected stellar equatorial velocity \vsini\, for which we used a Normal prior to account for the spectral analysis constraints. We also added 15 new free parameters related to the five new light curves modelled here: the contamination, the flux out-of-transit, and the extra source of white noise (jitter), all with large and non-informative priors. This leads to a total of 50 free parameters for this analysis (see Table \ref{PriorModel}, online). \\

To account for the relatively long exposure time (1800s) of the HARPS-N data compared with the transit duration, which should smooth down the amplitude of the Rossiter-McLaughlin effect, we oversampled the model of HARPS-N observations by a factor of 10, as for the \textit{Kepler} long-cadence data. By doing so, we ensure that the amplitude of the Rossiter-McLaughlin effect is correctly driven by the \vsini\, of the star and not affected by the long exposure time. Not oversampling the Rossiter-McLaughlin effect model might lead to a biased solution, as is the case for the modelling of the transit light curve \citep{2010MNRAS.408.1758K}. However, the present data are not accurate enough for this effect to significantly affect the result.\\

We ran another 20 MCMC chains. We analysed the chains as previously and we derived the median values and the 68.3\% confidence intervals of the parameters that we give in the online table \ref{ModelResult}. The posterior distribution of the spin-orbit angle shows a bi-Normal distribution with nearly symmetric values of $\lambda$ = -76 $\pm$ 42\degr\, and $\lambda$ = 71 $\pm$ 47\degr. The absolute value of the spin-orbit angle $| \lambda |$\, therefore has a value of 74$^{_{+32}}_{^{-46}}$\degr. Figure \ref{1257ProAm} displays the five ground-based transit light curves superimposed with their best model. The sixteen transits observed by \textit{Kepler} are shifted in time for comparison. In this figure, the HARPS-N data obtained during the transit night of 2013 September 30 are displayed with the two median-fit models of the Rossiter-McLaughlin effect that correspond to $\lambda$ = -76 \degr\, and $\lambda$ = 71 \degr. The aligned model of the Rossiter-McLaughlin effect (with $\lambda$ = 0 \degr) is also displayed for comparison.\\

\begin{figure}[h!]
\begin{center}
\includegraphics[width=\figw]{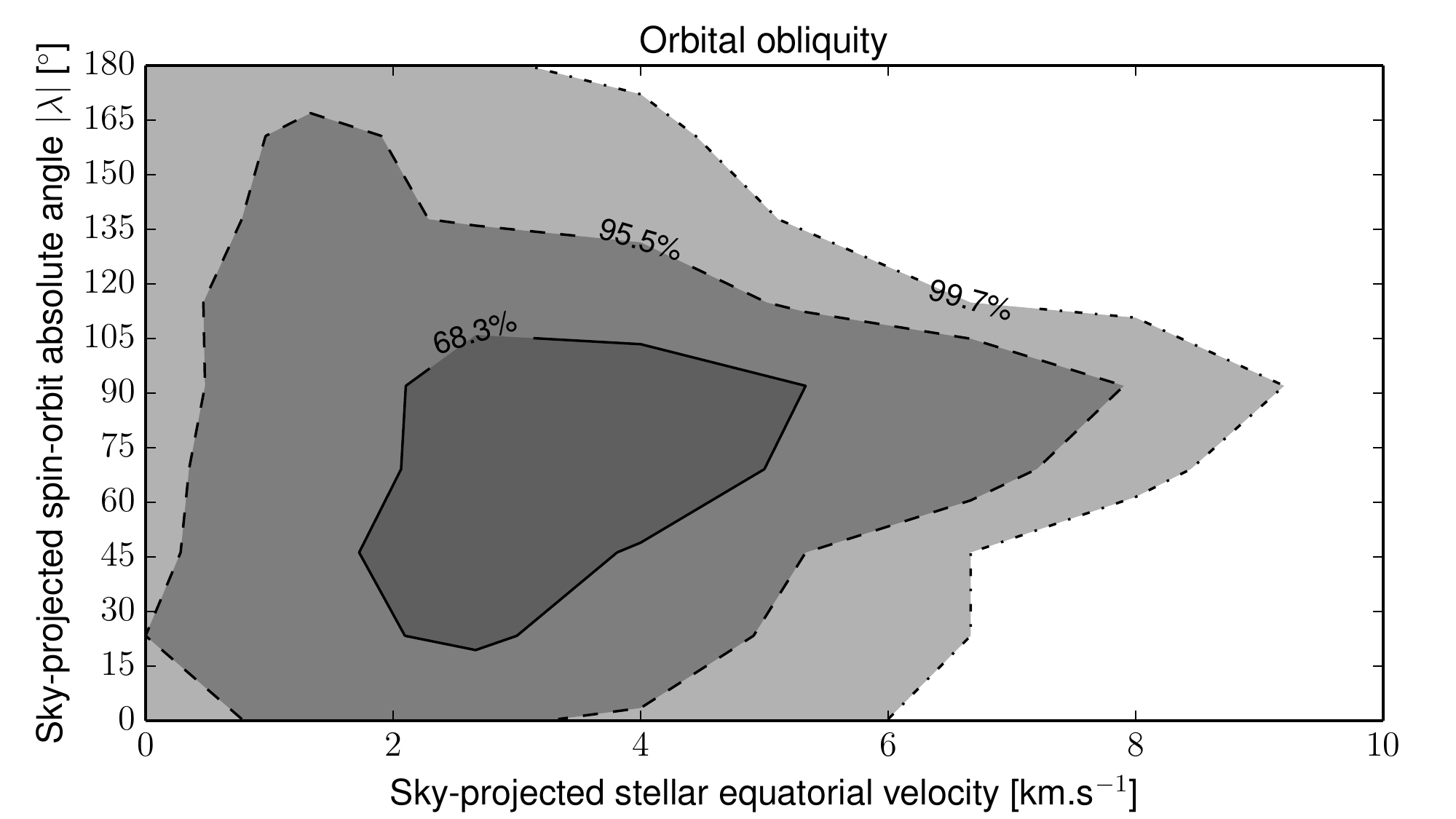}
\caption{Posterior distribution of the sky-projected spin - orbit angle as a function of the stellar rotational velocity. The different regions represent the 68.3\%, 95.5\%, and 99.7\% confidence intervals (from dark to light grey).}
\label{spinorbit}
\end{center}
\end{figure}

Figure \ref{spinorbit} shows the 68.3\%, 95.5\%, and 99.7\% confidence intervals of the posterior distribution of the absolute value of spin-orbit angle as a function of the stellar rotation velocity. Since the Rossiter-McLaughlin effect is not clearly detected, the obliquity of the planet is poorly constrained. The maximum of likelihood indicates a nearly polar orbit for the planet, but an orbit aligned with the stellar spin cannot be rejected from the current data. These nearly polar-orbit solutions might be explained by the absence of clear variation during the Rossiter-McLaughlin effect and a central transit, as in the case of HAT-P-32 \citep{2012ApJ...757...18A}. The deduced impact parameter of the transit of KOI-1257.01 is $b = 0.11 \pm 0.11$, hence compatible with a central transit, which explains those obliquity constraints.\\

\subsection{Constraints on the outer companion}
\label{OuterCompanion}

We repeated the analysis of the radial velocities described in section \ref{ModelB} (Model B), but instead of fitting  a quadratic drift, we fitted a second, outer companion assuming a circular orbit. The choice of this circular orbit is counterintuitive given the range of orbital periods considered for this outer companion (longer than a few hundred days). However, since only a small fraction of its orbit has been observed so far, it is not possible to constrain the eccentricity and the argument of periastron. Future follow-up observations of this target will permit the orbital solution of this companion to be improved. We assumed here non-informative priors on the periastron epoch (uniform prior), orbital period (Jeffreys prior), and the radial velocity semi-amplitude (Jeffreys prior) of the outer companion.\\

\begin{figure}[h!]
\begin{center}
\includegraphics[width=\figw]{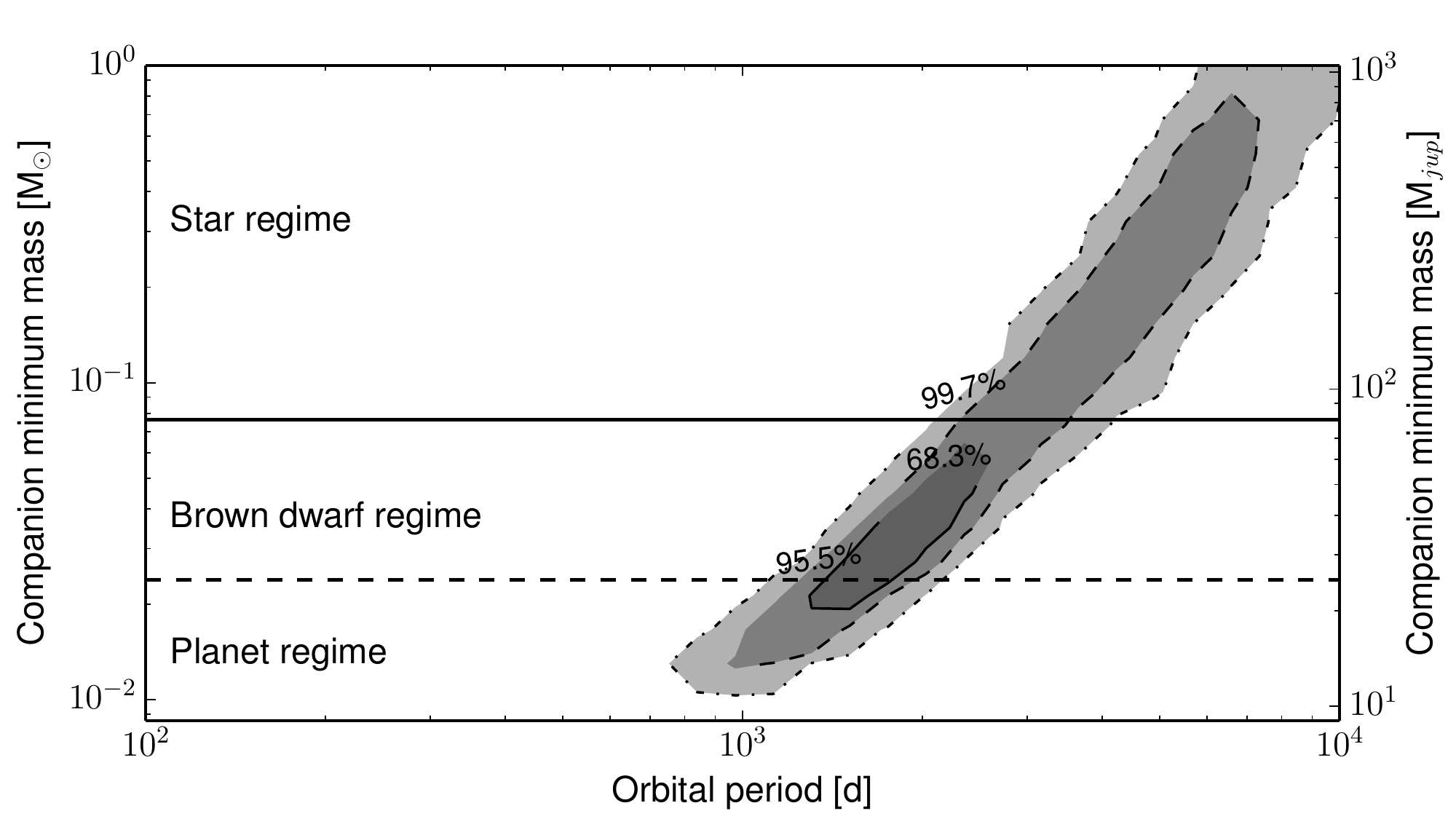}
\caption{Posterior distribution of the second companion minimum mass as a function of its orbital period, assuming a circular orbit. The three regions represent the 68.3\%, 95.5\%, and 99.7\% confidence intervals from the MCMC analysis (from dark to light grey). The dashed line indicates the planet -- brown dwarf limit as suggested by \citet{2011A&A...532A..79S}. The solid line indicates the hydrogen-burning limit.}
\label{OuterComp}
\end{center}
\end{figure}

We ran 40 MCMC chains for this analysis. From their analysis (see section \ref{PASTIS}), we computed the minimum mass for this outer companion. We show in Fig. \ref{OuterComp} the 68.3\%, 95.5\%, and 99.7\% confidence intervals of the posterior distribution of the companion minimum mass as function of its orbital period. The quadratic drift observed in the radial velocity data is compatible with a circular, massive outer planet, but a brown dwarf and a star cannot be rejected from this analysis. The reason of the upper-limit constraint in mass of this outer companion is discussed in Appendix \ref{RVdrift}.

\subsection{Search for transit time variations}
\label{TTVs}

\begin{figure}[h!]
\begin{center}
\includegraphics[width=\figw]{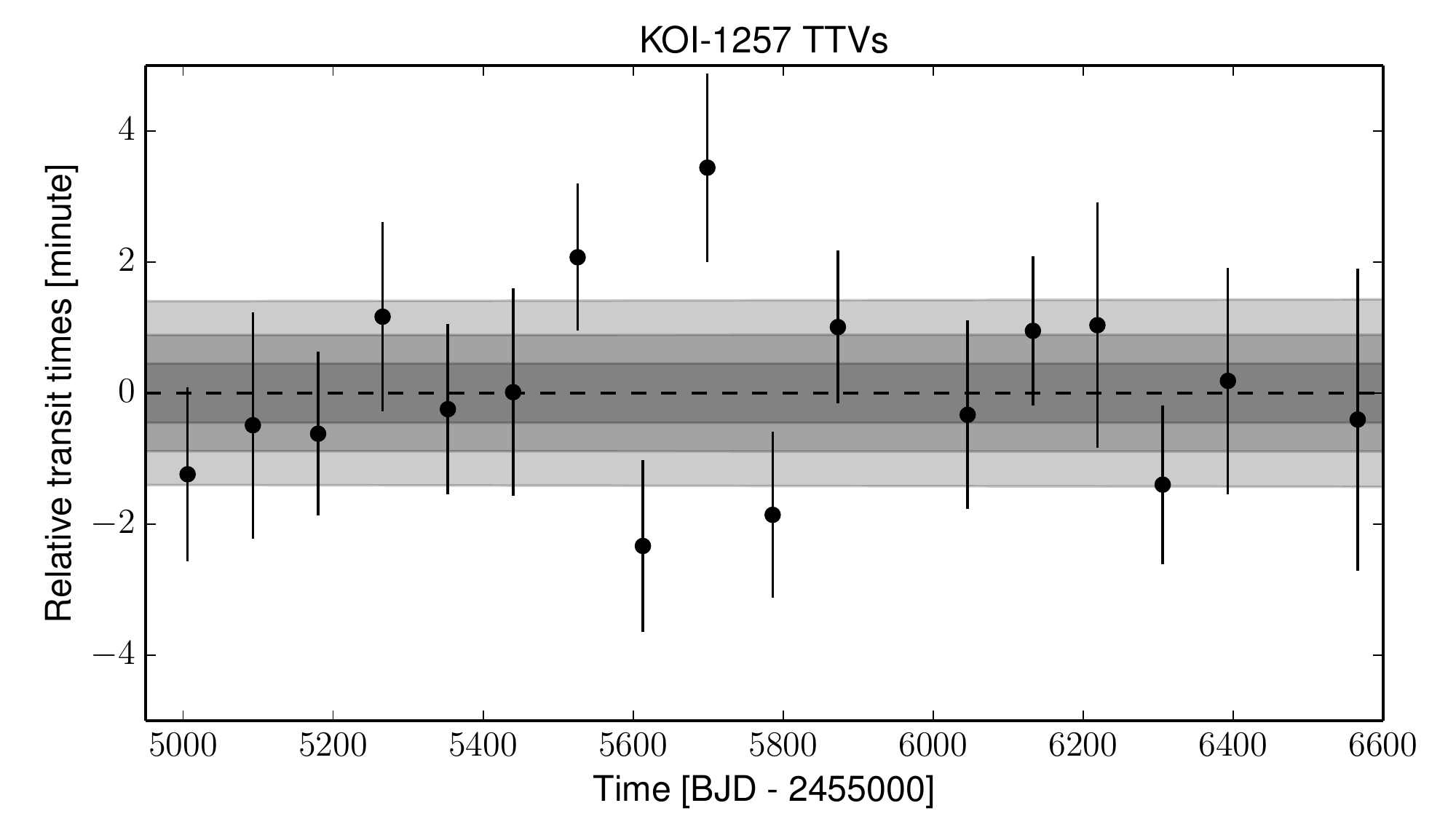}
\caption{Transit times of KOI-1257.01 compared to the best linear ephemeris found in Model C (dashed line). The grey regions represent the 63.8\%, 95.5\%, and 99.7\% confidence regions of the best linear ephemeris (from dark to light grey). The confidence regions account for the covariance between the orbital period and the epoch of first transit, but this covariance is too small to be visible in this plot.}
\label{1257TTVs}
\end{center}
\end{figure}

To derive the transit times of KOI-1257.01 that might be induced by the outer companion, we cut the \textit{Kepler} light curve into chunks centred on each transit. We then repeated the analysis done in section \ref{ModelA} (Model A) but on the individual transits. We fixed all the parameters to the best-fit values found in model C (see section \ref{ModelC}) following the recommendation of \citet{2013A&A...556A..19O} and \citet{2013MNRAS.430.3032B}, except for $T_{0}$, the contamination, the flux out-of-transit, and the jitter value. We used non-informative priors for each of these four parameters. We also analysed together the five transit light curves that were obtained from the ground on the night 2013 September 30 (see section \ref{ProAmObs}). We modelled these five transits simultaneously allowing only the time of transit, the contamination, the flux out-of-transit, and the jitter for each light curve to vary. We ran one chain for each of the seventeen light curves (sixteen from \textit{Kepler} and one from the ground) and analysed them as before. The derived transit times are displayed in Fig. \ref{1257TTVs} and listed in Table \ref{TTVsdata}. The transit times are compatible with a linear ephemeris, as already reported by \citet{2011ApJS..197....2F} and \citet{2013ApJS..208...16M}. Our uncertainties are, however, slightly larger than those derived by \citet{2013ApJS..208...16M}.\\

\begin{table}[h!]
\caption{Transit times of KOI-1257.01. The times are relative to the best ephemeris derived in Model C.}
\begin{center}
\setlength{\tabcolsep}{4pt}
\begin{tabular}{cccc}
\hline
\hline
Transit epoch & Uncertainty & Relative time & Uncertainty \\

[BJD$_\mathrm{TDB}$] & [BJD$_\mathrm{TDB}$] & [min] & [min]\\
\hline
2455006.79362  &  0.00092  &  -1.24  &  1.32\\
2455093.44181  &  0.00120  &  -0.49  &  1.73\\
2455180.08938  &  0.00087  &  -0.62  &  1.25\\
2455266.73829  &  0.00100  &  1.17  &  1.44\\
2455353.38497  &  0.00090  &  -0.24  &  1.30\\
2455440.03282  &  0.00110  &  0.01  &  1.58\\
2455526.68192  &  0.00078  &  2.07  &  1.12\\
2455613.32652  &  0.00091  &  -2.33  &  1.31\\
2455699.97820  &  0.00100  &  3.44  &  1.44\\
2455786.62218  &  0.00088  &  -1.86  &  1.27\\
2455873.27184  &  0.00081  &  1.01  &  1.17\\
2456133.21480  &  0.00079  &  0.95  &  1.14\\
2456219.86252  &  0.00130  &  1.04  &  1.87\\
2456306.50850  &  0.00084  &  -1.40  &  1.21\\
2456393.15727  &  0.00120  &  0.19  &  1.73\\
2456566.45219  &  0.00160  &  -0.40  &  2.30\\
\hline
\hline
\end{tabular}
\end{center}
\label{TTVsdata}
\end{table}

\section{\texttt{PASTIS} validation}
\label{Blend}

Even if the quadratic drift observed in the SOPHIE data is compatible with a circular, coplanar, massive planet (see section \ref{OuterCompanion}), the likelihood of this scenario is smaller that of having a brown dwarf or another star in the system, which might be coplanar or not with KOI-1257.01. If this hypothetical second star in the system is bright enough, it would significantly affect the spectral parameters of the host star. It would also dilute both the transit depth, the radial velocity amplitude, and the amplitude of the Rossiter-McLaughlin effect. Therefore, the derived parameters of the system would be affected. Moreover, if there is an unresolved star in the system, it is not clear on which star the transit occurs. We can therefore assume the four following scenarios to describe this system.

\begin{itemize}
\item Scenario 0 -- the system KOI-1257 is composed of a primary star with two substellar objects: the transiting planet KOI-1257~b and the outer companion KOI-1257~c.
\item Scenario 1 -- the candidate KOI-1257.01 is a planet transiting the main star of a binary system: KOI-1257~Ab.
\item Scenario 2 -- the candidate KOI-1257.01 is a planet transiting the secondary star of a binary system: KOI-1257~Bb.
\item Scenario 3 -- the candidate KOI-1257.01 is a third, low-mass star eclipsing the secondary star of a binary system: KOI-1257~C.
\end{itemize}

A fifth scenario can be imagined which is a low-mass star eclipsing the main star of the system. This system would, however, produce a large radial velocity variation, which was not observed by SOPHIE. We also rejected the background eclipsing binary and background transiting planet scenarios. Indeed, such background systems need to have nearly the same systemic radial velocity as KOI-1257 to reproduce the radial velocity variations observed by SOPHIE. It is therefore extremely unlikely to have a foreground binary system blended with a background eclipsing binary or transiting planet which have nearly the same systemic radial velocity.\\

Scenarios 0 to 3 can, in principle, be constrained thanks to the high precision of the \textit{Kepler} transit light curve alone \citep{2014arXiv1403.6725D}. This is especially true for scenarios 2 and 3 for which the stellar density constrained from the light curve is expected to be different from the one derived by the spectral analysis, except in the particular case of two stars with similar masses and density. However, if the system is composed of a circular planet (KOI-1257~Bb) transiting a low-mass secondary star, it might mimic the same stellar density as constrained by the light curve as an eccentric planet transiting a larger primary star. Thus, to improve the constraints on the system, we used both the \textit{Kepler} transit light curve, the SOPHIE data and the spectral energy distribution. If the system is composed of two unresolved stars, the observed CCF should present either variation of the bisector or of the CCF width, or both. We did not use here the HARPS-N data because (1) the time span of the observations is short (10 days) compared with the orbital period of the outer companion (more than 1000 days), (2) the parameters of the CCF have not yet been calibrated, and (3) we are not yet able to model the Rossiter-McLaughlin effect of a planet in a blended binary star system. \\

\begin{figure}[h!]
\begin{center}
\includegraphics[width=\figw]{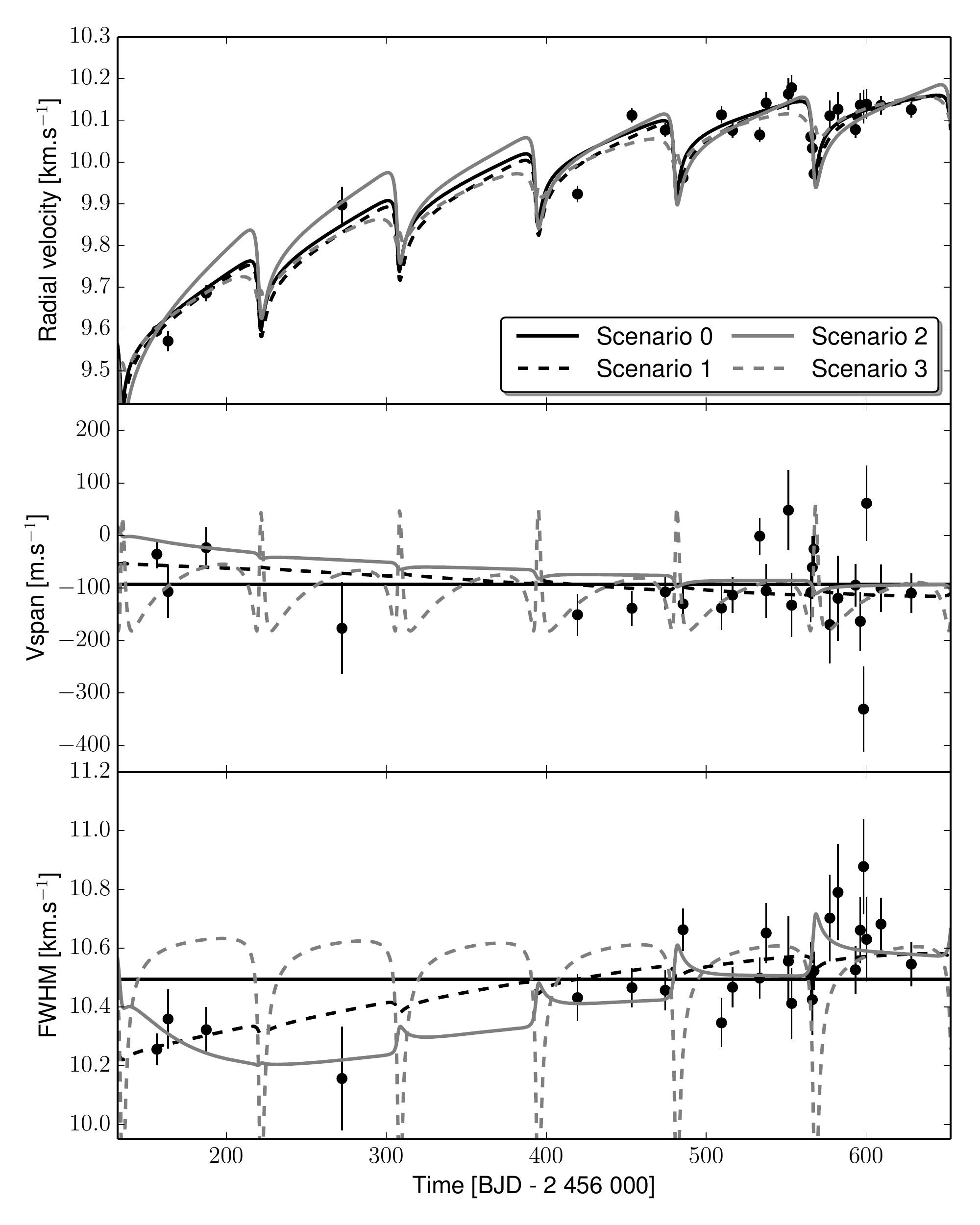}
\caption{SOPHIE radial velocities (upper panel), bisector (middle panel), and FWHM (lower panel) as function of time superimposed with the best model of scenarios 0, 1, 2, and 3.}
\label{1257Blend}
\end{center}
\end{figure}

The radial velocity, bisector, and FWHM observed by SOPHIE are displayed in Fig. \ref{1257Blend}. In this plot, one can see that the FWHM present a drift in time with an amplitude similar to the radial velocity drift (i.e. about half a \kms\, in one year). However, the bisector does not show any clear drift with an amplitude larger than about 100\ms. For this analysis, we used the \texttt{PASTIS} software \citep{2014arXiv1403.6725D}. The modelling of the SOPHIE CCF using the \texttt{PASTIS} software is described in Santerne et al. (in prep.). It consists in estimating the parameters of the CCF using the equations (B.3), (B.4), and (B.5) of \citet{2010A&A...523A..88B}, assuming a (B$-$V) computed from the stellar atmosphere models and the stellar evolution tracks and a \met\, drawn from the prior distribution. The CCF of the various stellar objects are then blended together and normalised after accounting for the relative luminosity of each object in the Johnson-V band. The expected radial velocity, FWHM and bisector are then fitted to the total CCF as is done in the SOPHIE pipeline. The modelling of the SOPHIE CCF is quite expensive in terms of CPU computation. For example, each MCMC chain computed in this section took between two and three weeks (except for those of scenario 0) on the multi-CPU cluster at the Laboratoire d'Astrophysique de Marseille. Since the radial velocity products are measured on the same CCF, they might not be independent. However, to simplify the computation of the global likelihood, we assumed these three datasets to be independent from each other. We assumed that this non-independence of the dataset is a second-order effect and that it does not change significantly the results and the conclusions.\\

Bisector analyses were first used in \citet{2002A&A...392..215S} to resolve a brown dwarf companion in an unresolved binary mimicking a radial velocity giant planet. Bisector analyses were also used in radial velocity detections to detect nearly face-on binaries by \citet[using a preliminary version of the \texttt{PASTIS} tool]{2012A&A...538A.113D} and \citet{2013ApJ...770..119W}. Finally, \citet{2012A&A...541A.149O} modelled the HARPS CCF together with the CoRoT light curve and the SED to validate the detection of the transiting planet CoRoT-16~b. However, this is the first time that we have modelled both the bisector and the FWHM together with the radial velocities, the transit light curve, and the spectral energy distribution to resolve a potential blended system.

\subsection{Scenario 0: KOI-1257 is a multi-substellar objects system}
\label{Scenario0}

Scenario 0 was deeply explored in section \ref{Analysis} and section \ref{OuterCompanion} for the outer companion. We repeated the analysis performed in model C without the HARPS-N data but with the SOPHIE diagnosis (bisector and FWHM). Since the outer companion is a substellar object in this scenario, it should not produce a variation of the observed bisector, nor of the FWHM. We therefore modelled the bisector V$_\mathrm{span}$ and FWHM as constant with time. An offset for the bisector was fitted in order to account for unmodelled line asymmetry such as the convective blue shift (assumed to be constant with time at this precision). The list of parameters that describe the models of this scenario is given in the online table \ref{ScenarioPriors}, together with the prior distributions used in this analysis. The best model found by a MCMC procedure of 20 chains of 10$^{6}$ iterations each is displayed in Fig. \ref{1257Blend}. Median values and their uncertainties are given in the online table \ref{ScenarioResult}. As shown in Section \ref{Analysis}, this scenario is able to reproduce appropriately the \textit{Kepler} transits, the SOPHIE radial velocities, and the spectral energy distribution. However, this scenario is not expected to produce a drift in the FWHM. To illustrate this, we computed a weighted RMS (without accounting for the jitter) of the residuals from the best model (wRMS) of each dataset and found a wRMS of 39 \ms\, for the radial velocities, a wRMS of 70 \ms\, for the bisector, a wRMS of 161 \ms\, for the FWHM. We note that these wRMS support our choice of uncertainty for the bisector and the FWHM mentioned in section \ref{SOPHIEObs}: the bisector and the FWHM have a wRMS of about twice and four times, respectively, the wRMS of the radial velocities. The wRMS of the four \textit{Kepler} seasons is of 302 ppm, and the wRMS of the spectral energy distribution is 128 mmag. All these wRMS are given in Table \ref{wRMS}, together with the mean uncertainty of the different datasets.

\subsection{Scenario 1: KOI-1257.01 is KOI-1257~Ab}
\label{Scenario1}

We simulated a planet orbiting the main component of an unresolved binary star system. The model is described by the following free parameters.
\begin{itemize}
\item Primary star: effective temperature \teff, iron abundance \met, surface gravity \logg, and equatorial velocity $\upsilon \sin i_{\star_{1}}$.
\item Planet: mass $m_{p}$, radius $r_{p}$, orbital period $P_{in}$, transit epoch $T_{0}$, eccentricity $e_{in}$, argument of periastron $\omega_{in}$, and orbital inclination $i_{in}$.
\item Secondary star: initial mass $m_{init_{2}}$ and equatorial velocity $\upsilon \sin i_{\star_{2}}$.
\item Binary: orbital period $P_{out}$, periastron epoch $T_{p}$, eccentricity $e_{out}$, argument of periastron $\omega_{out}$, and inclination $i_{out}$.
\end{itemize}
We assumed here that both stars have the same age and the same metallicity. In comparison with \ref{OuterCompanion}, here we allow the eccentricity and the argument of periastron of the outer orbit to vary. Even if the data do not constrain these parameters well, they are accounted for in the error budget. We set as free parameters the background contamination, the out-of-transit flux, and the jitter of each of the four \textit{Kepler} transit light curves. We attributed jitter values to the radial velocity, bisector, and FWHM that were fitted in the analysis. Finally, the model also assumes an offset value for the bisector to account for constant line-profile asymmetry that is not modelled, such as the convective blue shift. This results in a total of 38 free parameters. For the parameters of the primary star, we used as prior the result from the stellar analysis assuming that the secondary star does not affect significantly the results. For the parameters of the planet, the secondary star and the binary orbit, we assumed non-informative priors, allowing us to explore a mass domain up to 90 \Mjup\, for the planet and down to 0.1 \Msun\, for the secondary star. The exhaustive list of parameters and prior distributions used in this analysis is provided in the online table \ref{ScenarioPriors}.\\

As previously, we use the Dartmouth stellar evolution tracks of \citet{2008ApJS..178...89D} and the BT-SETTL stellar atmosphere models of \citet{2012IAUS..282..235A} to model both stars within the \texttt{PASTIS} software \citep[; Santerne et al., 2014, in prep.]{2014arXiv1403.6725D}. For the limb darkening coefficients, we used the values of \citet{2011A&A...529A..75C}. The radial velocity data were fitted using Keplerian orbits, neglecting the dynamical interactions between them. Finally, we constrained the secondary star to be less bright than the primary star. We ran 20 MCMC chains and analysed them as described in section \ref{PASTIS}. From this posterior distribution, we derived the 68.3\% confidence interval of the 38 free parameters that are listed in the online table \ref{ScenarioResult}. \\

Thanks to the constraints provided by the SOPHIE FWHM and bisector, the MCMC converged toward a secondary star with a mass of 0.70 $\pm$ 0.07\,\Msun. This secondary star orbits in the system with an inclination of 18.2$^{_{+18.0}}_{^{-5.4}}$ \degr\, and a period of 3430 $ \pm $ 1200 days. This uncertainty is only the statistical uncertainty and does not include the stellar models errors. This secondary star contributes to only 8.9\% of the total flux of the system (in the V band). If this second star is real, it would not affect significantly the spectral analysis (see in section \ref{SpACheck}). The \textit{Kepler} transit light curve, the SOPHIE radial velocities and bisector are slightly better fitted by this scenario, compared with scenario 0 (see the wRMS in Table \ref{wRMS}). However, the spectral energy distribution is slightly worse fitted (the wRMS is 142 mmag) by considering a stellar companion of $\sim$ 0.7 \Msun, but this turns out to be not significant (see section \ref{Bayes}). However, this scenario better fits the SOPHIE FWHM, compared with scenario 0 as one can see in Fig. \ref{1257Blend}. The corresponding wRMS is 113 \ms. 

\subsection{Scenario 2: KOI-1257.01 is KOI-1257~Bb}
\label{Scenario2}

We simulated scenario 2 in the same way as scenario 1, but assuming that a planet is transiting the secondary (fainter) star of the system. The parameters and their respective prior distributions are listed in the online Table \ref{ScenarioPriors}. We give in the online table \ref{ScenarioResult} the median values and their uncertainties on the parameters that describe this scenario, based on a 20-chain posterior distribution. The best-fit model of scenario 2 is displayed in Fig.~\ref{1257Blend}.\\

This scenario is able to reproduce the \textit{Kepler} transit light curve almost equally well as the two previous scenarios because the fit converged toward a secondary star with a mass of 1.00 $\pm$ 0.05 \Msun\, i.e. similar to the primary mass. Once again, this uncertainty on the mass of the secondary star does not account for the stellar models uncertainties. The wRMS of the \textit{Kepler} data is 303 ppm, which is close to the previous scenarios. Therefore, both components of this binary have similar density, producing similar transit shapes and durations. With two similar stars, the spectral energy distribution is better fitted (the wRMS is 136 mmag) than with the two different stars of scenario 1 but slightly worse fitted than assuming the single star of scenario 0. In this scenario, the system has a distance of nearly twice that found in scenario 0. The SOPHIE FWHM are slightly better fitted (wRMS of 109 \ms) by this scenario than in the case of scenarios 0 and 1. However, the SOPHIE radial velocities and bisector are slightly worse fitted in this case (see Fig. \ref{1257Blend}), with a wRMS of 43 \ms\, and 70 \ms, respectively.\\ 

With two blended stars of similar brightness, the only possibility to observe no bisector variation is when the stars have a similar FWHM, thus, in that case, a similar \vsini\, \citep[Santerne et al., in prep ; see also the Figs 3 and 12 of][which show a blind zone for secondary stars with a FWHM similar to the target one]{2012A&A...538A.113D}. The \vsini\, of the primary star is constrained thanks to the median value of the observed FWHM. The \vsini\, of the secondary star is constrained by the amplitude of the bisector variation. This explains why we derive a relatively small uncertainty on the $\upsilon \sin i_{\star_{2}}$ as well as $m_{init_{2}}$ (about 5\%). All the other parameters have uncertainties at the same level as for scenario 1.\\

This scenario reproduces the observed data almost equally well as scenario 1. However, scenario 1 and 2 do not converge toward the same physical parameters for the planet (see Table \ref{ScenarioResult}, online): a mass of 1.45 $\pm$ 0.35 \Mjup\, (scenario 1) or 3.92 $\pm$ 0.88 \Mjup\, (scenario 2) and a radius of 0.94 $\pm$ 0.12 \Rjup\, (scenario 1) or 1.56 $\pm$ 0.13 \Rjup\, (scenario 2). The statistical comparison between these two scenarios and the estimation of their respective probability are presented in section \ref{Bayes}. 

\subsection{Scenario 3: KOI-1257.01 is KOI-1257~C}
\label{Scenario3}

Scenario 3 was simulated as for scenario 2, but the transiting planet was replaced by a third star that eclipse the secondary star of the system. The system is therefore described with the following parameters.
\begin{itemize}
\item Primary star: effective temperature \teff, iron abundance \met, surface gravity \logg, and equatorial velocity $\upsilon \sin i_{\star_{1}}$.
\item Secondary star: initial mass $m_{init_{2}}$ and equatorial velocity $\upsilon \sin i_{\star_{2}}$.
\item Tertiary star: initial mass $m_{init_{3}}$ and equatorial velocity $\upsilon \sin i_{\star_{3}}$.
\item inner orbit: orbital period $P_{in}$, transit epoch $T_{0}$, eccentricity $e_{in}$, argument of periastron $\omega_{in}$, and orbital inclination $i_{in}$.
\item outer orbit: orbital period $P_{out}$, periastron epoch $T_{p}$, eccentricity $e_{out}$, argument of periastron $\omega_{out}$, and inclination $i_{out}$.
\end{itemize}

The parameters and their respective prior distributions are listed in the online table \ref{ScenarioPriors}. We ran 20 chains and we list in the online table \ref{ScenarioResult} the median values and 68.3\% confidence intervals of the parameters. We display in Fig. \ref{1257Blend} the best-fit model of the SOPHIE data. This scenario of a triple stellar system is able to reproduce the observed radial velocities only if the secondary star is a fast rotator. Indeed, the large radial velocity variation of the inner-binary would be more easily diluted because the main contaminating star is rotating fast (its observed line contrast being much lower than the target star contrast). However, such a fast-rotating secondary star would have produced relatively large FWHM variations, which are incompatible with the data. This scenario is therefore not able to reproduce the observed FWHM. The resulting wRMS is as large as 319 \ms, which excludes this model. The radial velocities are also worse fitted with a wRMS of 51 \ms. Surprisingly, the bisector is better fitted than the other scenarios, with a wRMS of 65 \ms, but this might be explained by a relatively sparse sampling near the periastron. The wRMS of the spectral energy distribution is 139 mmag, which is better than scenario 1, but worse than scenarios 0 and 2. The wRMS of the \textit{Kepler} data is of 303 ppm, and thus similar to the other scenarios.\\
 
We note that our CCF model assumed a Gaussian profile for the stellar line. In the present scenario, it would be more rigorous to model the line of the main contaminant by a rotation profile convolved by the SOPHIE instrumental resolution as done in \citet{2012A&A...544L..12S}. We assumed, however, that this approximation does not change significantly the result of this analysis.

\begin{table}[h!]
\caption{Weighted RMS from the best-fit model found for each scenario 0 to 3 for the different datasets. These RMS have to be compared with the mean uncertainty $<\sigma>$ of the dataset, computed without including the jitter.}
\begin{center}
\setlength{\tabcolsep}{4pt}
\begin{tabular}{lccccc}
\hline
\hline
wRMS & \textit{Kepler} & SED & RV & V$_\mathrm{span}$ & FWHM\\
 & [ppm] & [mmag] & [\ms] & [\ms] & [\ms]\\
\hline
Scenario 0 & 302 & 128 & 39 & 70 & 161\\
Scenario 1 & 301 & 142 & 37 & 68 & 113 \\
Scenario 2 & 303 & 136 & 43 & 70 & 109 \\
Scenario 3 & 303 & 139 & 51 & 65 & 319 \\
\hline
$<\sigma>$ & 264 & 60 & 26 & 52 & 103 \\
\hline
\hline
\end{tabular}
\end{center}
\label{wRMS}
\end{table}

\subsection{Bayesian statistical comparison of the scenarios and planet validation}
\label{Bayes}

We analysed in Sects. \ref{Scenario0}, \ref{Scenario1}, \ref{Scenario2}, and \ref{Scenario3} the same dataset considering four different scenarios. To quantify which scenario is best supported by the data, we computed for each pair of scenarios the odds ratio $\mathcal{O}_{ij}$ between the scenarios $i$ and $j$, as defined in the Bayesian statistics.

\begin{eqnarray}
\mathcal{O}_{ij} &=& \frac{\mathcal{P}\left(\mathcal{S}_{i} \,\Big\vert\, \mathcal{D}, \mathcal{I}\right)}{\mathcal{P}\left(\mathcal{S}_{j} \,\Big\vert\, \mathcal{D}, \mathcal{I}\right)} \\
\mathcal{O}_{ij} &=& \frac{\pi\left(\mathcal{S}_{i} \,\Big\vert\, \mathcal{I}\right)}{\pi\left(\mathcal{S}_{j} \,\Big\vert\, \mathcal{I}\right)}\cdot\frac{\mathcal{P}\left(\mathcal{D} \,\Big\vert\, \mathcal{S}_{i}, \mathcal{I}\right)}{\mathcal{P}\left(\mathcal{D} \,\Big\vert\, \mathcal{S}_{j}, \mathcal{I}\right)} \\
\mathcal{O}_{ij} & = & \frac{\pi\left(\mathcal{S}_{i} \,\Big\vert\, \mathcal{I}\right)}{\pi\left(\mathcal{S}_{j} \,\Big\vert\, \mathcal{I}\right)}\cdot \frac{\bigints_{\theta_{i}}{\pi\left(\theta_{i}\,\Big\vert\, \mathcal{S}_{i},\mathcal{I}\right)\cdot \mathcal{P}\left(\mathcal{D}\,\Big\vert\,\theta_{i},\mathcal{S}_{i}, \mathcal{I}\right)d\theta_{i}}}{\bigints_{\theta_{j}}{\pi\left(\theta_{j}\,\Big\vert\, \mathcal{S}_{j},\mathcal{I}\right)\cdot \mathcal{P}\left(\mathcal{D}\,\Big\vert\, \theta_{j},\mathcal{S}_{j}, \mathcal{I}\right)d\theta_{j}}}\, , \label{eq2}
\end{eqnarray}
where $\mathcal{P}\left(\mathcal{S}_{i} \,|\, \mathcal{D}, \mathcal{I}\right)$ is the probability of the scenario $\mathcal{S}_{i}$ given the data $\mathcal{D}$ and the information $\mathcal{I}$. The symbol $\pi$ represents the a priori information and $\theta_{i}$ is the parameter space which described the models of scenario $\mathcal{S}_{i}$.\\

The first term of equation \ref{eq2} is called the prior ratio. It represents the a priori probability that a given scenario occurs. In the case studied here, we want to compare four scenarios of triple systems. Even if these scenarios have a different number of stars and planets, hence different a priori probabilities to occur \citep{2010ApJS..190....1R, 2014arXiv1401.6825T, 2014arXiv1401.6827T}, it is not straightforward to estimate those probabilities outside the solar neighbourhood. The statistics based on observations for scenarios 1 and 2 are indeed poor. We therefore assumed that the difference of a priori probabilities between the various scenarios is relatively small and that $\pi(\mathcal{S}_{i} \,|\, \mathcal{I}) \,/\, \pi(\mathcal{S}_{j} \,|\, \mathcal{I}) \sim 1$ for all pairs of scenarios.\\

The second term of equation \ref{eq2} is the Bayes factor. It can be computed by marginalising the posterior distribution over all the parameters. Since our models have a relatively high number of free parameters, computing numerically this Bayes factor is quite challenging \citep[see e.g. the discussion in][]{2014MNRAS.437.3540F}. To estimate the Bayes factor, we used the truncated posterior mixture (TPM) as defined by \citet{2012A&A...544A.116T}. This estimator of the evidence has some limitations that are presented in \citet{2012A&A...544A.116T} and discussed in \citet{2014arXiv1403.6725D}. However, since the four scenarios we are testing here have nearly the same number of free parameters, we assume that these limitations do not significantly affect our results.\\

\begin{figure}[h!]
\begin{center}
\includegraphics[width=\figw]{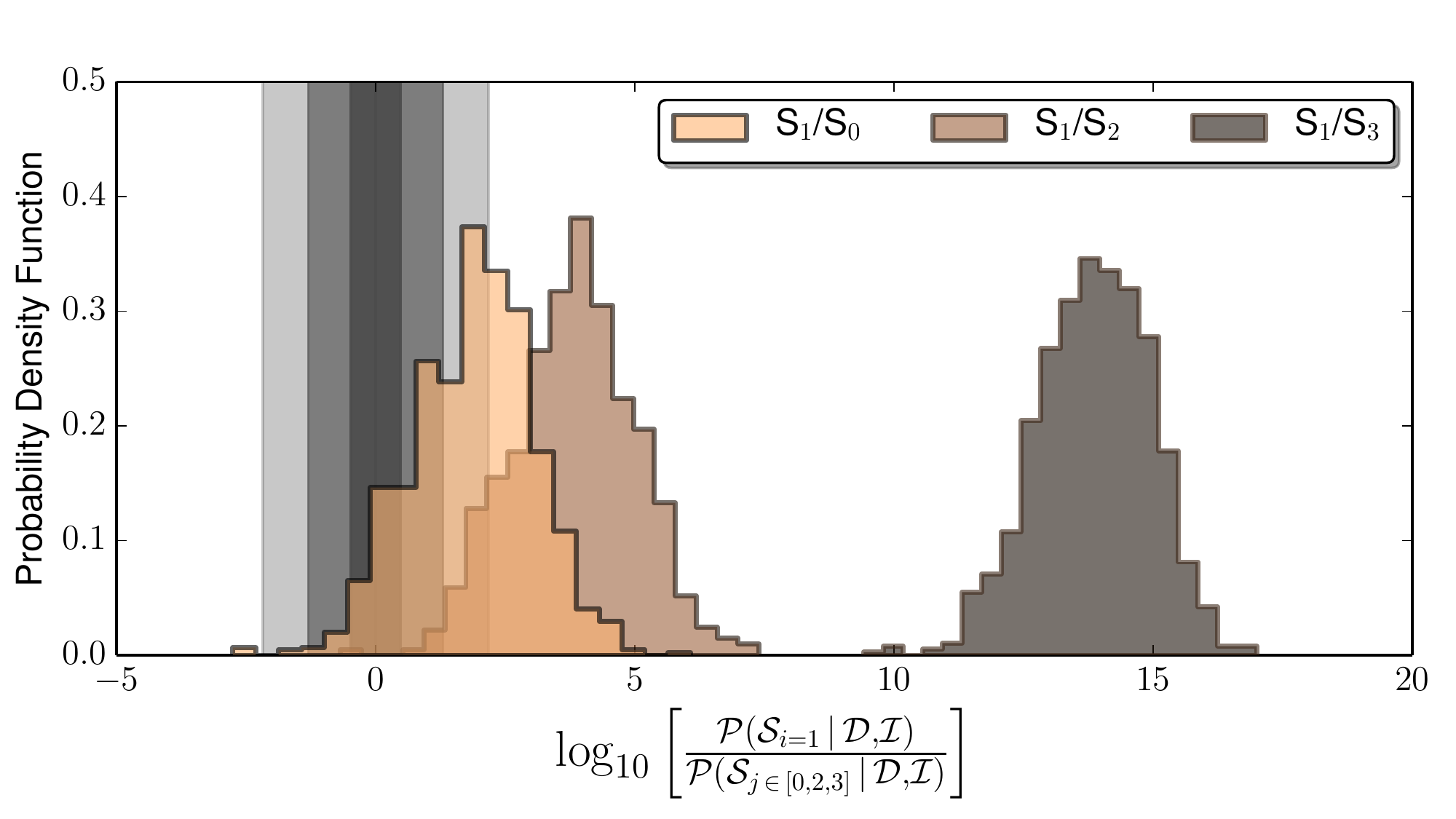}\\
\includegraphics[width=\figw]{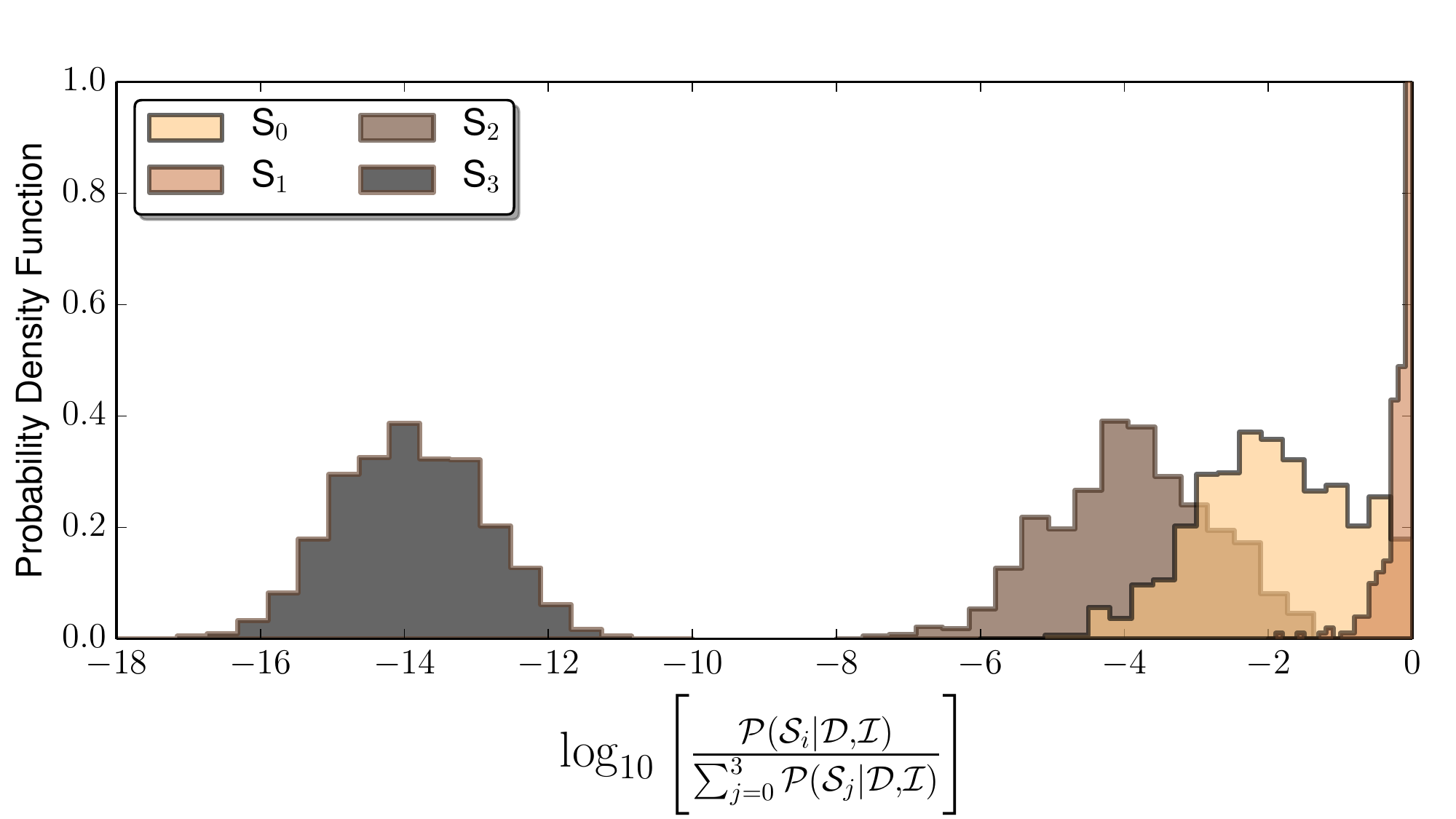}
\caption{\textit{Upper panel:} Odds ratio of scenario 1 against scenarios 0, 2, and 3. The grey regions indicate the ``not worth more than a bare mention'', ``positive'', and ``strong'' evidence as defined by \citet{KassRaftery}. The white region corresponds to ``very strong'' evidence for one model against the other one. \textit{Lower panel:} Probability of each of the four scenarios considered. We assume here that the other, untested scenarios are not significant (see section \ref{Blend}).}
\label{1257Odds}
\end{center}
\end{figure}

The probability distribution of the odds ratios, computed for scenario 1 against all other scenarios is displayed in Fig. \ref{1257Odds} (upper panel) and given in Table \ref{OddsTable}. There is strong to very strong evidence \citep[as defined by][for an odds ratio greater than 150]{KassRaftery}, or decisive evidence \citep[as defined by][for an odds ratio greater than 100]{Jeffreys61} for scenario 1 compared with scenarios 0, 2 and 3. Scenario 3 is clearly rejected in favour of scenario 1 (see Fig. \ref{1257Odds} and Table \ref{OddsTable}). This can easily be explained by the fact that scenario 3 cannot reproduce the observed FHWM. Scenario 2 is also much less supported by the data than scenario 1 (see Fig. \ref{1257Odds} and Table \ref{OddsTable}). This might be surprising since scenario 2 explains all the data quite well. However, to reproduce the data, especially the transit light curve and the bisector, scenario 2 needs to be fine-tuned, as illustrated by the uncertainty on the secondary stellar mass (nearly 5\%) or the secondary $\upsilon \sin i_{\star_{2}}$ (nearly 6\%). These statistical uncertainties are indeed small for an unresolved star. Since scenario 2 needs a fine-tuning of the parameters to reproduce the data, it is therefore less likely than scenario 1 which requires much less fine-tuning. Scenario 2 is therefore penalised by the Occam's razor, as explained in section 3.5 of \citet{2005blda.book.....G}. Finally, scenario 0 does not reproduce the observed drift in the FWHM and it is therefore less likely than scenario 1 which does reproduce this drift (see Fig. \ref{1257Odds} and Table \ref{OddsTable}). Moreover, we expect the odds ratio between these two scenarios to significantly increase as more SOPHIE data are obtained with a longer timespan, by increasing the significance, or not, of the FWHM variation.\\

The lower panel of figure \ref{1257Odds} shows the probability of each scenario, assuming that
\begin{equation}
\label{probasum}
\sum_{i=0}^{3} \mathcal{P}\left(\mathcal{S}_{i} \,|\, \mathcal{D}, \mathcal{I}\right) = 1\, .
\end{equation}
This is not the case, as we discussed in the introduction of section \ref{Blend}, but we assume that those untested scenarios are significantly rejected, and thus negligible. The table \ref{OddsTable} (lower part) gives the mode and the 63.8\% uncertainty on the probability of scenarios 0, 1, 2, and 3. Scenario 1 is clearly the most likely scenario, with a probability of 98.7$^{_{+1.2}}_{^{-13.3}}$ \%, much more supported by the data than the other scenarios (see Table \ref{OddsTable}). The transiting candidate detected by \textit{Kepler} is therefore validated as a planet transiting the primary star of a binary system. We therefore rename the transiting planet candidate KOI-1257.01 as the bona fide planet KOI-1257~b in the rest of the paper.

\begin{table}[h!]
\caption{\textit{Top}: common logarithm of the odds ratios between the scenario pairs. We show only half of the table since $\mathcal{O}_{ij} = 1 / \mathcal{O}_{ji}$. \textit{Bottom}: absolute probability of the four scenarios, assuming no other scenario can reproduce the data (eq. \ref{probasum}).}
\begin{center}
\begin{tabular}{l|ccc}
\hline
\hline
$\log_{10}\left(\mathcal{O}_{ij}\right)$ & $i=0$ & $i=1$ & $i=2$ \\
\hline
$j=0$  &  0 &  --  &  --  \\ 
$j=1$  &  -1.90  $\pm$  1.14  &  0  & --  \\ 
$j=2$  &  1.97  $\pm$  0.91  &  3.85  $\pm$  1.17  &  0  \\ 
$j=3$  &  12.01  $\pm$  0.79  &  13.86  $\pm$  1.06  &  10.03  $\pm$  0.89  \\ 
\hline
\hline
\end{tabular}

\vspace{0.5cm}

\setlength{\tabcolsep}{15pt}
\begin{tabular}{c|c}
\hline
\hline
Scenario & $\mathcal{P}\left(\mathcal{S}_{i} \,|\, \mathcal{D}, \mathcal{I}\right)$ \\
\hline
$\mathcal{S}_{0}$ & 0.93$^{_{+22.60}}_{^{-0.09}}$ \%\\
$\mathcal{S}_{1}$ & 98.7$^{_{+1.2}}_{^{-13.3}}$ \%\\
$\mathcal{S}_{2}$ & 0.01$^{_{+0.18}}_{^{-0.01}}$ \%\\
$\mathcal{S}_{3}$ & 1.2 10$^{-12}$ $^{_{+1.0\,10^{-11}}}_{^{-1.1\,10^{-12}}}$ \%\\
\hline
\hline
\end{tabular}
\end{center}
\label{OddsTable}
\end{table}%

\subsection{Frequentist and other Bayesian model-comparison methods}
\label{F-test}

The analyses performed here are based on a total of 1439 data points (1354 from \textit{Kepler}, 10 from the spectral energy distribution, and 25 from each of the SOPHIE RV, bisector, and FWHM). For each scenario, we find a reduced $\chi^{2}$ of $\chi^{2}_{\mathcal{S}_{0}} = 1.43$, $\chi^{2}_{\mathcal{S}_{1}} = 1.40$, $\chi^{2}_{\mathcal{S}_{2}} = 1.43$ and $\chi^{2}_{\mathcal{S}_{3}} = 1.60$. We can therefore compute an $F$-test, assuming the errors are perfectly normally distributed. We find that $F_{1/0} = 0.98$ has a $p$-value of 66.8\%, $F_{1/2} = 0.98$ with a $p$-value of 65.5\% and finally $F_{1/3} = 0.87$ with a $p$-value of 99.4\%. Therefore, this test does not allow us to conclude that scenario 0, 1, and 2 have a significantly different variance, but it shows that the variances of scenarios 1 and 3 are significantly different, with a probability of 0.6\% of being similar.\\

One can also compute the Bayesian information criterion (BIC)\footnote{BIC = $k\ln n - 2\ln \mathcal{L}_{\rm max}$, where $\mathcal{L}_{\rm max}$ is the maximum of likelihood found, $k$ is the number of free parameters and $n$ is the number of data points.} and the Akaike information criterion (AIC)\footnote{AIC = $2k - 2\ln \mathcal{L}_{\rm max}$, where $\mathcal{L}_{\rm max}$ is the maximum of likelihood found and $k$ is the number of free parameters.} for the four scenarios from the likelihood of the best models. We find $\Delta$BIC$_{1/0} = 6.14$ while $\Delta$AIC$_{1/0} = -9.67$. This means that the BIC supports\footnote{For both the BIC and the AIC, \citet{KassRaftery} defined \textit{strong} evidence as values between 6 and 10, and \textit{very strong} evidence as values larger than 10.} strong evidence of scenario 0, while the AIC supports strong evidence for scenario 1. Scenarios 2 and 3 are clearly rejected against scenario 1 by both criteria which give the same results: $\Delta$BIC$_{1/2} = \Delta$AIC$_{1/2} = -20.23$ and $\Delta$BIC$_{1/3} = \Delta$AIC$_{1/3} = -67.44$. For the comparison between scenario 1 and scenarios 2 and 3, the BIC and the AIC result in the same values because those scenarios have the same number of degrees of freedom. In that case, the value of the BIC and AIC is only based on the difference of $\mathcal{L}_{\rm max}$. Those tests are therefore able to reject scenario 2 and 3, but they either support scenario 0 or scenario 1 depending on the criterion used.\\

As already discussed in \citet{2014arXiv1403.6725D}, the frequentist and Bayesian criteria only use the information contained in the best-fit model, either the minimum of the $\chi^{2}$ or the maximum of the likelihood $\mathcal{L}_{\rm max}$. These criteria therefore do not take into account the uncertainties on the model parameters which are included in the posterior distribution. In the model comparison we performed in section \ref{Bayes}, we estimated the probability of each scenario by marginalising the posterior distribution. By doing that, we took into account not only the best model, but also the entire distribution of the parameters that describes the data. Therefore, the odds ratios presented in section \ref{Bayes} are more robust than the ones presented here.

\subsection{Caveats}
\label{caveats}

The Bayesian analysis and the estimation of the probability scenarios performed in this section are based on several hypotheses which might affect the result and conclusions. We discuss the impact of those hypothesis on the results below.\\

The a priori probability of each scenario was set to be equal because of a lack of information about the relative occurrence rate of those scenarios. If one assumes that actually the a priori probability of the scenarios decreases with the increasing number of stars in the system, scenario 0 would have a higher a priori probability compared with scenario 1. This assumption would counterbalance the Bayes factor which favours scenario 1. This would decrease the significance of scenario 1, and could even provide a higher probability to scenario 0.\\

The analyses used the Dartmouth stellar tracks \citep{2008ApJS..178...89D}, the BT-SETTL stellar atmosphere models \citep{2012IAUS..282..235A}, as well as the SOPHIE CCF calibration from \citet{2010A&A...523A..88B}. The systematic errors from these models are not taken into account in the global error budget. This results in an overestimation of the penalisation of the scenarios through Occam's razor. Because scenario 1 relies more on the stellar models, accounting for such systematic errors in the odds ratio computation would therefore increase the relative probability of scenario 1 against scenario 0.\\

We assumed that no other scenario can explain the data (see eq. \ref{probasum}). Accounting for other scenarios would decrease the probability of scenario 1.\\

Models that are dynamically unstable are not self-penalised in the analyses (see discussion in section \ref{Dynamic}) ; those for which the transiting companion would present significant TTVs variations are also not penalised. However, it is not clear to us which scenarios would be more affected by this effect and how it would change the relative probability of the scenarios.\\

The best models of scenarios 0, 1, and 2 fit almost equally well all the data (see section \ref{F-test}). The differences found in the evidence therefore come from Occam's razor when computing the evidence. However, this is related to the size of the joint prior distribution. If one scenario has priors that are too narrow or too wide compared with other scenarios, it will be too favoured or too penalised, respectively. In the analyses performed here, we chose exactly the same prior distribution for all the parameters that are common in the four scenarios (see Table \ref{ScenarioPriors}, online). By doing this, any of the scenarios is incorrectly penalised by Occam's razor when computing the odds ratio, at least for the parameters in common. Only a few parameters are not common to all the scenarios, and therefore limit the impact of this effect. Scenario 0 has fewer free parameters (35) than scenario 1 (38) and thus should be less penalised by Occam's razor than scenario 1. The fact that scenario 1 turns out to be the most likely scenario means that the data actually support this scenario more reliably in such a way that it balances the penalisation of its wider parameter space.\\

To test the dependance of our results on the choice of priors of the unshared parameters, we reran the analysis of scenario 0 as described in section \ref{Scenario0} but with different priors. We decreased the size of the priors for the following parameters: the linear and quadratic term of the radial velocity drift (the uniform prior size was decreased by a factor of 100 and 1000, respectively), the radial velocity semi-amplitude, and the radius ratio (the upper limit of both Jeffreys priors was decreased by a factor of 5). However, we increased the size of the prior on the systemic radial velocity for scenario 0 to correspond to the one chosen for scenarios 1, 2, and 3 (see Table \ref{ScenarioPriors}, online). The result is that the probability of scenario 1 slightly, but non-significantly, increases to a value of 99.98$^{_{+0.02}}_{^{-0.23}}$\%, while the probability of scenario 0 decreases to a value of 0.020$^{_{+0.400}}_{^{-0.016}}$\%, which is the same level as the probability of scenario 2. This test therefore shows that our results do not depend significantly on the choice of priors for the unshared parameters.\\

The probabilities of the various scenarios strongly depend on the method used to estimate the evidence. In this work, we used the TPM evidence estimator proposed by \citet{2012A&A...544A.116T}. This choice is driven by the fact that it is easy to compute from the result of a MCMC analysis. However, as noticed by \citet{2012A&A...544A.116T} in section 5 of their paper (see also their Table 4), the TPM estimator overestimates the Bayes factor (not penalising the scenario with the widest parameter space as it should) compared with a direct numerical integration of the posterior distribution. The authors present this overestimation as a strength of the TPM in detecting weak signals. This might also be seen as a higher sensitivity of the TPM estimator to false positives: see for example the discussions of the number of planets in the GJ667C system \citep{2013A&A...553A...8D,2013A&A...556A.126A,2014MNRAS.437.3540F}, and the HD41248 system \citep{2013ApJ...771...41J, 2014arXiv1404.6135S}. As a conclusion, using the TPM method might overestimate the actual odds ratios, leading to an overestimation of the relative probability of scenario 1 against the other scenarios. However, our scenarios have nearly the same number of free parameters and most of them have the same priors in the various scenarios. This should limit the impact of the overestimation of the evidence estimation from the TPM method. \\

In this section, we want to point out that the derived probability of scenario 1 might be slightly under- or overestimated by several aforementioned effects. One can also note that the median value of its probability (98.7\%) corresponds to a marginal detection, being slightly smaller than the widely assumed $3\sigma$-detection threshold which corresponds to a probability of 99.7\% (equivalent to an odds ratio of 370). The presence of this binary star thus needs to be independently confirmed (see discussion in section \ref{checkB}). However, even considering the limitations discussed here, scenarios 2 and 3 seems to be significantly rejected. The scepticism resides only between scenario 0 and 1, which have no significant impact on the nature of the candidate and the derived physical properties of the transiting planet.

\section{Physical properties of the system}
\label{SystemProp}


We presented in section \ref{Analysis} a careful and in-depth analysis of the photometric and spectroscopic data of the KOI-1257 system. The planet in the KOI-1257 system has an orbital period of 86.647661 d $\pm$ 3 s. It has a highly eccentric orbit of $0.772 \pm 0.045$. This high eccentricity is supported by both the \textit{Kepler} light curve assuming a prior on the stellar density and the SOPHIE radial velocities as illustrated in Fig. \ref{PhotEccEffect}. The HARPS-N spectroscopic measurements obtained during a transit do not allow a clear detection of the Rossiter-McLaughlin anomaly and do not provide strong constraints on the orbital obliquity of the planet. The SOPHIE radial velocities present a significant quadratic drift that reveals the presence of an outer companion in this system.\\

\begin{figure}[h!]
\begin{center}
\includegraphics[width=\figw]{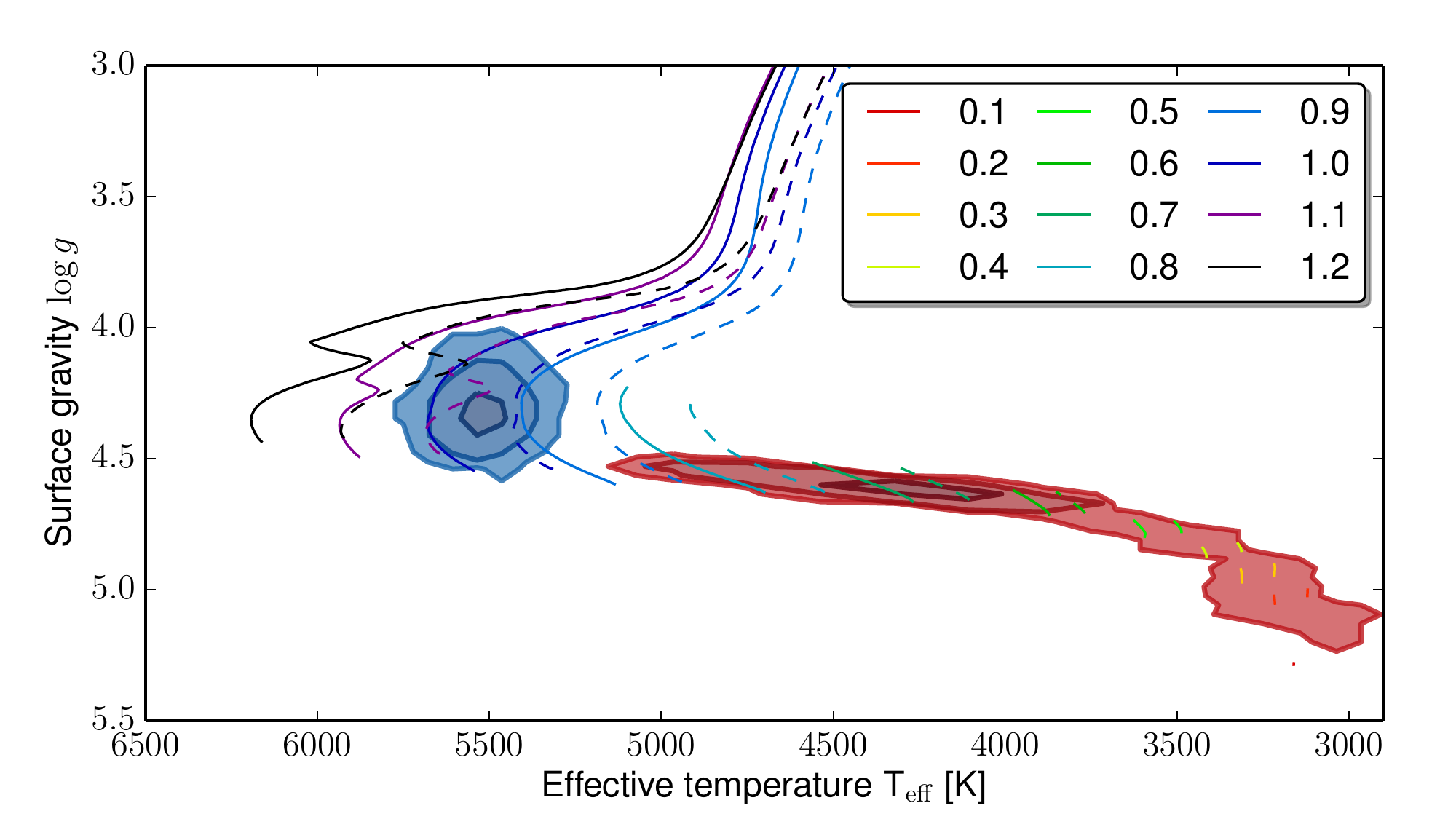}
\caption{Main- and post-main-sequence Dartmouth evolution tracks for a \met\, of 0.15 dex (solid lines) and 0.5 dex (dashed lines). The 68.3\%, 95.5\%, and 99.7\% confidence regions (from scenario 1) of the host star (KOI-1257~A) are superimposed in blue and those of the secondary star (KOI-1257~B) are superimposed in red. The tracks are not displayed after an age corresponding to the age of the Universe.}
\label{1257tracks}
\end{center}
\end{figure}

We presented in section \ref{Blend} a detailed analysis of four scenarios to explain the nature of the quadratic drift observed by SOPHIE. By performing a Bayesian model comparison, we found that the KOI-1257 system is a nearly face-on binary system with an orbital inclination of $18.2^{_{+18.0}}_{^{-5.5}}$ degrees, with regard to the line of sight. It has an orbital period within the range 1030 -- 5830 days at the 99\% level. This binary is composed of two main-sequence stars with respective masses of M$_{\star_{1}}$ = 0.99 $\pm$ 0.05 \Msun\, and M$_{\star_{2}}$ = 0.70 $ \pm $ 0.07 \Msun. However, a low-mass secondary star (down to 0.1 \Msun) is not excluded within 99\% of confidence. Both stars are assumed to have the same metallicity of +0.27 $ \pm $ 0.09 dex. They have an orbital separation of 5.3 $ \pm $ 1.3 AU. The system has an age of 9.3 $\pm$ 3.0 Gyr and is located at 900 $\pm$ 110 pc from the solar system. Thus, the two stars are separated by only 5.8 $ \pm $ 1.6 mas. The confidence regions of the \logg\, and \teff\, of both stars, as derived by the analysis described in section \ref{Scenario1} are displayed in Fig. \ref{1257tracks}, together with the Dartmouth stellar evolution tracks. \\

The planet KOI-1257~b is transiting the primary star of the system, KOI-1257~A. It has a mass of 1.45 $ \pm $ 0.35 \Mjup, a radius of 0.94 $ \pm $ 0.12 \Rjup, hence, it has a planetary bulk density of 2.1 $ \pm $ 1.2 g.cm$^{-3}$. Assuming a perfect redistribution of the heat in the atmosphere and a zero albedo, we estimated its time-averaged equilibrium temperature to be 511 $ \pm $ 50 K. KOI-1257~b therefore belongs to the population of the warm jupiters. All the fitted and deduced physical properties of the KOI-1257 system are shown in Table \ref{params}. These parameters are the ones derived in the analysis of scenario 1. This table also lists the 68.3\% uncertainty on the parameters as well as their 99\% confidence interval. We note that the derived physical properties of the transiting planet are not significantly different if the second companion is a substellar object (scenario 0).\\

It is interesting to note that the actual period of the transiting planet differs from the observed one. Indeed, as observed for the first time by \citet{Romer1677} in the system Io -- Jupiter, because of the finite speed of light, the true period of the transiting planet is shifted by the relative systemic velocity of the star along the line of sight. The true orbital period $P_{\rm true}$ is therefore (neglecting relativistic effects)

\begin{equation}
P_{\rm true} = P_{\rm obs}\times \left(1 - \frac{\gamma}{c}\right),
\end{equation}
where $P_{\rm obs}$ is the observed period and $c$ is the speed of light in the vacuum. In the case of KOI-1257~b, the term $P_{\rm obs}\times (\gamma / c)$ is of 220 $\pm$ 10 s. The true period of KOI-1257~b is therefore $P_{\rm true}$ = 86.64511 d $\pm$ 10 s which is significantly different from the observed one ($P_{\rm obs}$ = 86.647661 d $\pm$ 3 s). This true period should not be used for ephemeris computation. For this reason, we give in table \ref{params} the observed period. The difference between the true and the observed periods being relatively small, this effect should not affect the interpretation or future studies of this system.

\begin{table*}[t!]
\caption{Physical properties of the KOI-1257 systems}
\begin{minipage}[t]{\textwidth} 
\begin{center}
\setlength{\tabcolsep}{5pt}
\begin{tabular}{lcc}
\hline
\hline
Parameter & Median value and 68.3\% uncertainty & 99\% confidence interval \\
\hline
\multicolumn{3}{l}{\it Planet orbital parameters}\\
& & \\
Orbital period $P_{in}$ [d] & 86.647661 $ \pm $ 3.4 10$^{-5}$  & [86.647576 -- 86.647746]\\
Semi-major axis $a_{in}$ [AU] & 0.382 $ \pm $ 0.006 & [0.366 -- 0.398]\\
Transit epoch $T_{0}$ [BJD$_\mathrm{TDB}$ - 2455000] & 6.79454 $ \pm $ 3.0 10$^{-4}$ & [6.79370 -- 6.79538]\\
Orbital inclination $i_{in}$ [$^{\circ}$] & 89.66$^{_{+0.24}}_{^{-0.38}}$ & [88.46 -- 89.97]\\
Orbital eccentricity $e_{in}$ & 0.772 $ \pm $ 0.045 & [0.60 -- 0.88] \\
Argument of periastron $\omega_{in}$ [$^{\circ}$] & 141.3 $ \pm $ 17 & [31 -- 171] \\
Spin-orbit absolute angle $|\lambda |$ [$^{\circ}$] & 74$^{_{+32}}_{^{-46}}$ & [0 -- 180] \\
& & \\
\hline
\multicolumn{3}{l}{\it Binary orbital parameters}\\
& & \\
Orbital period $P_{out}$ [d] & 3430 $ \pm $ 1200 & [1030 -- 5830]\\
Semi-major axis $a_{out}$ [AU] & 5.3 $ \pm $ 1.3 & [2.7 -- 7.9]\\
Periastron epoch $T_{p}$ [BJD$_\mathrm{TDB}$ - 2450000] & 8000$^{_{+3500}}_{^{-2300}}$ & [5145 -- 14545]\\
Orbital inclination $i_{out}$ [$^{\circ}$] & 18.2$^{_{+18.0}}_{^{-5.4}}$ & [8.2 -- 85.2] \\
Orbital eccentricity $e_{out}$ & 0.31$^{_{+0.37}}_{^{-0.21}}$ & [0.02 -- 0.90] \\
Argument of periastron $\omega_{out}$ [$^{\circ}$] & 180 $ \pm $ 110 & [0 -- 360] \\
& & \\
\hline
\multicolumn{3}{l}{\it Primary star parameters}\\
& & \\
Effective temperature \teff$_{\star_{1}}$ [K] & 5520 $ \pm $ 80 & [5320 -- 5720] \\
Surface gravity \logg$_{\star_{1}}$ [\cmss] & 4.32 $ \pm $ 0.10 & [4.03 -- 4.51] \\
Iron abundance \met$_{\star_{1}}$ [dex] & +0.27 $ \pm $ 0.09 & [0.02 -- 0.52] \\
Microturbulence velocity [\kms] & 0.80 & -- \\ 
Macroturbulence velocity [\kms] & 1.7 & -- \\ 
Stellar mass M$_{\star_{1}}$ [\Msun] & 0.99 $ \pm $ 0.05 & [0.86 -- 1.12]\\
Stellar radius R$_{\star_{1}}$ [\Rsun] & 1.13 $ \pm $ 0.14 & [0.90 -- 1.63] \\
Stellar bulk density $\rho_{\star_{1}}$ [$\rho_{\odot}$] & 0.70 $ \pm $ 0.25 & [0.05 -- 1.35]\\
Sky -- projected rotational velocity $\upsilon \sin i_{\star_{1}}$ [\kms] & 4.6 $ \pm $ 0.2 & [4.0 -- 4.9] \\
Spectral Type & G5V & -- \\
& & \\
\hline
\multicolumn{2}{l}{\it Secondary star parameters}\\
& & \\
Effective temperature \teff$_{\star_{2}}$ [K] & 4270 $ \pm $ 290 & [3070 -- 4960] \\
Surface gravity \logg$_{\star_{2}}$ [\cmss] & 4.62 $ \pm $ 0.05 & [4.51 -- 5.20] \\
Stellar mass M$_{\star_{2}}$ [\Msun] & 0.70 $ \pm $ 0.07 & [0.11 -- 0.84] \\
Stellar radius R$_{\star_{2}}$ [\Rsun] & 0.68 $ \pm $ 0.07 & [0.13 -- 0.83] \\
Stellar bulk density $\rho_{\star_{2}}$ [$\rho_{\odot}$] & 2.2 $ \pm $ 0.5 & [1.8 -- 40.2]\\
Sky -- projected rotational velocity $\upsilon \sin i_{\star_{2}}$ [\kms] & 2.6 $ \pm $ 2.0 & [0.3 -- 17.6]\\
Spectral Type & K6V / K7V & --\\
& & \\
\hline
\multicolumn{3}{l}{\it Planet physical parameters}\\
& & \\
Planet mass m$_{p}$ [\Mjup] & 1.45 $ \pm $ 0.35 & [0.05 -- 2.36]\\
Planet radius r$_{p}$ [\Rjup] & 0.94 $ \pm $ 0.12 & [0.75 -- 1.34] \\
Planet bulk density $\rho_{p}$ [g.cm$^{-3}$] & 2.1 $ \pm $ 1.2 & [0.1 -- 5.1] \\
Equilibrium temperature T$_{eq}$ [K] & 511 $ \pm $ 50 & [380 -- 640] \\ 
& & \\
\hline
\multicolumn{3}{l}{\it System parameters}\\
& & \\
Distance from Earth $D$ [pc] & 900 $ \pm $ 110 & [710 -- 1220] \\
Interstellar absorption E(B$-$V)& 0.097 $ \pm $ 0.040 & [0.000 -- 0.200] \\
Systemic radial velocity $\gamma$ [km.s$^{-1}$] & 8.8 $ \pm $ 0.4 & [7.7 -- 9.9]\\ 
System age $\tau$ [Gyr] & 9.3 $\pm$ 3.0 & $>$ 1.55\\
\hline
\hline
\end{tabular}
\tablefoot{Adopted values: \Msun\, = 1.98842 10$^{30}$ kg ; \Rsun\, = 695 508 km ; \Mjup\, = 1.89852 10$^{27}$ kg ; \Rjup\, = 71 492 km ; 1 AU = 149 597 870 700 m. 
}
\end{center}
\end{minipage}
\label{params}
\vspace{-0.3cm}
\end{table*}%

\subsection{Testing the robustness of the stellar parameters in the presence of a contaminant}
\label{SpACheck}
The stellar analysis performed in section \ref{SpA} assumes that the target's spectrum is not significantly blended by another star. However, we showed in section \ref{Blend} that the KOI-1257 system is most likely composed of two stars separated by a few mas. Therefore, the spectrum of KOI-1257 contains the lines from both stars, one contributing to only $\sim$ 9\% of the total flux. To test if this contaminating star does not affect significantly the derived stellar parameters, then used as priors of our analyses, we blended the observed spectrum of a G-dwarf, HD100777, together with the one of a K dwarf, HD32147. For HD100777, \citet{2008A&A...487..373S} found a \teff\, of 5536 $\pm$ 26 K, a \logg\, of 4.33 $\pm$ 0.05 \cmss\, and a \met\, of 0.25 $\pm$ 0.02 dex, which are similar to the assumed stellar parameters of KOI-1257~A. For HD32147, \citet{2006A&A...458..873S} found a \teff\, of 4705 $\pm$ 51 K, a \logg\, of 4.44 $\pm$ 0.31 \cmss\, and a \met\, of 0.30 $\pm$ 0.08 dex, which is similar to the derived parameters of KOI-1257~B. We blended together the HARPS spectrum of these two stars assuming a flux ratio of 9\% for HD32147. The HARPS spectra were chosen because they have the same spectral resolution and spectral range as the HARPS-N spectrum of KOI-1257 used for the spectral analysis. From the analysis of scenario 1, we estimated that the two stars were separated in radial velocity by 1.0$^{_{+2.5}}_{^{-4.5}}$ \kms\, at the time of the HARPS-N observations (early October 2013). We therefore assumed that the two stars are shifted by less than one HARPS-N pixel in our simulation. \\

After normalising the blended spectrum, we used the ARES procedure described in \citet{2007A&A...469..783S} and found that by blending the spectrum of a G-dwarf target with a K-dwarf contaminant, it affects the measured \teff\, by 50 K, the \logg\, by 0.02 \cmss, and the \met\, by 0.01 dex. This impact on the stellar atmospheric parameters is, however, at the same level, or even smaller than the uncertainty on the measurement. We therefore conclude that the derived parameters of the KOI-1257~A star are not significantly affected if the secondary star KOI-1257~B exists or not.

\section{Planetary system evolution}

\subsection{Planet interior structure and composition}

In this section, we investigate the interior structure and composition of KOI-1257~b. With a mass of $\sim$ 1.45 \Mjup, a radius of $\sim$ 0.94 \Rjup, and a time-averaged equilibrium temperature of $\sim$ 511\,K, KOI-1257~b is a warm Jupiter-like planet. Planets with mid to low temperatures are important for testing and improving the theoretical evolution models \citep[eg.][]{2010Natur.464..384D}. They fill the observational gap between the two solar giant planets and the large population of hot jupiters.\\

We used CEPAM \citep{1995A&AS..109..109G, 2010A&A...520A..27G} to build a proper planetary evolution models grid. The planet is assumed to be made of a central rocky core and a solar-composition envelope (the so-called standard models). Of course, we do not know whether the heavy elements are concentrated in a core, dispersed in the envelope, or a mix of both. However, as \citet{2008A&A...482..315B} have shown, dispersing heavy elements in the envelope will tend to produce, at a given age, a smaller planet compared to a core-only model. Hence our models should provide a lower-limit for the total mass of elements heavier than helium.\\

Since the absolute planetary parameters are fully dependent on that of the parent star, and both are model-dependent, we combined stellar \citep[PARSEC: ][]{2012MNRAS.427..127B} and planetary evolution model using SET \citep[see][]{2011A&A...527A..20G, 2011A&A...531A...3H, 2013A&A...555A.118A}. Through the use of SET's statistical algorithm and using the observed values only, we thus obtained posterior probability distributions of the bulk composition of the planet (i.e. its core mass), as well as independent results for the fundamental parameters of both the star and planet, the latter being entirely consistent with those presented in section \ref{SystemProp}.\\

\begin{figure}[h!]
\begin{center}
\includegraphics[width=\figw]{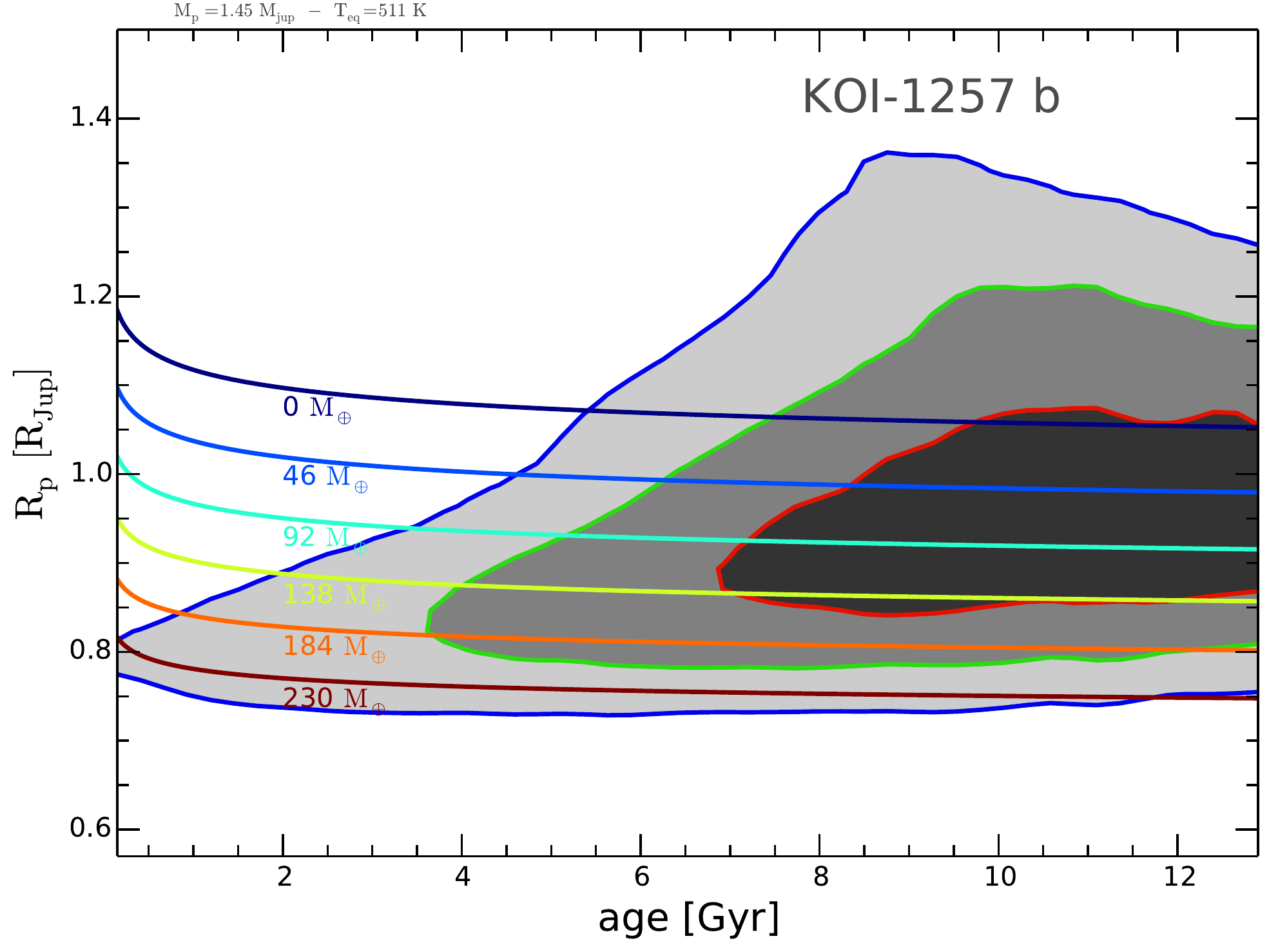}
\caption{Evolution of the KOI-1257b radius as a function of the age. The 68.3\%, 95.5\%, and 99.7\% confidence regions are denoted by black, dark grey, and light grey areas, respectively. The curves represent the thermal evolution of a 1.45 \Mjup\, planet with a time-averaged equilibrium temperature of 511\,K. Text labels indicate the amount of heavy elements in the planet (its core mass, in Earth masses). No dissipation of incoming stellar flux was considered (standard models), as they do not significantly depart from the ones shown above.}
\label{InterStruct}
\end{center}
\end{figure}

The results are presented in terms of planetary radii as a function of age in Fig. \ref{InterStruct}: the regions show the 68.3\%, 95.5\%, and 99.7\% confidence regions from the modelling of the star and transit, while the lines show a subset of planetary models for the nominal mass and equilibrium temperature of the planet at different compositions as labelled. Standard models indicate that the planet has a core mass of $68^{_{+76}}_{^{-68}}$ \Mearth\footnote{results from independent 1-D posterior distributions}, which translates into a heavy elements mass fraction $Z$ of $0.15^{_{+0.16}}_{^{-0.15}}$. The planet could be coreless, but the high metallicity of the parent star suggest that there is a significant amount of heavy elements in its interior \citep[e.g.][]{2006A&A...453L..21G}.\\

Finally, other unknown model parameters could affect the results by making the planet slightly bigger for a given age (and hence implies a larger $Z$ or core mass value): (i) opacities \citep[see eg.][]{2010A&A...520A..27G, 2013MNRAS.434.3283V}, (ii) different equation of state \citep[e.g.][]{2013ApJ...774..148M}, (iii) the dissipation of a fraction of the incoming stellar flux (or other dissipation mechanism) could slow down the planet's contraction \citep[e.g.][]{2002A&A...385..156G, 2013ApJ...772...76S}. However, in this last case, assuming a common value of 1\% for the dissipation of the incoming flux, one would have $5.23 \times 10^{24}$ ergs.s$^{-1}$ of additional energy. Our models show this has a small effect on the radius evolution of the planet with less than 2\%, compared to the almost 13\% from the observational uncertainties.\\

Here we only consider the primary star, KOI-1257~A, and its transiting planet, KOI-1257~b, thus completely ignoring the possible effects of the outer stellar companion, KOI-1257~B on this two-body system. Even though it is not correct, we assume it is a sensible approach for the purpose of getting the composition of the planet, the impact of KOI-1257~B being presumably negligible.\\

KOI-1257~b thus appears to be a Jupiter-like planet with a core-mass estimate consistent with other planets of similar mass and equilibrium temperature. A more detailed study is necessary to uncover the effects of its high eccentricity on its evolution, and the role of the outer binary KOI-1257~B on the formation of such an object. Such a study is beyond the scope of this paper.

\subsection{Dynamical evolution}
\label{Dynamic}
The orbital parameters for the KOI-1257 system given in Table \ref{params} almost fully characterise the geometry of the orbits, since we were able to estimate the inclinations of the orbital planes with respect to the plane of the sky. However, some uncertainty is still present, not only because the inclination of the binary orbit is poorly constrained ($i_{out} = 18.2^{_{+18.0}}_{^{-5.4}}\,^{\circ}$), but mainly because the difference between the longitude of the nodes of the two orbits, $\Delta \Omega$, is unknown.
As a consequence, the mutual inclination, $i$, and the argument of the pericentre of the inner orbit measured from the line of nodes of the two orbits, $\omega$, are also undetermined.
These two parameters are critical to understanding the full dynamics of the system, but we can still place some constraints on their values \citep{2012A&A...541A.151G}:
\begin{equation}
\cos i \approx  \cos \Delta \Omega \sin i_{out} \ , \label{140404a}
\end{equation}
and
\begin{equation}
\cos (\omega - \omega_{in}) \approx \frac{\cos i_{out}}{\sin i} \ , \label{140404b}
\end{equation}
where we assumed $i_{in} \approx 90\,^\circ$.
Therefore, we have that 
\begin{equation}
 i = 90\,^\circ \pm i_{out} \quad \mathrm{and} \quad \omega = \omega_{in} \pm i_{out}  \ . \label{140404c}
\end{equation}
Adopting $\omega_{in}=141\,^\circ$, and the maximum value for the outer orbit inclination, $i_{out} = 36\,^\circ$, we get $54\,^\circ \le i \le 126\,^\circ$, and $105\,^\circ \le \omega \le 177\,^\circ$.
Thus, in spite of all the uncertainty, we conclude that this system has a high mutual inclination. 
In particular, we have that $\cos i < \sqrt{3/5}$, the critical value that allows Lidov-Kozai cycles, where mutual inclination is exchanged with the inner orbit eccentricity \citep[e.g.][]{1962P&SS....9..719L, 1962AJ.....67..591K}.\\

\begin{figure}[h!]
    \includegraphics[width=\figw]{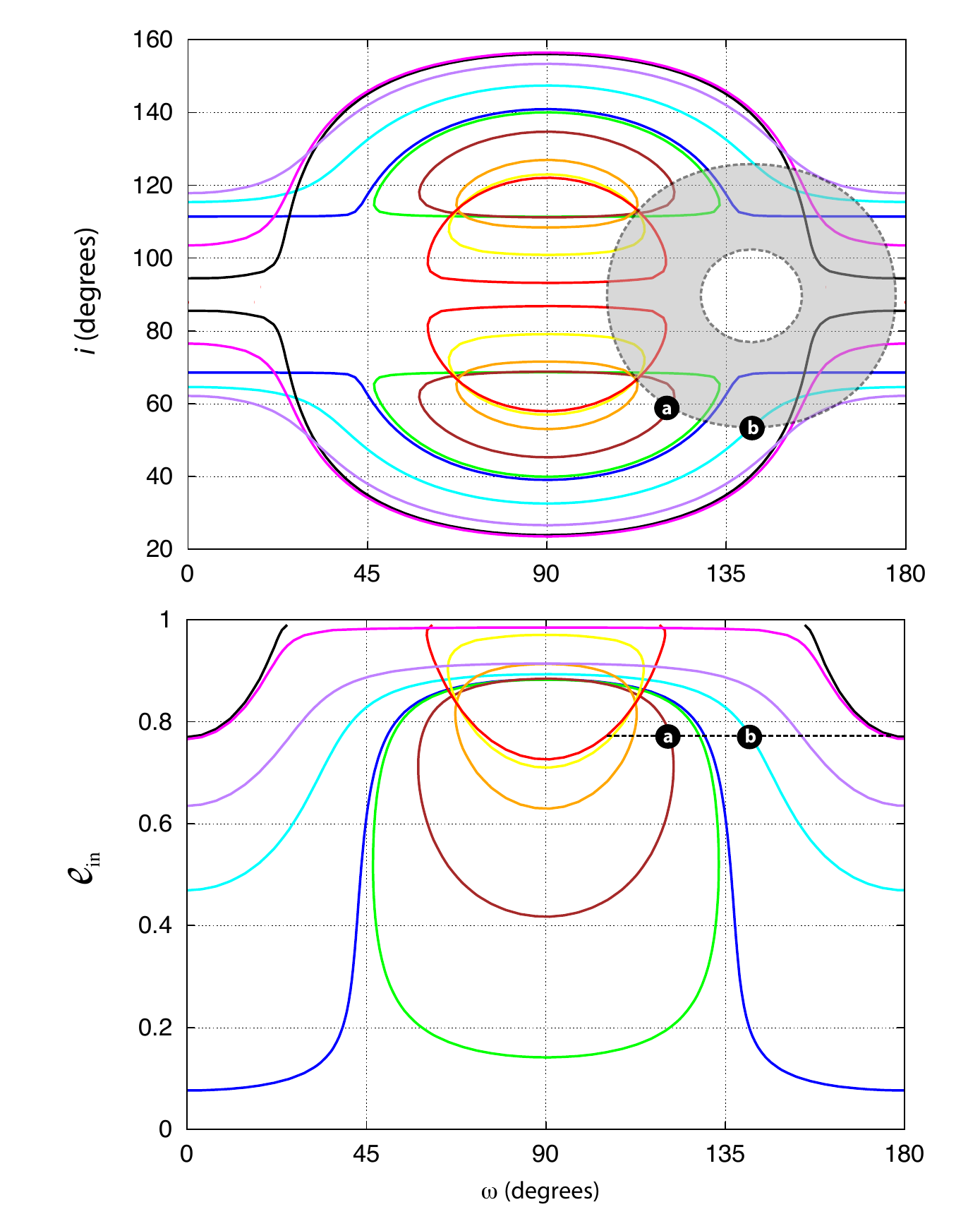} 
  \caption{Possible secular trajectories for the KOI-1257 system seen in the $(\omega,i)$ plane ({\it top}), and in the $(\omega, e_{in})$ plane ({\it bottom}). The colours are preserved in both pictures, each one corresponding to the same orbital evolution. The trajectories correspond to the level curves of constant energy using the quadrupolar approximation combined with general relativity.  The dashed lines define the regions of initial conditions that are compatible with the observational data (Table \ref{params}), namely $i_{out} = 18.2^{_{+18.0}}_{^{-5.4}}$~$^{\circ}$ ({\it top}) and $e_{in} = 0.772$ ({\it bottom}), while $105\,^\circ \le \omega \le 177\,^\circ$ (Eqs.\,\ref{140404a}$-$\ref{140404c}).   \label{dyn1}   }
\end{figure}

Since the system is hierarchical ($a_{in}/a_{out} \approx 0.07$) with high mutual inclination, we can limit the expansion of the potential energy to the order of two in the ratio of semi-major axes, i.e. we can use a quadrupolar approximation to study its secular dynamics \citep[e.g.][]{1962AJ.....67..591K, 2013A&A...553A..39C}. As a result, the orbital evolution becomes integrable and easy to understand.
In Figure~\ref{dyn1} we show all the possible paths that are compatible with the observational data listed in Table \ref{params}.
We observe that there are two possible dynamical regimes: (1) libration around the equilibrium points at $\omega=90\,^\circ$; (2) Lidov-Kozai cycles around the libration region.
More interestingly, we see that for both regimes the eccentricity of the inner orbit undergoes large variations, whose maximum value is always above the presently observed one, $e_{in}=0.772$.
For the orbits starting with a mutual inclination close to $90\,^\circ$, the maximum eccentricity is close to one, meaning that the planet can be engulfed by the central star.\\

Near the pericentre of the orbit, the planet is always close enough to the star to undergo tidal dissipation, which modifies the present orbit.
In order to test this scenario, we performed some numerical simulations of the secular equations of the motion combined with tidal effects, as described in \citet{2011CeMDA.111..105C}.
For the planet dissipation time-lag we adopt $\Delta t = 40$~s, which is equivalent to $Q = 3 \times 10^{4}$, where $Q$ is the quality factor.
For all paths shown in Figure~\ref{dyn1} we observe that the eccentricity is damped in less than 1.5~Gyr, time after which the planet becomes a regular hot jupiter in a circular orbit.
The mutual inclination is also reduced, so that the system becomes nearly coplanar (prograde or retrograde).
The evolution timescale mostly depends on the proximity to the star at the pericentre, that is, 
the orbits that attain higher eccentricities evolve faster.\\

\begin{figure}[h!]
  \includegraphics[width=\figw]{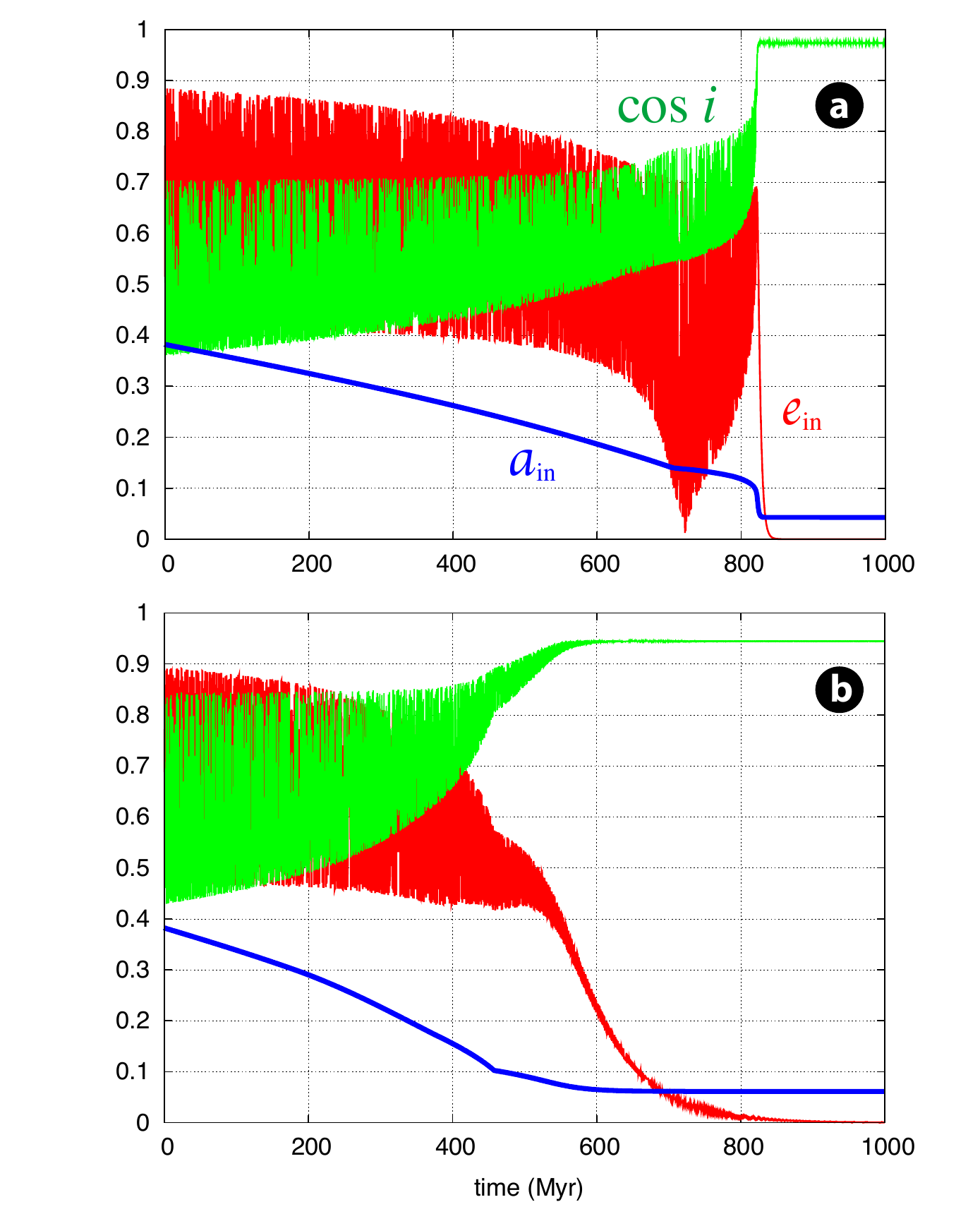} \\    
  \caption{Long-term tidal evolution of the inner orbit of the KOI-1257 system with $Q = 3 \times 10^4$. We show the evolution for two sets of initial conditions taken from Figure~\ref{dyn1}, one starting in libration (a) and the other starting in circulation (b). In both simulations the present configuration can be maintained for at least 0.5~Gyr, after which time the planet evolves into a regular hot jupiter in a close-in circular orbit. \label{dyn2}}   
\end{figure}

In Figure~\ref{dyn2} we show the orbital evolution of two different initial configurations, libration (a) and circulation (b) around the Lidov-Kozai equilibria at $\omega = 90\,^\circ$.
Both configurations correspond to situations where the eccentricity grows to a minimal upper limit of about $e_{in,max} = 0.89$, ensuring that the timescale for tidal evolution is on the low side.
However, we observe that the present configuration cannot be maintained for more than 0.5~Gyr if the planets start in circulation (Fig.~\ref{dyn2}b).
The configuration initially in libration (Fig.~\ref{dyn2}a) holds an eccentric orbit a bit longer ($0.8$~Gyr), because before the eccentricity is damped the planet needs to move outside the libration area.
For initial conditions starting between (a) and (b), the lower limit for the eccentricity oscillations decreases, and for the initial conditions near the separatrix between the circulation and libration regimes it can be as small as 0.1 (Fig.~\ref{dyn1}).
As a consequence, in the frame of a quadrupolar model the tidal evolution can be delayed to a maximum of 1.5~Gyr.
However, this is not very realistic, because near the separatrix the inner orbit becomes chaotic, and the planet becomes a hot jupiter in a few Myr \citep[see][]{2013A&A...553A..39C}.\\

Smaller values for the maximum eccentricity variations could be obtained if the present inclination of the binary orbit is underestimated, that is, if $i_{out} > 36.2\,^\circ$.
The evolution timescale is also nearly inversely proportional to $Q$, so we can delay the evolution by a factor of 10 if we assume $Q = 3 \times 10^{5}$. 
This is still reasonable, and allows the system to stay for additional 5~Gyr near the present configuration (instead of only 0.5~Gyr). 
Given the estimated age of the star ($\sim 9$~Gyr), a delayed scenario is in better agreement with the past history of this system.
We therefore conclude from this analysis that the inclination of the binary companion must be closer to its higher limit, in order to decrease the mutual inclination of the system and the inner orbit maximal eccentricity variations.\\

To account for this dynamical result in the analyses, we reran the \texttt{PASTIS} analysis of scenario 1 as described in section \ref{Scenario1} changing only the prior on the outer-companion orbit to have $i_{out} > 36.2\,^\circ$. The derived parameters are fully compatible with those given in the online table \ref{ScenarioResult}, except for the outer-companion orbital inclination and its orbital eccentricity. The orbital inclination of the binary orbit is $i_{out}$ = 52 $\pm$ 11 $^{\circ}$. The eccentricity of the binary orbit is now surprisingly well constrained with $e_{out}$ = 0.8 $ \pm $ 0.1. This also slightly increases the probability of scenario 1 to $\mathcal{P}\left(\mathcal{S}_{1} \,|\, \mathcal{D}, \mathcal{I}\right)$ = 99.63$^{_{+0.35}}_{^{-8.89}}$ \%, under the hypothesis of eq. \ref{probasum}. This increase might be surprising since the most likely inclination found in section \ref{Scenario1} is now rejected by the new prior. If the maximum of likelihood decreased with this new prior, this scenario would be less probable. However, the size of the prior of scenario 1 decreased, decreasing also its penalisation through Occam's razor. Since the inclination $i_{out}$ is poorly constrained in section \ref{Scenario1} (see the 99\% confidence interval in table \ref{params}), the value of $\mathcal{L}_{max}$ is only slightly changed by rejecting the nearly face-on solutions (the relative decrease of $\ln \mathcal{L}_{max}$ is only of 4 10$^{-5}$). However, the gain due to the decrease of the prior size is slightly higher (the relative increase of the median of the log of the posterior is of 1.5 10$^{-4}$), which explains this counter-intuitive increase in probability.

\section{Discussions}
\label{Discut}

\subsection{Confirming the binary star KOI-1257~B}
\label{checkB}
The presence of a secondary star in the KOI-1257 system was shown in section \ref{Bayes}. However, there are some limitations to the analyses performed (see section \ref{caveats}) and the probability of scenario 0 might be significantly non-zero. Thus there is a low, but non-zero, probability than the secondary star KOI-1257~B does not exist. In this case, the system would be composed of another substellar companion. Moreover, this is the first time here that we infer the presence of a blended star based on the joint analysis of the RV, bisector, and FWHM variation. One might be skeptical about this detection and need some independent confirmation to validate the analysis and the results performed here. There are several ways to independently confirm the presence of KOI-1257~B which we discuss below.\\

The two binary stars are separated by 5.8 $ \pm $ 1.6 mas, which is smaller than the resolution of the Robo-AO observation performed by \citet{2013arXiv1312.4958L}. To resolve these two stars, it is necessary to have either a high-throughput interferometer operating in the visible with a baseline of at least 23.8 $ \pm $ 7.1 m, or an extremely large telescope. Given the magnitude-limit of current interferometers, the confirmation of this binary would be challenging. \\

Another way to confirm the binary would be to detect its astrometric signal. With a mass ratio of 0.71 $ \pm $ 0.06 and an angular separation of 5.8 $ \pm $ 1.6 mas, the astrometric signal of this binary is expected to be of less than 4.1 $ \pm $ 1.1 mas (assuming that the flux contribution from the secondary star is negligible). Theoretically, it should be possible to confirm this binary using the astrometric data obtained by the \textit{Kepler} telescope \citep{2010arXiv1001.0305M}, which has a precision of about 1 mas for KOI-1257. However, \textit{Kepler} astrometric measurements are affected by systematics \citep{2013AAS...22230201B} that make this astrometric detection extremely challenging. The confirmation of the KOI-1257 binary from \textit{Kepler} astrometry is beyond the scope of this paper. The \textit{Gaia} astrometric observatory is expected to reach an astrometric precision at the level of the $\mu$as. This precision will be, by far, enough to detect and confirm the presence of KOI-1257 B.\\

Finally, another way to confirm the presence of KOI-1257~B would be to observe with high-resolution spectrographs when the two stars are at their maximum of radial velocity separation. If they are well separated, the spectrum will look like the one of a double-line binary. However, the current constraints on the orbital parameters of the system do not allow us to predict with a reasonable precision (1) if the stars can be resolved with a high-resolution spectrograph such as HARPS-N or SOPHIE and (2) when this maximum of RV-separation will happen. Future spectroscopic observations of this system will be useful to improve the constraints on this system and improve the predictions. Techniques to disentangle spectra might also be used to confirm the K dwarf companion, such as TODMOR \citep[see for example][ and references therein]{2011A&A...534A..67T}. A detailed analysis of a near-infrared high-resolution spectrum of the system might reveal some spectral lines that are unique to K dwarfs and thus confirm the scenario 1.\\

We note that the derived parameters of the planet do not change significantly between scenario 0 and 1. Thus, if the secondary star KOI-1257~B is not confirmed by future observations, this would not change significantly the physical properties of the transiting planet. In that case, the best parameters for the transiting planet would be those derived by the Model C in section \ref{ModelC}.

\subsection{Magnetic cycle mimicking a binary system ?}

The secondary star KOI-1257 B was highlighted thanks to a drift in both the radial velocity and the FWHM as observed by SOPHIE. However, magnetic cycles are also known to produce radial velocity and FWHM variations with time \citep{2011arXiv1107.5325L}. The target star KOI-1257 is a relatively old G dwarf and does not present variability in the \textit{Kepler} light curve. As illustrated by figure 19 in \citet{2011arXiv1107.5325L}, low-activity G dwarfs present a magnetic-related RV variation at the level of a few \ms. It is therefore extremely unlikely to have observed a low-activity G dwarf for which the magnetic cycle produces an effect on the radial velocities and the FWHM at the level of several hundred of \ms.

\subsection{Other candidate transiting planets in binary}

Among the \textit{Kepler} planet population, other objects were found to transit one of the components of a binary system. This is the case of the KOI-13 / Kepler-13 system where a giant planet is transiting the main component of a A-dwarf binary \citep{2011AJ....142...19H, 2011ApJ...736L...4S, 2012A&A...544L..12S} and the Kepler-14 system \citep{2011ApJS..197....3B}. The planet in the KOI-42 / Kepler-410 system \citep{2014ApJ...782...14V} is also orbiting the main component of a binary revealed by asteroseismology. Recently, \citet{2014ApJ...784...44L} and \citet{2014ApJ...784...45R} reported the interesting cases of KOI-284 / Kepler-132 and KOI-1422 / Kepler-296, which are two systems with several planets transiting each member of a binary system. Other planet-candidate hosts were found by ground-based speckle \citep{2011AJ....142...19H} or high-resolution imaging observations \citep{2013arXiv1312.4958L} to have close contaminant that might or might not be bound with the \textit{Kepler} target.\\ 

With nearly half of the FGK dwarfs being members of a multiple-stellar system \citep{2010ApJS..190....1R, 2014arXiv1401.6825T, 2014arXiv1401.6827T}, it would be surprising not to find a large population of planets in binary. Based on false-positive population synthesis in the \textit{Kepler} field of view, \citet{2013ApJ...766...81F} concluded that planets in binary should indeed be the dominant source of false positives among \textit{Kepler} transit candidates\footnote{Even if they are planets located in the target system, \citet{2013ApJ...766...81F} considered planets transiting the secondary star of a binary system as false positive because the physical parameters derived for those planets (specially their radius) would be strongly affected by the dilution from the binary primary star, if it is not taken into account. A large planet transiting a secondary star of a binary star might therefore mimic an Earth-sized planet, affecting the statistics based on the list of planet candidates.}. One might therefore expect more planets in binary to be found among the \textit{Kepler} planet population and the candidates catalogue.\\

Recently, \citet{2014ApJS..210...20M} reported the results of the ground-based radial velocity follow-up with the Keck telescope of 22 \textit{Kepler} transit hosts. Among them, a few targets present a significant radial velocity drift, as is observed in the KOI-1257 system or non-transiting long-period objects for which only the minimum mass is known. Those targets are KOI-69 / Kepler-93, KOI-104 / Kepler-94, KOI-148 / Kepler-48, KOI-244 / Kepler-25, KOI-246 / Kepler-68, KOI-292 / Kepler-97, and KOI-1442 / Kepler-407. Some of them might therefore be member of binary systems. This is especially true for KOI-292 / Kepler-97 for which a stellar companion at 0.37'' was detected by high-resolution imaging. This might also be the case of the negative-mass planets (KOI-41 / Kepler-100, KOI-82 / Kepler-102, KOI-116 / Kepler-106, and KOI-153 / Kepler-113) reported by \citet{2014ApJS..210...20M}, that most likely can be explained by low-amplitude drifts in the data or stellar activity that are not modelled, as stated by the authors. These drifts might be of planetary or stellar origin. However, if these planets are transiting in binary and without additional constraints, it is not clear whether they are transiting the primary or the secondary star of the system. If they actually orbit the secondary star of a binary system, one might expect their reported physical parameters to be strongly affected. Radial velocity follow-up of these systems with spectrographs such as SOPHIE and HARPS-N that allow the measurement of the bisector and the FWHM might be useful to constrain this kind of scenario, as done in this work for KOI-1257. High-precision astrometric observations from the \textit{Gaia} telescope will also provide useful constraints.\\

Other transiting planet hosts were found to present radial-velocity drifts revealing long-period companions \citep[and references therein]{2014ApJ...785..126K}. Thanks to adaptive optic observations, the authors constrained some of the reported accelerations to be compatible with a stellar companion. However, based on their constraints on the outer-companion \textit{minimum} mass, it is not possible to exclude this companion from the stellar regime. Once again, long-timespan FWHM and bisector measurements on these systems would provide additional constraints on the nature of the outercompanion, as we showed here.

\subsection{KOI-1257~b: a highly-eccentric period valley giant exoplanet}

\begin{figure}[h!]
\begin{center}
\includegraphics[width=\figw]{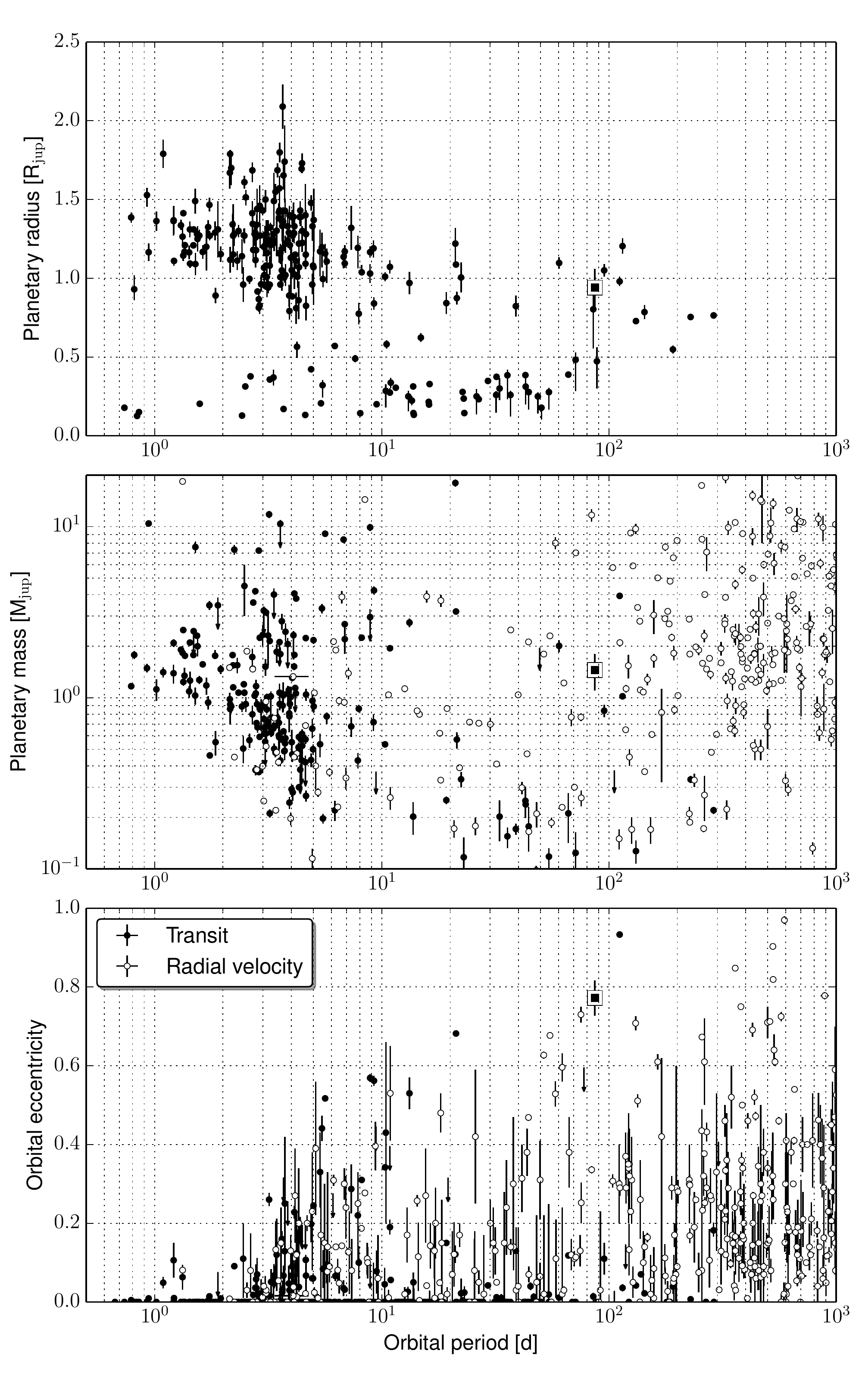}
\caption{From top to bottom: Planetary radius, mass, and orbital eccentricity of confirmed planets as functions of their orbital periods. The black filled dots represent the transiting planets and the black open dots the radial velocity planets. The large, black and white square is KOI-1257~b. Arrows indicate upperlimits in mass or eccentricity. For radial velocity planets, the minimum mass is displayed (source: \href{http://exoplanetarchive.ipac.caltech.edu}{NASA Exoplanet Archive}).}
\label{plotparams}
\end{center}
\end{figure}

Figure \ref{plotparams} displays the mass, the radius, and the eccentricity of KOI-1257~b as functions of its orbital period together with all the transiting and radial velocity planets confirmed so far (source: \href{http://exoplanetarchive.ipac.caltech.edu}{NASA Exoplanet Archive}). KOI-1257~b is among the few known giant planets that have an orbital period longer than 10 days, but less than about 100 days (see Fig. \ref{plotparams}, upper panel). This region is known as the period valley \citep{2003A&A...407..369U}. The relatively few giant planets that are found in this period valley should have a different formation process, a different migration process, or different efficiency than the population of hot jupiters (with orbital periods of less than $\sim$ 10 days) and the population of cold giants (with orbital periods of more than 100 days, like Jupiter).\\

It was proposed by \citet{2013A&A...560A..51A} that giant planets in metal-poor systems might have formed farther out from their host and/or later than in metal-rich systems which makes the migration process less efficient. However, KOI-1257 is a metal-rich system (\met\, = +0.27 $\pm$ 0.09 dex). This hypothesis cannot explain its formation. \citet{2013ApJ...767L..24D} showed that valley-period giant planets orbiting metal-rich stars have much more eccentric orbits than metal-poor stars. They suggested that this is due to planet -- planet scattering that occurs more efficiently in metal-rich systems since they formed planets more efficiently. These planets then interact gravitationally and produce high eccentricity planets. Being metal-rich and highly eccentric, the system KOI-1257 supports this hypothesis. Recently, \citet{2014ApJ...782..113T} proposed another explanation for this metallicity trend in valley-period giant planets, without the need of planet -- planet scattering. Their solution is based on the fact that at the dust sublimation radius ($\sim$ 0.1 AU) the shape of the disk creates a shadow region up to 1 AU which allows disk interactions to damp eccentricity. However, according to the authors, at high metallicity (as in the case of KOI-1257), the optical depth of the disk is higher which decreases the self-shadowing effect. In this case, high eccentricity of planets in metal-rich systems might be explained by a planet -- disk interaction as suggested by \citet{2003ApJ...585.1024G}, without the need of planet -- planet scattering. Such planet -- disk interaction might therefore explain the high eccentricity observed for KOI-1257~b. However, since this planet is most-likely in a binary system, or at least has an outer, massive companion, the interaction with this companion should have played a role in the formation and the dynamical evolution history of KOI-1257~b. If the Lidov-Kozai mechanism and tides are too strong, they would have circularised this planet in a hot-jupiter orbit within a few hundred Myr, as discussed in section \ref{Dynamic}. \\

The KOI-1257 system is similar to the HD80606 system. The latter is also composed of two bounded stars (the second star being HD80607) which host a transiting 111-day period highly-eccentric ($e \sim$ 0.93) giant exoplanet \citep{2001A&A...375L..27N, 2009A&A...498L...5M}. Both are located at the edge of the eccentricity -- period distribution (see Fig. \ref{plotparams}, lower panel). However, the binary separation is much larger in the case of HD80606 (about 1000 UA) than the one found for KOI-1257. The planet HD80606~b has an orbital obliquity of $\lambda = 42 \pm 8\,\degr$\, \citep{2009A&A...498L...5M, 2010A&A...516A..95H}, which helps to constrain its dynamical evolution. Unfortunately, the faintness of the KOI-1257 host does not allow us to significantly detect the Rossiter-McLaughlin effect then used to determine its orbital obliquity.\\

As illustrated in the top panel of figure \ref{plotparams}, the planet KOI-1257~b does not show evidence of an inflated radius, as is common in the population of hot jupiters. This lack of inflated radius was already suggested by \citet{2011ApJS..197...12D} since longer-period planets received much less irradiation from their host stars than the hot jupiters. Taking this into account makes scenario 2 even more unlikely since the planet would have an unexpected inflated radius of $1.56 \pm 0.13$ \Rjup.

\section{Conclusions}
\label{Conclu}

In this paper, we present a new transiting giant planet: KOI-1257~b. It was first detected by the \textit{Kepler} space telescope and then validated thanks to ground-based follow-up observations with the SOPHIE and HARPS-N spectrographs (see section \ref{Obs&Data}). This planet has a relatively long orbital period of $P_{in}$ = 86.647661 d $\pm$ 3 s and a high eccentricity $e_{in}$ = 0.772 $ \pm $ 0.045. This eccentricity is independently supported by both the \textit{Kepler} transit light curves and the SOPHIE and HARPS-N radial velocities (knowing the planet ephemeris), as discussed in section \ref{photecc}. The HARPS-N data during a transit of KOI-1257~b does not allow us to detect significantly the Rossiter-McLaughlin effect, leading to poor constraints on the orbital obliquity of the planet (see section \ref{ModelE}). \\

The SOPHIE radial velocities present a long-term drift that revealed the presence of an outer, massive companion in the system (see section \ref{OuterCompanion}). This companion does not perturb significantly the orbit of the transiting planet within the timespan of the \textit{Kepler} observations since no TTVs were detected (see section \ref{TTVs}). The drift present in the radial velocity data is compatible with a stellar companion, revealing that the KOI-1257 system is actually a binary system. Thus, in such a situation, it is not clear on which star of the binary the planet is transiting. It is also difficult to derive accurate fundamental parameters of the planet since they might be affected by the presence of the other star. Indeed, it is well known that in the case of a binary star system, the observed transit depth would be diluted, which would lead to an underestimation of the planetary radius. Moreover, if both stars of the binary are also spectroscopically blended, the measured radial velocity amplitude would be diluted, leading to an underestimation of the planet mass \citep[as shown for the first time in][]{2002A&A...392..215S}. \\

To solve this uncertainty, we simulated all the SOPHIE data (including the radial velocity, the bisector and the FWHM), the \textit{Kepler} transit light curve and the spectral energy distribution of the system using the \texttt{PASTIS} software \citep{2014arXiv1403.6725D}. We tested four scenarios: the system is composed of a star orbited by two substellar objects ($\mathcal{S}_{0}$), the planet transits the primary star of a binary ($\mathcal{S}_{1}$), the planet transits the secondary star of a binary ($\mathcal{S}_{2}$), and finally, a low-mass star eclipses the secondary star of a triple system ($\mathcal{S}_{3}$). By computing the probability $\mathcal{P}\left(\mathcal{S}_{i} \,|\, \mathcal{D}, \mathcal{I}\right)$ of each scenario $\mathcal{S}_{i}$, we find that the scenario involving a planet transiting the primary star of a binary system ($\mathcal{S}_{1}$) is the most likely scenario, with a probability of 98.7$^{_{+1.2}}_{^{-13.3}}$ \%. As discussed in section \ref{caveats}, this probability might be either over- or underestimated, which might change the conclusion in favour of scenario 0. This scenario 1 is mainly constrained thanks to the variations of both the RV and FWHM and the non-variation of the bisector observed by SOPHIE. \\

By accounting for the secondary star in the system, we find that the planet KOI-1257~b has a mass of m$_{p}$ = 1.45 $ \pm $ 0.35 \Mjup\, and a radius of r$_{p}$ = 0.94 $ \pm $ 0.12 \Rjup\, which give it a bulk density of $\rho_{p}$ = 2.1 $ \pm $ 1.2 g.cm$^{-3}$. With an equilibrium temperature of T$_{eq}$ = 511 $\pm$ 50 K, it belongs to the population of the warm jupiters. KOI-1257~b is the fifth transiting warm jupiter known with both a measured mass and radius (not accounting for the circumbinary planets, see Fig. \ref{plotparams}). Using the CEPAM and SET planet evolution models, we find that the planet has a core mass of $68^{_{+76}}_{^{-68}}$ \Mearth, which corresponds to a heavy elements mass fraction $Z$ of $0.15^{_{+0.16}}_{^{-0.15}}$.\\

The main component of the binary star system has a mass of M$_{\star_{1}} = 0.99  \pm  0.05$, a radius of R$_{\star_{1}}$ = 1.13 $ \pm $ 0.14 \Rsun, and an iron abundance of \met\, = +0.27 $ \pm $ 0.09 dex. The second component of the binary star system has an estimated mass of M$_{\star_{2}}$ = 0.70 $ \pm $ 0.07 \Msun\, and a radius of R$_{\star_{2}}$ = 0.68 $ \pm $ 0.07 \Rsun. It orbits the primary star with a semi-major axis of $a_{out}$ = 5.3 $\pm$ 1.3 AU and an inclination of $i_{out}$ = 18.2$^{_{+18.0}}_{^{-5.4}}\, ^{\circ}$. However, the dynamical evolution analysis presented in section \ref{Dynamic} suggests that the binary system is not nearly face-on, otherwise the Lidov-Kozai mechanism would have circularised the orbit of the transiting planet into a hot jupiter in a few hundred Myr. This not being compatible with the estimated age of the system ($\tau = 9.3 \pm 3.0$ Gyr), the inclination of the binary star system should be $i_{out} > 36.2\, ^{\circ}$. This is compatible with our analysis within the 99 \% confidence interval (see table \ref{params}).\\

This is the first time that a blended stellar system is constrained thanks to a joint analysis of the RV, bisector, and FWHM variation together with a transit light curve and the spectral energy distribution. It was well known that a stellar blended system mimicking a planetary transit might produce a bisector variation \citep[e.g.][]{2002A&A...392..215S, 2004A&A...421L..13B, 2004A&A...426L..15P, 2004ApJ...614..979T, 2009A&A...506..321M, 2009A&A...506..287L}. However, if the contaminating star has a similar \vsini, it might not produce a significant bisector variation. In that case, the FWHM turns out to be more efficient at revealing blended stellar contaminant. \\

In the case of the KOI-1257 system, the contaminating star (the secondary star of the binary) is not the source, nor the host of the transit. In other configuration systems, it would be possible to have a false positive that does not produce significant bisector variation, but does produce significant FWHM variation. To validate the establishment by radial velocity of a transiting planet it is therefore necessary to check not only for bisector variation, but also for FWHM variation. This is especially true if a drift is detected in radial velocity since it might be imprinted by the transit host to the main star of a binary star system. This type of scenario is expected to be the largest source of false positives among the \textit{Kepler} candidates \citep{2013ApJ...766...81F}. This study shows that high-precision bisector and FWHM are therefore useful to constrain this type of false-positive scenario.\\

The HARPS-N observations obtained during a transit of KOI-1257~b do not permit us to detect clearly the Rossiter-McLaughlin effect (see section \ref{ModelE}). Using the Arome tool \citep{2013A&A...550A..53B} implemented into the \texttt{PASTIS} software, we find that $|\lambda |$ = 74$^{_{+32}}_{^{-46}}\, ^{\circ}$. However, this modelling of the Rossiter-McLaughlin effect does not account for the contamination from the contaminating star (KOI-1257~B). By comparing the diluted radial velocity amplitude found in Model C (see section \ref{ModelC}), with the expected radial velocity amplitude of the KOI-1257~A system (after correcting for the dilution from KOI-1257~B), we find that the undiluted amplitude of the Rossiter-McLaughlin effect would be $\sim$ 13\% larger. This dilution factor is too small to change significantly the derived constraints on the orbital obliquity of the transiting planet.\\

This paper also demonstrates that amateur facilities might participate in the follow-up of giant transiting planets that present large and not-well-understood TTV, as already suggested by \citet{2013arXiv1305.3647M}. An efficient collaboration between professional and amateur astronomers will also be extremely useful for the ground-based photometric follow-up of the future \textit{TESS} and \textit{PLATO} space missions. These two missions will have large photometric masks \citep[see][for the case of \textit{PLATO}]{2013arXiv1310.0696R} in which a significant number of background eclipsing or transiting systems might reside \citep{2013sf2a.conf..555S}. Ground-based photometry with small telescopes, such as amateur ones, could efficiently rule out background false positives. It could also significantly decrease the exclusion radius of background contaminants, then be used to validate planets with tools such as \texttt{PASTIS} \citep{2014arXiv1403.6725D}.

\subsection*{System name}

At the time of writing this paper, the system was attributed only two \textit{Kepler} names: the \textit{Kepler} Input Catalog \citep[KIC8751933 ;][]{2011AJ....142..112B} because it is located in the \textit{Kepler} field of view, and the \textit{Kepler} Object of Interest \citep[KOI-1257 ;][]{2011ApJ...736...19B} because a transiting planet candidate (KOI-1257.01) was detected on this target. This catalogue of \textit{Kepler} Objects of Interest has a significantly non-zero false-positive rate \citep{2011ApJ...738..170M, 2012A&A...545A..76S, 2013ApJ...766...81F, 2013A&A...557A.139S}, therefore the planetary nature of candidates has to be established. \\

This study is the first one to show that this transiting planet candidate is a bona fide planet and not a false positive. Following the recommendation of \citet{2011ApJ...736...19B}, we should name this planet KIC8751933~b. Since this KIC identification is not convenient, and since we are the first ones to claim the planetary nature of this candidate, we propose renaming the transiting candidate KOI-1257.01 as the bona fide planet KOI-1257~b. \\

For homogeneity and consistency with the other planets validated in the \textit{Kepler} field of view, a \textit{Kepler} identification will be attributed to this system\footnote{Following the recommendation provided at \url{http://exoplanetarchive.ipac.caltech.edu/docs/KeplerNumbers.html}} after the acceptance of this paper for publication\footnote{This new identification of the system is Kepler-420}. However, even if this study is the first one that validates the planetary nature of the candidate KOI-1257.01, its discovery was only possible thanks to the work done by \citet{2011ApJ...736...19B}. To clearly make the distinction between our characterisation and validation and the complete planet-discovery process (from its detection in the photometric data to its validation and characterisation) made by the \textit{Kepler} team, we decided to name this planet based on its KOI identification, as widely used in the scientific community \citep[e.g.][]{2011ApJ...736L...4S, 2012ApJ...747..144M, 2012AJ....143..111J, 2013ApJ...768...14W}.


\begin{acknowledgements}
We thank the technical team at the Observatoire de Haute-Provence for their support with the SOPHIE instrument and the 1.93 m telescope. In particular, we acknowledge the essential work of the night assistants. Financial support for the SOPHIE observations from the Programme National de Planetologie (PNP) of CNRS/INSU, France is gratefully acknowledged. We also thank the staff of the TNG for their support, especially G.~Andreuzzi and M.~Pedani to have performed the HARPS-N observations.

ACMC acknowledges support by Funda\c{c}\~ao para a Ci\^encia e a Tecnologia, Portugal (PEst-C/CTM/LA0025/2011). NCS and AS acknowledge the support from the European Research Council/European Community under the FP7 through Starting Grant agreement number 239953. NCS was supported by FCT through the Investigador FCT contract reference IF/00169/2012 and POPH/FSE (EC) by FEDER funding through the programme ``Programa Operacional de Factores de Competitividade'' - COMPETE. AS also acknowledges the whole ExoEarth research group at Centro de Astrof\'isaca da Universidade do Porto for fruitful discussions about this system, especially M.~Oshagh, V.~Adibekyan, L.~Benamati, S.~Sousa, P.~Figueira, C.~M.~Correia, J.~Faria, and M.~Montalto. AS also thanks J.~Souchay, G.~Bou\'e, and S.~Eggl for an interesting discussion about the impact of the finite speed of light on the derivation of orbital periods. RA acknowledges support from Spanish Ministry through its Ram\'on y Cajal programme RYC-2010-06519. DV acknowledges G. Brabant and O. Chirpaz. We are grateful to the anonymous referee for his/her fruitful comments and suggestions that greatly improve the quality of the manuscript.

This research was made possible through the use of the AAVSO Photometric All-Sky Survey (APASS), funded by the Robert Martin Ayers Sciences Fund. This publication makes use of data products from the Two Micron All Sky Survey, which is a joint project of the University of Massachusetts and the Infrared Processing and Analysis Center/California Institute of Technology, funded by the National Aeronautics and Space Administration and the National Science Foundation. This publication makes use of data products from the Wide-field Infrared Survey Explorer, which is a joint project of the University of California, Los Angeles, and the Jet Propulsion Laboratory/California Institute of Technology, funded by the National Aeronautics and Space Administration. This research has made use of the VizieR catalogue access tool, CDS, Strasbourg, France. This research has made use of the NASA Exoplanet Archive, which is operated by the California Institute of Technology, under contract with the National Aeronautics and Space Administration under the Exoplanet Exploration Program. It has also made use of the Exoplanet Orbit Database and the Exoplanet Data Explorer at exoplanets.org.
\end{acknowledgements}


\newpage

\begin{appendix}

\section{Radial velocity correction from the constant star HD185144}
\label{constcorr}

\begin{figure}[h!]
\begin{center}
\includegraphics[width=\figw]{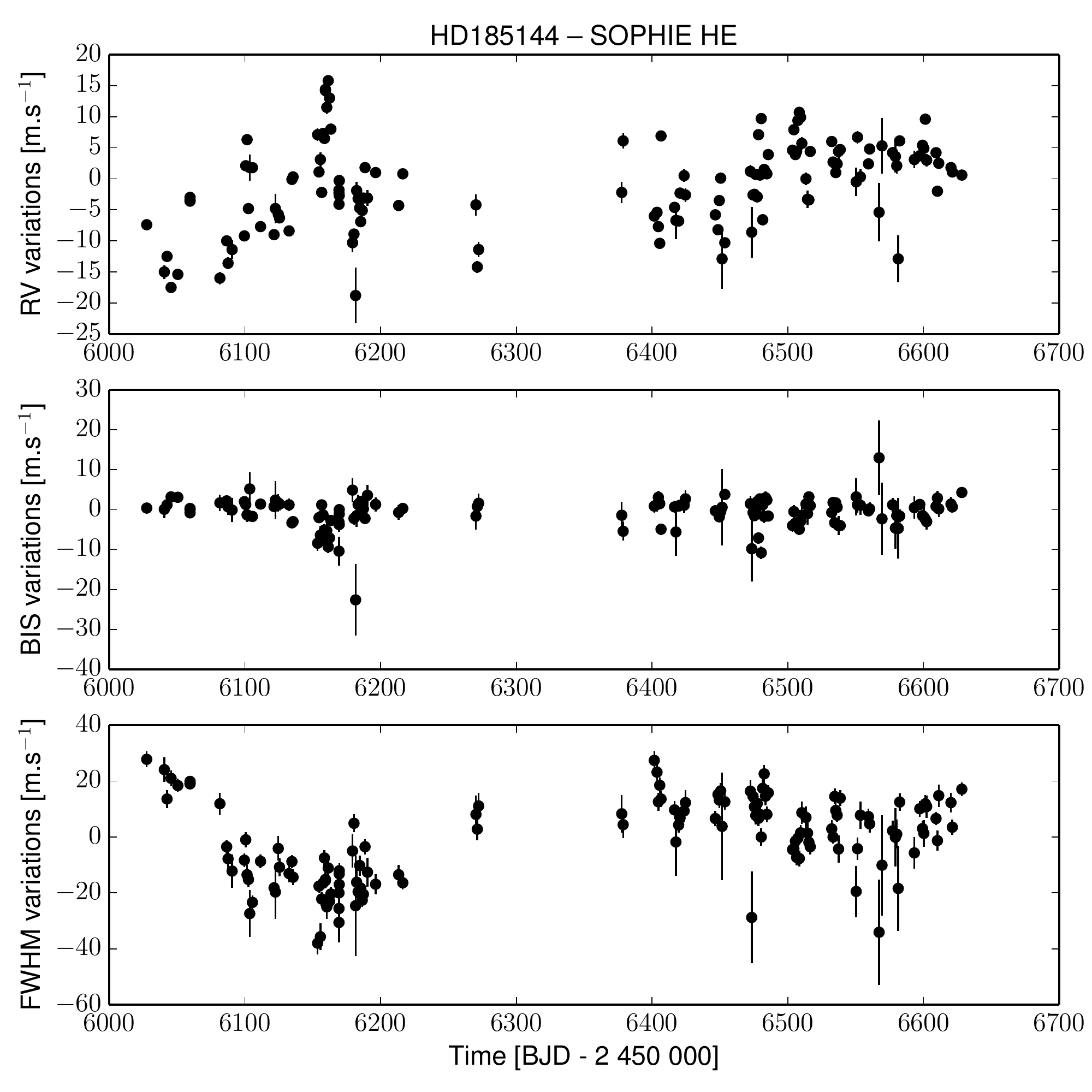}
\caption{Radial velocities, bisector, and FWHM variations (from top to bottom) of the constant star HD185144 observed by SOPHIE in HE during 2012 and 2013. We corrected these measurements by using their median value: about 27.79 \kms\, for the RVs, about 8.8 \kms\, for the FWHM, and 3 \ms\, for the bisector.}
\label{185144}
\end{center}
\end{figure}

We present in this Appendix the SOPHIE radial velocities of the constant star HD185144. During 2012 and 2013, this constant star was observed systematically on the same nights as the \textit{Kepler} targets. It was observed using the same instrumental mode, i.e. the High-Efficiency mode (HE), but in ThoSimult mode (with the calibration ThoAr lamp observed simultaneously). This star was also observed in High-Resolution mode (HR), less systematically, in order to control the performance and stability of the spectrograph to search for low-mass planets \citep{2013A&A...549A..49B}. We selected here only the observations that have reached a signal-to-noise per pixel in the spectra of at least 50 at 550nm. The star HD185144 was observed 137 times in HE during 2012 and 2013. Figure \ref{185144} shows its radial velocity, bisector, and FWHM variations. Those measurements are listed in the online table \ref{HD185144RVtable}.\\

Radial velocities of HD185144 present a RMS of 6.9 \ms\, in HE during both seasons. During 2012 only, this RMS was of 8.0 \ms\, while in 2013 it was of 5.5 \ms. These RMS are much smaller than the radial velocity uncertainty of the target KOI-1257 ($<\sigma_{\rm RV}>\, = 26$ \ms). However, even if the RMS is relatively small for the required precision of KOI-1257, the constant star was observed to vary in HE by a maximum of 34 \ms\, in 2012 during a timescale of 20 days, and by 24 \ms\, in 2013 with a timescale of four days. This corresponds to about one third and one fourth of the radial velocity amplitude of the transiting planet (K = 94 $\pm$ 21 \ms). Those variations are not well understood but seem to be correlated with the temperature outside the dome. During the summers of 2012 and 2013 at Observatoire de Haute-Provence there was a fast drop in the temperature due to a storm after a week of relatively hot days. This fast drop of the local temperature at the observatory is observed for both seasons in HD185144 data by a radial velocity variation at the level of $\sim$ 20 \ms\, with a timespan of about one week. This effect is not observed in the HR mode. Before the implementation of octagonal fibres in June 2011 \citep{2011SPIE.8151E..37P}, the same effect was observed with an amplitude about five times larger \citep[see][]{2013A&A...549A.134H}. \\

The bisector of HD185144 does not show variation within $\sim$ 4 \ms\, during 2012 and 2013. We concluded that SOPHIE bisectors are stable and we did not correct the observed bisectors of KOI-1257.\\

The observed FWHM of HD185144 presents a RMS of about 15 \ms\, for 2012 and 2013. The observed pattern is not well understood, but is correlated with the flux of the ThoAr lamp. The peak-to-valley amplitude is of about 60 \ms, which is one order of magnitude less than the variation observed for KOI-1257 (at the level of about 600 \ms). Since the observations of KOI-1257 were not performed with the simultaneous ThoAr lamp, we decided not to apply a correction to its FWHMs.\\

\section{Upper-limit constraint in mass from a radial velocity drift.}
\label{RVdrift}

\begin{figure}[h!]
\begin{center}
\includegraphics[width=\figw]{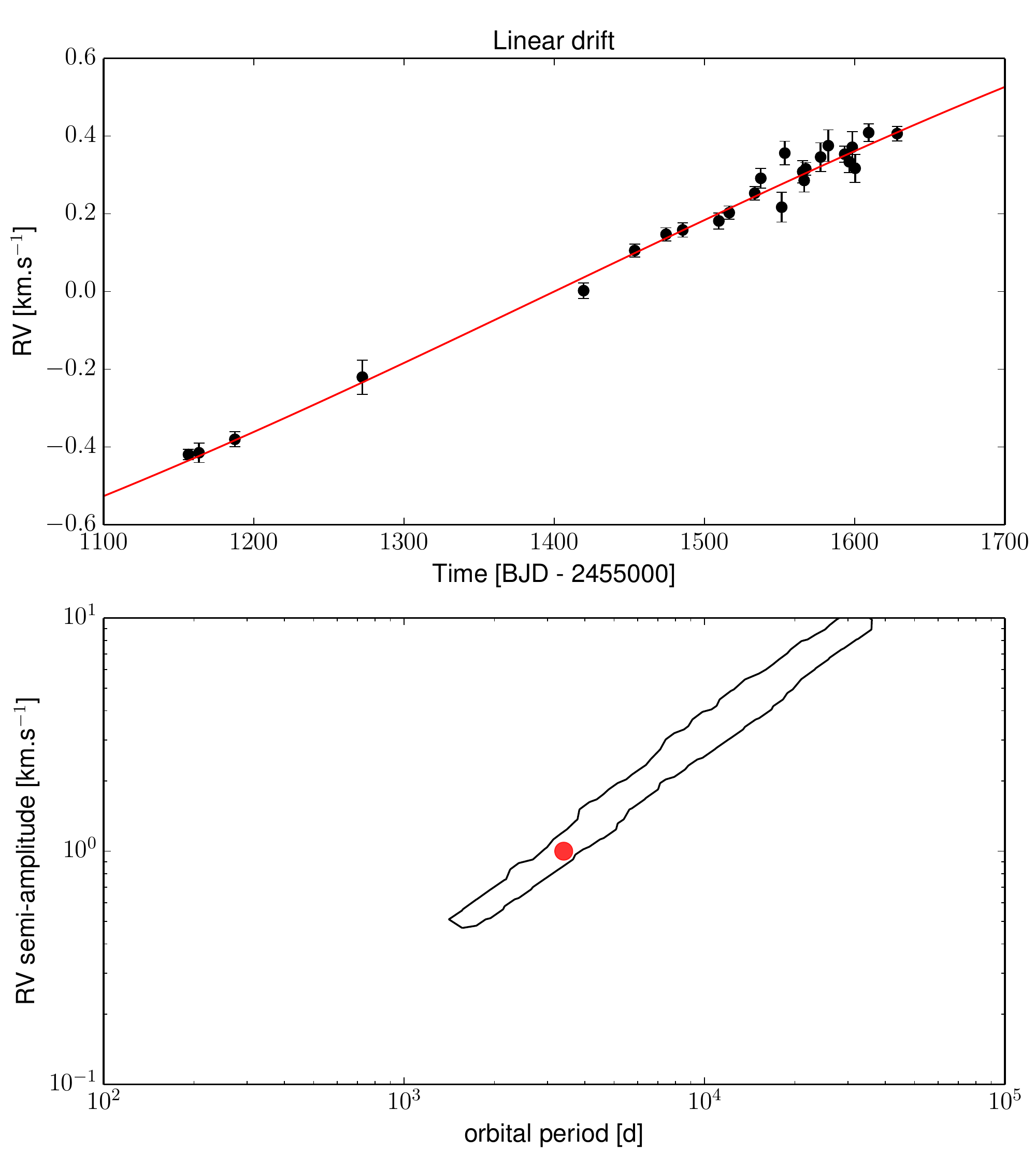}
\caption{\textit{Top panel:} Synthetic radial velocity dataset (black points) superimposed with the circular orbit model (red line) used to generate the data. This model shows only a linear drift during the timespan of the observations. \textit{Bottom panel:} The 99.7\% confidence region of the posterior distribution for the orbital period versus the radial velocity semi-amplitude (black line). The modelled orbit is marked with the red circle. The upper limit in the radial velocity amplitude (at 10\kms) comes from the prior.}
\label{testRVlin}
\end{center}
\end{figure}

\begin{figure}[h!]
\begin{center}
\includegraphics[width=\figw]{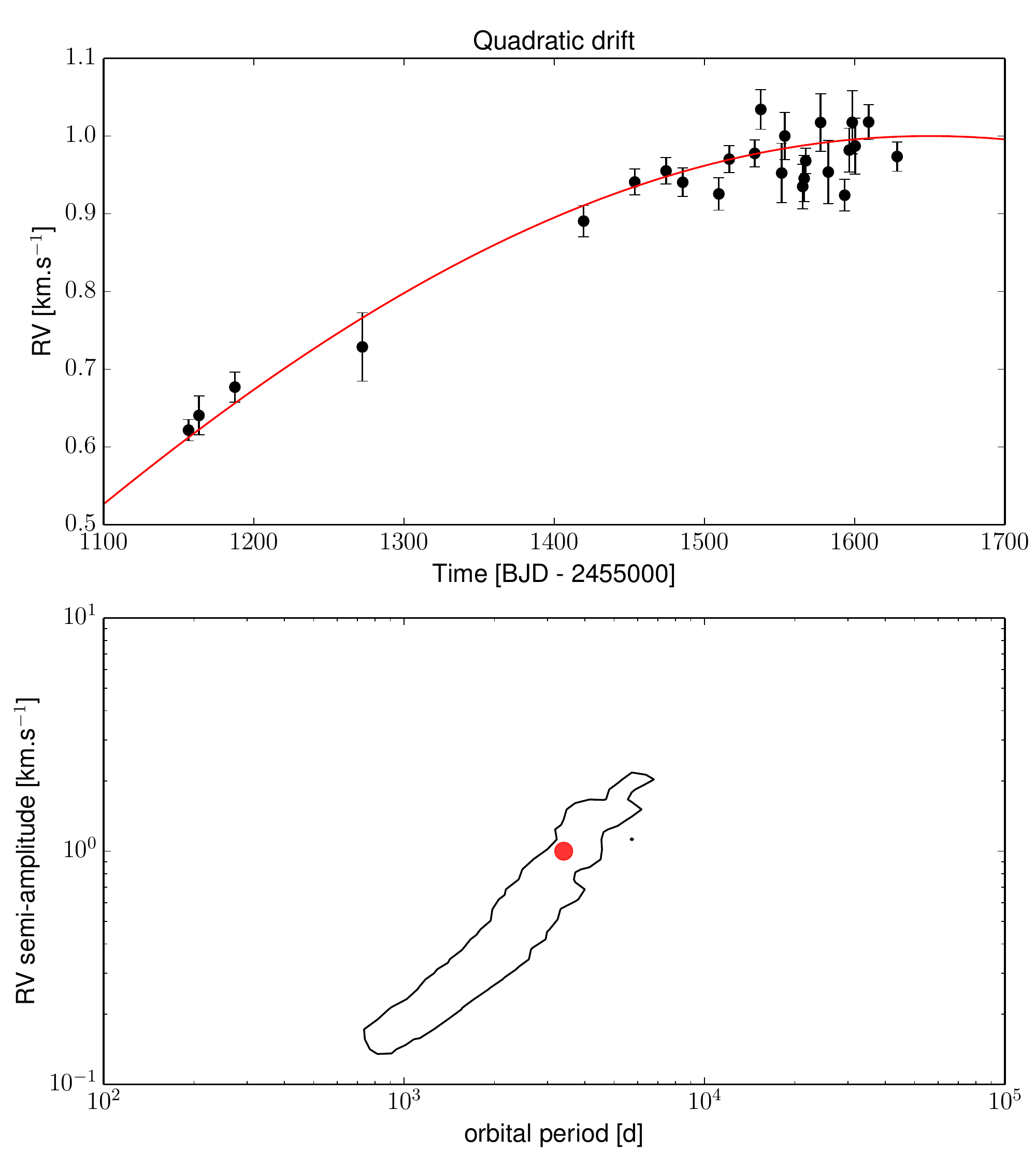}
\caption{Same as Fig. \ref{testRVlin}, but for the quadratic drift dataset.}
\label{testRVq}
\end{center}
\end{figure}

The quadratic radial velocity drift observed in SOPHIE data of KOI-1257 was analysed in section \ref{OuterCompanion} assuming a circular orbit. The MCMC analysis converged toward a higher probability of having a short-period brown dwarf rather than a long-period solar-like star. However, it is well known that from a radial velocity drift, it is possible to constrain only the lower-limit in mass of a companion, but not its upper-limit in mass. To understand the result obtained in section \ref{OuterCompanion}, we performed the following test. We generated synthetic radial velocity data using the SOPHIE observations (the observing times and the radial velocity uncertainty) assuming a pure white noise. We modelled two circular orbits with a period of 3400 days (which corresponds to the most-likely period of the outer orbit in the KOI-1257 system) and a radial-velocity semi-amplitude of 1 \kms. The epoch of periastron of the two orbits were chosen in order that the data display either a linear drift or a quadratic drift, the latter being the same as observed in the case of KOI-1257. The synthetic data for the linear and quadratic drift are displayed in the top panels of figures \ref{testRVlin} and \ref{testRVq}, respectively. \\

We analysed both datasets as done in section \ref{OuterCompanion} but assuming that the system is only described with a circular orbit. We used exactly the same priors for both analyses, using Jeffreys priors for both the orbital period and the radial-velocity semi-amplitude. The other priors were chosen as large and uniform distributions. We show in the bottom panels of figures \ref{testRVlin} and \ref{testRVq} the 99.7\% confidence region from the posterior distribution computed through the same MCMC procedure as in section \ref{OuterCompanion}. The two posterior distributions present a different shape: while the linear-drift dataset (Fig. \ref{testRVlin}) provides only a lower-limit in mass of the companion, the quadratic-drift dataset (Fig. \ref{testRVq}) provides both a lower- and an upper-limit in mass. This can be explained easily by considering that it is more likely to observe a significant curvature in the radial velocity data if the orbital period is relatively short. On the other hand, for orbital periods much longer than the timespan of the observations, it is more likely to observe a nearly linear drift. This effect might also be explained by the fact that the second-order polynomial, needed to describe the radial-velocity data, more accurately constrains the orbit of the companion than a first-order polynomial. Then, the constraints in period translate into constraints in radial-velocity amplitude (or companion mass) thanks to the assumption of purely circular orbit.\\

The constraints on the upper-mass of the companion found in section \ref{OuterCompanion} therefore come from the assumption of a perfectly circular orbit and the curvature of the drift observed by SOPHIE.

\end{appendix}

\Online
\onllongtab{
\begin{landscape}
\begin{longtable}{lcccccc}
\caption{Priors used for the analysis of models A, B, C, D, and E: $\mathcal{U}(a,b)$ represents a Uniform prior between $a$ and $b$; $\mathcal{J}(a,b)$ represents a Jeffreys distribution between $a$ and $b$; $\mathcal{N}(\mu,\sigma^{2})$ represents a Normal distribution with a mean of $\mu$ and a width of $\sigma^{2}$; $\mathcal{N_{A}}(\mu,\sigma_{-}^{2}, \sigma_{+}^{2})$ represents an asymmetric Normal distribution with mean $\mu$, upper width $\sigma_{+}^{2}$ and lower width $\sigma_{-}^{2}$; $\mathcal{N_{U}}(\mu,\sigma^{2}, a, b)$ represents a Normal distribution with a mean of $\mu$ and a width of $\sigma^{2}$ and limited by a Uniform distribution between $a$ and $b$; and finally $\mathcal{S}(a,b)$ represents a Sine distribution between $a$ and $b$.}\\
\hline
\hline
Parameter & Model A & Model B & Model C & Model D & Model E\\
\hline
\endfirsthead
\caption{Continued.} \\
\hline
Parameter & Model A & Model B & Model C & Model D & Model E\\
\hline
\endhead
\hline
\endfoot
\hline
\hline
\endlastfoot
Number of free parameters & 23 & 9 & 33 & 28 & 50 \\
\hline
\multicolumn{6}{l}{\it Orbital parameters}\\
& & & & &  \\
Orbital period $P$ [d] & $\mathcal{N}(86.64774, 0.001)$ & -- & $\mathcal{N}(86.64774, 0.001)$ & $\mathcal{N}(86.64774, 0.001)$ & $\mathcal{N}(86.64774, 0.001)$ \\
Transit epoch $T_{0}$ [BJD$_\mathrm{TDB}$ - 2455000] & $\mathcal{N}(6.79341, 0.02)$ & -- & $\mathcal{N}(6.79341, 0.02)$ & $\mathcal{N}(6.79341, 0.02)$ & $\mathcal{N}(6.79341, 0.02)$ \\
Orbital inclination $i_p$ [$^{\circ}$] & $\mathcal{S}(80,90)$ & -- & $\mathcal{S}(80,90)$ & $\mathcal{S}(80,90)$ & $\mathcal{S}(80,90)$ \\
Orbital eccentricity $e_p$ & $\mathcal{U}(0,1)$ & $\mathcal{U}(0,1)$ & $\mathcal{U}(0,1)$ & $\mathcal{U}(0,1)$ & $\mathcal{U}(0,1)$ \\
Argument of periastron $\omega_p$ [$^{\circ}$] & $\mathcal{U}(0,360)$ & $\mathcal{U}(0,360)$ & $\mathcal{U}(0,360)$ & $\mathcal{U}(0,360)$ & $\mathcal{U}(0,360)$ \\
Spin-orbit angle $\lambda$ [$^{\circ}$] & -- & -- & -- & -- & $\mathcal{U}(0,360)$ \\
& & & & & \\
\hline
\multicolumn{6}{l}{\it Transit parameters}\\
& & & & & \\
Radius ratio $r_{p}/R_{\star}$ & $\mathcal{J}(0.01, 0.5)$ & -- & $\mathcal{J}(0.01, 0.5)$ & $\mathcal{J}(0.01, 0.5)$ & $\mathcal{J}(0.01, 0.5)$ \\
System scale $a/R_{\star}$ & -- & -- & -- & $\mathcal{U}(1, 1000)$ & -- \\
Linear limb darkening coefficient $u_{a}$ & $\mathcal{U}(-0.5,1.2)$ & -- & $\mathcal{U}(-0.5,1.2)$ & $\mathcal{U}(-0.5,1.2)$ & $\mathcal{U}(-0.5,1.2)$ \\
Quadratic limb darkening coefficient $u_{b}$ & $\mathcal{U}(-0.5,1.2)$ & -- & $\mathcal{U}(-0.5,1.2)$ & $\mathcal{U}(-0.5,1.2)$ & $\mathcal{U}(-0.5,1.2)$ \\
& & & & &\\
\hline
\multicolumn{6}{l}{\it Velocimetric parameters}\\
& & & & & \\
Systemic velocity $\gamma$ [km.s$^{-1}$] & -- & $\mathcal{U}(-10,20)$ & $\mathcal{U}(-10,20)$ & $\mathcal{U}(-10,20)$ & $\mathcal{U}(-10,20)$ \\ 
Linear drift $d_{1}$ [km.s$^{-1}$.$d^{-1}$] & -- & $\mathcal{U}(-1, 1)$ & $\mathcal{U}(-1, 1)$ & $\mathcal{U}(-1, 1)$ & $\mathcal{U}(-1, 1)$ \\ 
Quadratic drift $d_{2}$ [km.s$^{-1}$.$d^{-2}$] & -- & $\mathcal{U}(-1, 1)$ & $\mathcal{U}(-1, 1)$ & $\mathcal{U}(-1, 1)$ & $\mathcal{U}(-1, 1)$ \\
Radial velocity semi-amplitude $K$ [m.s$^{-1}$] & -- & $\mathcal{J}(1, 1000)$ & $\mathcal{J}(1, 1000)$ & $\mathcal{J}(1, 1000)$ & $\mathcal{J}(1, 1000)$ \\
& & & & & \\
\hline
\multicolumn{6}{l}{\it Stellar parameters}\\
& & & & & \\
Effective temperature \teff\, [K] & $\mathcal{N}(5540, 90)$ & -- & $\mathcal{N}(5540, 90)$ & -- & $\mathcal{N}(5540, 90)$ \\
Iron abundance \met\, [dex] & $\mathcal{N}(0.26, 0.10)$ & -- & $\mathcal{N}(0.26, 0.10)$ & -- & $\mathcal{N}(0.26, 0.10)$ \\
Bulk density $\rho_{\star}$ [$\rho_{\odot}$] & $\mathcal{N_{A}}$(0.33, 0.10, 0.36) & -- & $\mathcal{N_{A}}$(0.33, 0.10, 0.36) & $\mathcal{N_{A}}$(0.33, 0.10, 0.36) & $\mathcal{N_{A}}$(0.33, 0.10, 0.36) \\
Projected rotational velocity \vsini\, [\kms] & -- & -- & -- & -- & $\mathcal{N_{U}}(4, 2, 0, 20)$ \\
& & & & &  \\
\hline
\multicolumn{6}{l}{\it System parameters}\\
& & & & &  \\
Distance from Earth $D$ [pc] & -- & -- & $\mathcal{U}(10, 5000)$ & -- & $\mathcal{U}(10, 5000)$ \\
Interstellar absorption E(B$-$V) [mag] & -- & -- & $\mathcal{U}(0,2)$ & -- & $\mathcal{U}(0,2)$ \\
& & & & &  \\
\hline
\multicolumn{6}{l}{\it Instrumental parameters}\\
& & & & &  \\
\multicolumn{6}{l}{\it \textit{Kepler} season 0}\\
Jitter [\%] & $\mathcal{U}(0, 10)$ & -- & $\mathcal{U}(0, 10)$ & $\mathcal{U}(0, 10)$ & $\mathcal{U}(0, 10)$ \\
Contamination [\%] & $\mathcal{N_{U}}(10.8, 5, 0, 100)$ & -- & $\mathcal{N_{U}}(10.8, 5, 0, 100)$ & $\mathcal{N_{U}}(10.8, 5, 0, 100)$ & $\mathcal{N_{U}}(10.8, 5, 0, 100)$ \\
Flux out-of-transit & $\mathcal{U}(0.999, 1.001)$ & -- & $\mathcal{U}(0.999, 1.001)$ & $\mathcal{U}(0.999, 1.001)$ & $\mathcal{U}(0.999, 1.001)$ \\
& & & & &  \\
\multicolumn{6}{l}{\it \textit{Kepler} season 1}\\
Jitter [\%] & $\mathcal{U}(0, 10)$ & -- & $\mathcal{U}(0, 10)$ & $\mathcal{U}(0, 10)$ & $\mathcal{U}(0, 10)$ \\
Contamination [\%] & $\mathcal{N_{U}}(14.6, 5, 0, 100)$ & -- & $\mathcal{N_{U}}(14.6, 5, 0, 100)$ & $\mathcal{N_{U}}(14.6, 5, 0, 100)$ & $\mathcal{N_{U}}(14.6, 5, 0, 100)$ \\
Flux out-of-transit & $\mathcal{U}(0.999, 1.001)$ & -- & $\mathcal{U}(0.999, 1.001)$ & $\mathcal{U}(0.999, 1.001)$ & $\mathcal{U}(0.999, 1.001)$ \\
& & & & &  \\
\multicolumn{6}{l}{\it \textit{Kepler} season 2}\\
Jitter [\%] & $\mathcal{U}(0, 10)$ & -- & $\mathcal{U}(0, 10)$ & $\mathcal{U}(0, 10)$ & $\mathcal{U}(0, 10)$ \\
Contamination [\%] & $\mathcal{N_{U}}(13.3, 5, 0, 100)$ & -- & $\mathcal{N_{U}}(13.3, 5, 0, 100)$ & $\mathcal{N_{U}}(13.3, 5, 0, 100)$ & $\mathcal{N_{U}}(13.3, 5, 0, 100)$ \\
Flux out-of-transit & $\mathcal{U}(0.999, 1.001)$ & -- & $\mathcal{U}(0.999, 1.001)$ & $\mathcal{U}(0.999, 1.001)$ & $\mathcal{U}(0.999, 1.001)$ \\
& & & & &  \\
\multicolumn{6}{l}{\it \textit{Kepler} season 3}\\
Jitter [\%] & $\mathcal{U}(0, 10)$ & -- & $\mathcal{U}(0, 10)$ & $\mathcal{U}(0, 10)$ & $\mathcal{U}(0, 10)$ \\
Contamination [\%] & $\mathcal{N_{U}}(12.1, 5, 0, 100)$ & -- & $\mathcal{N_{U}}(12.1, 5, 0, 100)$ & $\mathcal{N_{U}}(12.1, 5, 0, 100)$ & $\mathcal{N_{U}}(12.1, 5, 0, 100)$ \\
Flux out-of-transit & $\mathcal{U}(0.999, 1.001)$ & -- & $\mathcal{U}(0.999, 1.001)$ & $\mathcal{U}(0.999, 1.001)$ & $\mathcal{U}(0.999, 1.001)$ \\
& & & & & & \\
\multicolumn{6}{l}{\it SOPHIE}\\
Jitter [\ms] & -- & $\mathcal{U}(0, 1000)$ & $\mathcal{U}(0, 1000)$ & $\mathcal{U}(0, 1000)$& $\mathcal{U}(0, 1000)$ \\
& & & & &  \\
\multicolumn{6}{l}{\it HARPS-N}\\
Jitter [\ms] & -- & $\mathcal{U}(0, 1000)$ & $\mathcal{U}(0, 1000)$ & $\mathcal{U}(0, 1000)$ & $\mathcal{U}(0, 1000)$ \\
Offset [\ms] & -- & $\mathcal{U}(-300, 300)$ & $\mathcal{U}(-300, 300)$ & $\mathcal{U}(-300, 300)$ & $\mathcal{U}(-300, 300)$ \\
& & & & &  \\
\multicolumn{6}{l}{\it SED}\\
Jitter [mags] & -- & -- & $\mathcal{U}(0,1)$ & -- & $\mathcal{U}(0,1)$  \\
& & & & &  \\
\multicolumn{6}{l}{\it OHP-T120}\\
Jitter [\%] & -- & --  & -- & -- & $\mathcal{U}(0,10)$ \\
Contamination [\%] & -- & -- & -- & -- & $\mathcal{U}(0,1)$ \\
Flux out-of-transit & -- & -- & -- & -- & $\mathcal{U}(0.9, 1.1)$ \\
& & & & &  \\
\multicolumn{6}{l}{\it IAC80}\\
Jitter [\%] & -- & --  & -- & -- & $\mathcal{U}(0,10)$ \\
Contamination [\%] & -- & -- & -- & -- & $\mathcal{U}(0,1)$ \\
Flux out-of-transit & -- & -- & -- & -- & $\mathcal{U}(0.9, 1.1)$ \\
& & & & &  \\
\multicolumn{6}{l}{\it ROTAT}\\
Jitter [\%] & -- & --  & -- & -- & $\mathcal{U}(0,10)$ \\
Contamination [\%] & -- & -- & -- & -- & $\mathcal{U}(0,1)$ \\
Flux out-of-transit & -- & -- & -- & -- & $\mathcal{U}(0.9, 1.1)$ \\
& & & & &  \\
\multicolumn{6}{l}{\it MOOS}\\
Jitter [\%] & -- & --  & -- & -- & $\mathcal{U}(0,10)$ \\
Contamination [\%] & -- & -- & -- & -- & $\mathcal{U}(0,1)$ \\
Flux out-of-transit & -- & -- & -- & -- & $\mathcal{U}(0.9, 1.1)$ \\
& & & & &  \\
\multicolumn{7}{l}{\it Engarouines}\\
Jitter [ppt] & -- & --  & -- & -- & $\mathcal{U}(0,10)$ \\
Contamination [\%] & -- & -- & -- & -- & $\mathcal{U}(0,1)$ \\
Flux out-of-transit & -- & -- & -- & -- & $\mathcal{U}(0.9, 1.1)$ \\
\label{PriorModel}
\end{longtable}%
\end{landscape}
} 

\onllongtab{
\begin{landscape}
\begin{longtable}{lcccccc}
\caption{Results of the analysis for models A, B, C, D, and E.}\\
\hline
\hline
Parameter & SpA & Model A & Model B & Model C & Model D & Model E\\
\hline
\endfirsthead
\caption{Continued.} \\
\hline
Parameter & SpA & Model A & Model B & Model C & Model D & Model E\\
\hline
\endhead
\hline
\endfoot
\hline
\hline
\endlastfoot
Number of free parameters\,\tablefootmark{$\ast$} & 3 & 23 & 9 & 33 & 28 & 50 \\
\hline
\multicolumn{7}{l}{\it Orbital parameters}\\
& & & & & & \\
Orbital period $P$ [d] & -- & 86.647662 $\pm$ 3.4 10$^{-5}$ & 86.647662\,\tablefootmark{$\dag$} & 86.647666 $\pm$ 3.4 10$^{-5}$ & 86.647664 $\pm$ 3.5 10$^{-5}$ & 86.647663 $\pm$ 3.2 10$^{-5}$ \\
Transit epoch $T_{0}$ [BJD$_\mathrm{TDB}$ - 2455000] & -- & 6.79462 $\pm$ 3.2 10$^{-4}$ & 6.79462\,\tablefootmark{$\dag$} & 6.79448 $\pm$ 3.2 10$^{-4}$ & 6.79451 $\pm$ 2.9 10$^{-4}$ & 6.79451 $\pm$ 3.0 10$^{-4}$ \\
Orbital inclination $i_p$ [$^{\circ}$] & -- & 89.54$^{_{+0.28}}_{^{-0.59}}$ & -- & 89.60$^{_{+0.27}}_{^{-0.49}}$ & 89.70$^{_{+0.19}}_{^{-0.43}}$ & 89.63$^{_{+0.24}}_{^{-0.41}}$ \\
Orbital eccentricity $e_p$ & -- & 0.77 $\pm$ 0.08 & 0.72$^{_{+0.05}}_{^{-0.1}}$ & 0.75 $\pm$ 0.04 & 0.73$^{_{+0.04}}_{^{-0.07}}$ & 0.75 $\pm$ 0.03 \\
Argument of periastron $\omega_p$ [$^{\circ}$] & -- & 87$^{_{+31}}_{^{-47}}$ & 131$^{_{+11}}_{^{-17}}$ & 129 $\pm$ 8 & 131$^{_{8}}_{^{-14}}$ & 128$^{_{+6}}_{^{-9}}$ \\
Spin-orbit angle $| \lambda |$ [$^{\circ}$] & -- & -- & -- & -- & -- & 74$^{_{+32}}_{^{-46}}$ \\
& & & & & & \\
\hline
\multicolumn{7}{l}{\it Transit parameters}\\
& & & & & & \\
Radius ratio $r_{p}/R_{\star}$ & -- & 0.0821 $\pm$ 0.0015 & -- & 0.0822 $\pm$ 0.0014 & 0.0819 $\pm$ 0.0014 & 0.0830 $\pm$ 0.0014 \\
System scale $a/R_{\star}$ & -- & -- & -- & -- & 73 $\pm$ 13 & -- \\
Linear limb darkening coefficient $u_{a}$ & -- & 0.58 $\pm$ 0.07 & -- & 0.57 $\pm$ 0.07 & 0.57 $\pm$ 0.07 & 0.54 $\pm$ 0.07 \\
Quadratic limb darkening coefficient $u_{b}$ & -- & -0.02 $\pm$ 0.16 & -- & 0.00 $\pm$ 0.17 & 0.00 $\pm$ 0.15 & 0.06 $\pm$ 0.17 \\
& & & & & & \\
\hline
\multicolumn{7}{l}{\it Velocimetric parameters}\\
& & & & & & \\
Systemic velocity $\gamma$ [km.s$^{-1}$] & -- & -- & 10.101 $\pm$ 0.011 & 10.102 $\pm$ 0.010 & 10.101 $\pm$ 0.011 & 10.102 $\pm$ 0.009 \\ 
Linear drift $d_{1}$ [m.s$^{-1}$.$yr^{-1}$] & -- & -- & 135 $\pm$ 80 & 142 $\pm$ 77 & 135 $\pm$ 77 & 142 $\pm$ 73 \\ 
Quadratic drift $d_{2}$ [m.s$^{-1}$.$yr^{-2}$] & -- & -- & -283 $\pm$ 75 & -276 $\pm$ 73 & -283 $\pm$ 69 & -276 $\pm$ 71 \\
Radial velocity semi-amplitude $K$ [m.s$^{-1}$] & -- & -- & 86$^{_{+20}}_{^{-35}}$ & 94 $\pm$ 21 & 92$^{_{+16}}_{^{-26}}$ & 94 $\pm$ 22 \\
& & & & & & \\
\hline
\multicolumn{7}{l}{\it Stellar parameters}\\
& & & & & & \\
Effective temperature \teff\, [K] & 5540 $\pm$ 90 & 5542 $\pm$ 100 & -- & 5547 $\pm$ 95 & -- & 5548 $\pm$ 97 \\
Iron abundance \met\, [dex] & +0.26 $\pm$ 0.10 & 0.26 $\pm$ 0.10 & -- & 0.25 $\pm$ 0.11 & -- & 0.25 $\pm$ 0.10 \\
Surface gravity \logg\, [\cmss] & 4.30 $\pm$ 0.15 & -- & -- & -- & -- & -- \\
Bulk density $\rho_{\star}$ [$\rho_{\odot}$] & 0.33$^{_{+0.36}}_{^{-0.10}}$\,\tablefootmark{$\ddag$} & 0.50 $\pm$ 0.23 & -- & 0.58 $\pm$ 0.18 & 0.70 $\pm$ 0.37\,\tablefootmark{$\ddag$} & 0.59 $\pm$ 0.16 \\
Projected rotational velocity \vsini\, [\kms] & 4 $\pm$ 2 & -- & -- & -- & -- & 3.1 $\pm$ 1.8 \\
& & & & & & \\
\hline
\multicolumn{7}{l}{\it System parameters}\\
& & & & & & \\
Distance from Earth $D$ [pc] & -- & -- & -- & 848 $\pm$ 88 & -- & 846 $\pm$ 80 \\
Interstellar absorption E(B$-$V) [mag] & -- & -- & -- & 0.16 $\pm$ 0.04 & -- & 0.16 $\pm$ 0.04 \\
& & & & & & \\
\hline
\multicolumn{7}{l}{\it Instrumental parameters}\\
& & & & & & \\
\multicolumn{7}{l}{\it \textit{Kepler} season 0}\\
jitter [ppm] & --  & 168 $\pm$ 23 & -- & 168 $\pm$ 21 & 168 $\pm$ 23 & 169 $\pm$ 23 \\
contamination [\%] & -- & 11.7 $\pm$ 2.7 & -- & 12.0 $\pm$ 2.8 & 11.5 $\pm$ 2.7 & 14.3 $\pm$ 2.6 \\
& & & & & & \\
\multicolumn{7}{l}{\it \textit{Kepler} season 1}\\
jitter [ppm] & -- & 125 $\pm$ 26 & -- & 126 $\pm$ 27 & 126 $\pm$ 24 & 127 $\pm$ 28 \\
contamination [\%] & -- & 13.7 $\pm$ 2.5 & -- & 14.0 $\pm$ 2.6 & 13.5 $\pm$ 2.7 & 16.2 $\pm$ 2.5 \\
& & & & & & \\
\multicolumn{7}{l}{\it \textit{Kepler} season 2}\\
jitter [ppm] & -- & 156 $\pm$ 28 & -- & 157 $\pm$ 26 & 157 $\pm$ 25 & 156 $\pm$ 25 \\
contamination [\%] & -- & 11.9 $\pm$ 2.7 & -- & 12.1 $\pm$ 2.7 & 11.6 $\pm$ 2.7 & 14.4 $\pm$ 2.5 \\
& & & & & & \\
\multicolumn{7}{l}{\it \textit{Kepler} season 3}\\
jitter [ppm] & -- & 121 $\pm$ 33 & -- & 122 $\pm$ 48 & 120$^{_{+35}}_{^{-52}}$ & 119 $\pm$ 48 \\
contamination [\%] & -- & 11.3 $\pm$ 2.8 & -- & 11.5 $\pm$ 2.7 & 11.0 $\pm$ 2.8 & 13.8 $\pm$ 2.6 \\
& & & & & & \\
\multicolumn{7}{l}{\it SOPHIE}\\
jitter [\ms] & -- & -- & 34 $\pm$ 11 & 31 $\pm$ 9 & 32 $\pm$ 10 & 31 $\pm$ 9 \\
& & & & & & \\
\multicolumn{7}{l}{\it HARPS-N}\\
jitter [\ms] & -- & -- & 4.6 $\pm$ 4.6 & 4.6 $\pm$ 4.6 & 4.6$^{_{+5.2}}_{^{-3.6}}$ & 4.8 $\pm$ 4.8 \\
offset [\ms] & -- & -- & -37 $\pm$ 20 & -40 $\pm$ 16 & -41 $\pm$ 17 & -39 $\pm$ 16 \\
& & & & & & \\
\multicolumn{7}{l}{\it SED}\\
jitter [mags] & -- & -- & -- & 0.11 $\pm$ 0.06 & -- & 0.11 $\pm$ 0.06  \\
& & & & & & \\
\multicolumn{7}{l}{\it OHP-T120}\\
jitter [\%] & -- & -- & --  & -- & -- & 0.9$^{_{+0.4}}_{^{-0.6}}$ \\
contamination [\%] & -- & -- & -- & -- & -- & 1.9$^{_{+2.8}}_{^{-1.5}}$\\
& & & & & & \\
\multicolumn{7}{l}{\it IAC80}\\
jitter [\%] & -- & -- & --  & -- & -- & 0.9 $\pm$ 0.3 \\
contamination [\%] & -- & -- & -- & -- & -- & 25.1 $\pm$ 7.2 \\
& & & & & & \\
\multicolumn{7}{l}{\it ROTAT}\\
jitter [\%] & -- & -- & --  & -- & -- & 1.0 $\pm$ 1.0 \\
contamination [\%] & -- & -- & -- & -- & -- & 20 $\pm$ 16 \\
& & & & & & \\
\multicolumn{7}{l}{\it MOOS}\\
jitter [\%] & -- & -- & --  & -- & -- & 0.5$^{_{+0.6}}_{^{-0.4}}$ \\
contamination [\%] & -- & -- & -- & -- & -- & 17 $\pm$ 12 \\
& & & & & & \\
\multicolumn{7}{l}{\it Engarouines}\\
jitter [ppt] & -- & -- & --  & -- & -- & 0.9 $\pm$ 0.8 \\
contamination [\%] & -- & -- & -- & -- & -- & 45 $\pm$ 19 \\
\label{ModelResult}
\end{longtable}%
\tablefoot{\tablefoottext{$\ast$}{This number also accounts for the out-of-transit flux for each light curve which are not listed in this table} ; \tablefoottext{$\dag$}{Fixed parameter} ; \tablefoottext{$\ddag$}{Deduced parameter.}}
\end{landscape}
} 

\onllongtab{
\begin{landscape}
\begin{longtable}{lcccc}
\caption{Priors used in the analysis of the scenarios 0, 1, 2 and 3:  $\mathcal{U}(a,b)$ represents a Uniform prior between $a$ and $b$; $\mathcal{J}(a,b)$ represents a Jeffreys distribution between $a$ and $b$; $\mathcal{N}(\mu,\sigma^{2})$ represents a Normal distribution with a mean of $\mu$ and a width of $\sigma^{2}$; $\mathcal{N_{A}}(\mu,\sigma_{-}^{2}, \sigma_{+}^{2})$ represents an asymmetric Normal distribution with mean $\mu$, upper width $\sigma_{+}^{2}$ and lower width $\sigma_{-}^{2}$; $\mathcal{N_{U}}(\mu,\sigma^{2}, a, b)$ represents a Normal distribution with a mean of $\mu$ and a width of $\sigma^{2}$ and limited by a Uniform distribution between $a$ and $b$; and finally $\mathcal{S}(a,b)$ represents a Sine distribution between $a$ and $b$.}\\
\hline
\hline
Parameter & Scenario 0 & Scenario 1 & Scenario 2 & Scenario 3 \\
\hline
\endfirsthead
\caption{Continued.} \\
\hline
Parameter & Scenario 0 & Scenario 1 & Scenario 2 & Scenario 3 \\
\hline
\endhead
\hline
\endfoot
\hline
\hline
\endlastfoot
Number of free parameters & 35 & 38 & 38 & 38 \\
\hline
\multicolumn{5}{l}{\it Primary star}\\
& & & & \\
Effective temperature \teff\, [K] & $\mathcal{N}(5540, 90)$ & $\mathcal{N}(5540, 90)$ & $\mathcal{N}(5540, 90)$ & $\mathcal{N}(5540, 90)$ \\
Iron abundance \met\, [dex] & $\mathcal{N}(+0.26, 0.10)$  & $\mathcal{N}(+0.26, 0.10)$ & $\mathcal{N}(+0.26, 0.10)$ & $\mathcal{N}(+0.26, 0.10)$ \\
Surface gravity \logg\, [\cmss] & -- & $\mathcal{N}(4.30, 0.15)$ & $\mathcal{N}(4.30, 0.15)$ & $\mathcal{N}(4.30, 0.15)$ \\
Bulk density $\rho_{\star}$ [$\rho_{\odot}$] & $\mathcal{N_{A}}(0.33, 0.10, 0.36)$ & -- & -- & -- \\
Projected rotational velocity $\upsilon \sin i_{\star_{1}}$\, [\kms] & $\mathcal{N_{U}}(4.0, 2.0, 0.0, 10.0)$ & $\mathcal{N_{U}}(4.0, 2.0, 0.0, 10.0)$ & $\mathcal{N_{U}}(4.0, 2.0, 0.0, 10.0)$ & $\mathcal{N_{U}}(4.0, 2.0, 0.0, 10.0)$ \\
Linear limb darkening coefficient $u_{a}$ & $\mathcal{U}(-0.5, 1.2)$ & -- & -- & -- \\
Quadratic limb darkening coefficient $u_{b}$ & $\mathcal{U}(-0.5, 1.2)$ & -- & -- & -- \\
& & & & \\
\hline
\multicolumn{5}{l}{\it Secondary star}\\
& & & & \\
Initial mass $m_{init_{2}}$\, [\Msun] & -- & $\mathcal{U}(0.1, 1.2)$ & $\mathcal{U}(0.1, 1.2)$ & $\mathcal{U}(0.1, 1.2)$ \\
Projected rotational velocity $\upsilon \sin i_{\star_{2}}$\, [\kms] & -- & $\mathcal{U}(0.1, 20)$  & $\mathcal{U}(0.1, 20)$  & $\mathcal{U}(0.1, 30)$  \\
& & & & \\
\hline
\multicolumn{5}{l}{\it Tertiary star}\\
& & & & \\
Initial mass $m_{init_{3}}$\, [\Msun] & -- & -- & -- & $\mathcal{U}(0.1, 1.0)$ \\
Projected rotational velocity $\upsilon \sin i_{\star_{3}}$\, [\kms] & -- & --  & --  & $\mathcal{U}(0.1, 20)$  \\
& & & & \\
\hline
\multicolumn{5}{l}{\it Planet}\\
& & & & \\
Radius ratio $r_{p}/R_{\star}$ & $\mathcal{J}(0.01, 0.5)$ & -- & -- & -- \\
Planet radius r$_{p}$ [\Rjup] &-- & $\mathcal{U}(0.0, 2.2)$ & $\mathcal{U}(0.0, 2.2)$ & -- \\
Radial velocity semi-amplitude $K$ [m.s$^{-1}$] & $\mathcal{J}(1.0, 1000)$ & -- & -- & -- \\
Planet mass m$_{p}$ [\Mjup] & -- & $\mathcal{U}(0.0, 90.0)$ & $\mathcal{U}(0.0, 90.0)$ & -- \\
& & & & \\
\hline
\multicolumn{5}{l}{\it Inner orbit}\\
& & & & \\
Orbital period $P_{in}$ [d] & $\mathcal{N}(86.64774, 0.001)$ & $\mathcal{N}(86.64774, 0.001)$ & $\mathcal{N}(86.64774, 0.001)$ & $\mathcal{N}(86.64774, 0.001)$ \\
Transit / Eclipse epoch $T_{0}$ [BJD$_\mathrm{TDB}$ - 2455000] & $\mathcal{N}(6.79341, 0.02)$ & $\mathcal{N}(6.79341, 0.02)$ & $\mathcal{N}(6.79341, 0.02)$ & $\mathcal{N}(6.79341, 0.02)$ \\
Orbital inclination $i_{in}$ [$^{\circ}$] & $\mathcal{S}(80, 90)$ & $\mathcal{S}(80, 90)$ & $\mathcal{S}(80, 90)$ & $\mathcal{S}(80, 90)$ \\
Orbital eccentricity $e_{in}$ & $\mathcal{U}(0.0, 1.0)$ & $\mathcal{U}(0.0, 1.0)$ & $\mathcal{U}(0.0, 1.0)$ & $\mathcal{U}(0.0, 1.0)$ \\
Argument of periastron $\omega_{in}$ [$^{\circ}$] & $\mathcal{U}(0.0, 360.0)$ & $\mathcal{U}(0.0, 360.0)$ & $\mathcal{U}(0.0, 360.0)$ & $\mathcal{U}(0.0, 360.0)$ \\
& & & & \\
\hline
\multicolumn{5}{l}{\it Outer orbit}\\
& & & & \\
Linear drift $d_{1}$ [km.s$^{-1}$.$d^{-1}$] & $\mathcal{U}(-1.0, 1.0)$ & -- & -- & -- \\ 
Quadratic drift $d_{2}$ [km.s$^{-1}$.$d^{-2}$] & $\mathcal{U}(-1.0, 1.0)$ & -- & -- & -- \\
Orbital period $P_{out}$ [d] & -- & $\mathcal{J}(100, 10000)$ & $\mathcal{J}(100, 10000)$ & $\mathcal{J}(100, 10000)$ \\
Periastron epoch $T_{p}$ [BJD$_\mathrm{TDB}$ - 2455000] & -- & $\mathcal{U}(0, 10000)$ & $\mathcal{U}(0, 10000)$ & $\mathcal{U}(0, 10000)$ \\
Orbital inclination $i_{out}$ [$^{\circ}$] & -- & $\mathcal{S}(0, 90)$ & $\mathcal{S}(0, 90)$ & $\mathcal{S}(0, 90)$ \\
Orbital eccentricity $e_{out}$ & $\mathcal{U}(0.0, 1.0)$ & $\mathcal{U}(0.0, 1.0)$ & $\mathcal{U}(0.0, 1.0)$ & $\mathcal{U}(0.0, 1.0)$ \\
Argument of periastron $\omega_{out}$ [$^{\circ}$] & $\mathcal{U}(0.0, 360.0)$ & $\mathcal{U}(0.0, 360.0)$ & $\mathcal{U}(0.0, 360.0)$ & $\mathcal{U}(0.0, 360.0)$ \\
& & & & \\
\hline
\multicolumn{5}{l}{\it System}\\
& & & & \\
Distance from Earth $D$ [pc] & $\mathcal{U}(100, 5000)$ & $\mathcal{U}(100, 5000)$ & $\mathcal{U}(100, 5000)$ & $\mathcal{U}(100, 5000)$ \\
Interstellar absorption E(B$-$V) [mag] & $\mathcal{U}(0.0, 2.0)$ & $\mathcal{U}(0.0, 2.0)$ & $\mathcal{U}(0.0, 2.0)$ & $\mathcal{U}(0.0, 2.0)$ \\
Systemic velocity $\gamma$ [km.s$^{-1}$] & $\mathcal{U}(10.0, 20.0)$ & $\mathcal{U}(-800, 800)$ & $\mathcal{U}(-800, 800)$ & $\mathcal{U}(-800, 800)$ \\
& & & & \\
\hline
\multicolumn{5}{l}{\it Instrumental parameters}\\
& & & & \\
\multicolumn{5}{l}{\it \textit{Kepler} season 0}\\
Jitter & $\mathcal{U}(0.0, 0.1)$  & $\mathcal{U}(0.0, 0.1)$ & $\mathcal{U}(0.0, 0.1)$ & $\mathcal{U}(0.0, 0.1)$ \\
Background contamination [\%] & $\mathcal{N_{U}}(12.0, 2.8, 0.0, 100.0)$ & $\mathcal{N_{U}}(12.0, 2.8, 0.0, 100.0)$ & $\mathcal{N_{U}}(12.0, 2.8, 0.0, 100.0)$ & $\mathcal{N_{U}}(12.0, 2.8, 0.0, 100.0)$ \\
Flux out-of-transit & $\mathcal{U}(0.999, 1.001)$ & $\mathcal{U}(0.999, 1.001)$ & $\mathcal{U}(0.999, 1.001)$ & $\mathcal{U}(0.999, 1.001)$\\
& & & & \\
\multicolumn{5}{l}{\it \textit{Kepler} season 1}\\
Jitter & $\mathcal{U}(0.0, 0.1)$ & $\mathcal{U}(0.0, 0.1)$ & $\mathcal{U}(0.0, 0.1)$ & $\mathcal{U}(0.0, 0.1)$ \\
Background contamination [\%] & $\mathcal{N_{U}}(14.0, 2.6, 0.0, 100.0)$  & $\mathcal{N_{U}}(14.0, 2.6, 0.0, 100.0)$ & $\mathcal{N_{U}}(14.0, 2.6, 0.0, 100.0)$ & $\mathcal{N_{U}}(14.0, 2.6, 0.0, 100.0)$ \\
Flux out-of-transit & $\mathcal{U}(0.999, 1.001)$ & $\mathcal{U}(0.999, 1.001)$ & $\mathcal{U}(0.999, 1.001)$ & $\mathcal{U}(0.999, 1.001)$\\
& & & & \\
\multicolumn{5}{l}{\it \textit{Kepler} season 2}\\
Jitter & $\mathcal{U}(0.0, 0.1)$ & $\mathcal{U}(0.0, 0.1)$ & $\mathcal{U}(0.0, 0.1)$ & $\mathcal{U}(0.0, 0.1)$ \\
Background contamination [\%] & $\mathcal{N_{U}}(12.1, 2.7, 0.0, 100.0)$ & $\mathcal{N_{U}}(12.1, 2.7, 0.0, 100.0)$ & $\mathcal{N_{U}}(12.1, 2.7, 0.0, 100.0)$ & $\mathcal{N_{U}}(12.1, 2.7, 0.0, 100.0)$ \\
Flux out-of-transit & $\mathcal{U}(0.999, 1.001)$ & $\mathcal{U}(0.999, 1.001)$ & $\mathcal{U}(0.999, 1.001)$ & $\mathcal{U}(0.999, 1.001)$\\
& & & & \\
\multicolumn{5}{l}{\it \textit{Kepler} season 3}\\
Jitter & $\mathcal{U}(0.0, 0.1)$ & $\mathcal{U}(0.0, 0.1)$ & $\mathcal{U}(0.0, 0.1)$ & $\mathcal{U}(0.0, 0.1)$ \\
Background contamination [\%] & $\mathcal{N_{U}}(11.5, 2.7, 0.0, 100.0)$ & $\mathcal{N_{U}}(11.5, 2.7, 0.0, 100.0)$ & $\mathcal{N_{U}}(11.5, 2.7, 0.0, 100.0)$ & $\mathcal{N_{U}}(11.5, 2.7, 0.0, 100.0)$ \\
Flux out-of-transit & $\mathcal{U}(0.999, 1.001)$ & $\mathcal{U}(0.999, 1.001)$ & $\mathcal{U}(0.999, 1.001)$ & $\mathcal{U}(0.999, 1.001)$\\
& & & & \\
\multicolumn{5}{l}{\it SOPHIE}\\
Radial velocity jitter [\ms] & $\mathcal{U}(0, 1000)$ & $\mathcal{U}(0, 1000)$ & $\mathcal{U}(00, 1000)$ & $\mathcal{U}(0, 1000)$ \\
V$_{span}$ offset [\kms] & $\mathcal{U}(-1.0, 1.0)$ & $\mathcal{U}(-1.0, 1.0)$ & $\mathcal{U}(-1.0, 1.0)$ & $\mathcal{U}(-1.0, 1.0)$ \\
V$_{span}$ jitter [\kms] & $\mathcal{U}(0.0, 1.0)$ & $\mathcal{U}(0.0, 1.0)$ & $\mathcal{U}(0.0, 1.0)$ & $\mathcal{U}(0.0, 1.0)$ \\
FWHM jitter [\kms] &  $\mathcal{U}(0.0, 1.0)$ &  $\mathcal{U}(0.0, 1.0)$ &  $\mathcal{U}(0.0, 1.0)$ &  $\mathcal{U}(0.0, 1.0)$ \\
& & & & \\
\multicolumn{5}{l}{\it SED}\\
jitter [mags] &  $\mathcal{U}(0.0, 1.0)$ &  $\mathcal{U}(0.0, 1.0)$ &  $\mathcal{U}(0.0, 1.0)$ &  $\mathcal{U}(0.0, 1.0)$ \\
\label{ScenarioPriors}
\end{longtable}%
\end{landscape}
} 

\onllongtab{
\begin{landscape}
\begin{longtable}{lcccc}
\caption{Results of the analysis for the scenarios 0, 1, 2 and 3.}\\
\hline
\hline
Parameter & Scenario 0 & Scenario 1 & Scenario 2 & Scenario 3 \\
\hline
\endfirsthead
\caption{Continued.} \\
\hline
Parameter & Scenario 0 & Scenario 1 & Scenario 2 & Scenario 3 \\
\hline
\endhead
\hline
\endfoot
\hline
\hline
\endlastfoot
Number of free parameters\,\tablefootmark{$\ast$} & 35 & 38 & 38 & 38 \\
\hline
\multicolumn{5}{l}{\it Primary star}\\
& & & & \\
Effective temperature \teff\, [K] & 5545 $\pm$ 95 & 5518 $\pm$ 79 & 5576 $\pm$ 82 & 5612 $\pm$ 9.4 \\
Iron abundance \met\, [dex] & 0.25 $\pm$ 0.10 & 0.27 $\pm$ 0.10 & 0.21 $\pm$ 0.9 & 0.24 $\pm$ 0.10 \\
Surface gravity \logg\, [\cmss] & -- & 4.32 $\pm$ 0.10 & 4.06 $\pm$ 0.09 & 4.12 $\pm$ 0.16 \\
Bulk density $\rho_{\star}$ [$\rho_{\odot}$] & 0.59 $\pm$ 0.20 & -- & -- & -- \\
Projected rotational velocity $\upsilon \sin i_{\star_{1}}$\, [\kms] & 4.8 $\pm$ 1.0 & 4.6 $\pm$ 0.2 & 4.3 $\pm$ 0.2 & 3.3 $\pm$ 0.2 \\
Linear limb darkening coefficient $u_{a}$ & 0.57 $\pm$ 0.07 & -- & -- & -- \\
Quadratic limb darkening coefficient $u_{b}$ & 0.00 $\pm$ 0.16 & -- & -- & -- \\
& & & & \\
\hline
\multicolumn{5}{l}{\it Secondary star}\\
& & & & \\
Initial mass $m_{init_{2}}$\, [\Msun] & -- & 0.70 $\pm$ 0.07\,\tablefootmark{$\star$} & 1.00 $\pm$ 0.05\,\tablefootmark{$\star$} & 1.00 $\pm$ 0.07\,\tablefootmark{$\star$} \\
Projected rotational velocity $\upsilon \sin i_{\star_{2}}$\, [\kms] & -- & 2.6 $\pm$ 2.0 & 4.79 $\pm$ 0.32 & 29.3$^{_{+0.5}}_{^{-0.9}}$ \\
& & & & \\
\hline
\multicolumn{5}{l}{\it Tertiary star}\\
& & & & \\
Initial mass $m_{init_{3}}$\, [\Msun] & -- & -- & -- & 0.13 $\pm$ 0.02\,\tablefootmark{$\star$} \\
Projected rotational velocity $\upsilon \sin i_{\star_{3}}$\, [\kms] & -- & -- & -- & 9.9 $\pm$ 7.1 \\
& & & & \\
\hline
\multicolumn{5}{l}{\it Planet}\\
& & & & \\
Radius ratio $r_{p}/R_{\star}$ & 0.082 $\pm$ 0.001 & -- & -- & -- \\
Planet radius r$_{p}$ [\Rjup] & -- & 0.94 $\pm$ 0.12 & 1.56 $\pm$ 0.13 & -- \\
Radial velocity semi-amplitude $K$ [m.s$^{-1}$] & 99 $\pm$ 21 & -- & -- & -- \\
Planet mass m$_{p}$ [\Mjup] & -- & 1.45 $\pm$ 0.35 & 3.92 $\pm$ 0.88 & -- \\
& & & & \\
\hline
\multicolumn{5}{l}{\it Inner orbit}\\
& & & & \\
Orbital period $P_{in}$ [d] & 86.647662$\pm$3.4 10$^{-5}$ & 86.647661 $\pm$ 3.4 10$^{-5}$ & 86.647662 $\pm$ 3.4 10$^{-5}$ & 86.647660 $\pm$ 3.3 10$^{-5}$ \\
Transit / Eclipse epoch $T_{0}$ [BJD$_\mathrm{TDB}$ - 2455000] & 6.79447 $\pm$ 3.2 10$^{-4}$ & 6.79454 $\pm$ 3.1 10$^{-4}$ & 6.79458 $\pm$ 3.0 10$^{-4}$ & 6.79460 $\pm$ 3.1 10$^{-4}$ \\
Orbital inclination $i_{in}$ [$^{\circ}$] & 89.58$^{_{+0.27}}_{^{-0.51}}$ & 89.66$^{_{+0.23}}_{^{-0.38}}$ & 89.82$^{_{+0.12}}_{^{-0.19}}$ & 89.86$^{_{+0.10}}_{^{-0.17}}$ \\
Orbital eccentricity $e_{in}$ & 0.78 $\pm$ 0.05 & 0.77 $\pm$ 0.05 & 0.75 $\pm$ 0.04 & 0.78 $\pm$ 0.03 \\
Argument of periastron $\omega_{in}$ [$^{\circ}$] & 140 $\pm$ 11 & 141 $\pm$ 17 & 128 $\pm$ 11 & 151 $\pm$ 10 \\
& & & & \\
\hline
\multicolumn{5}{l}{\it Outer orbit}\\
& & & & \\
Linear drift $d_{1}$ [m.s$^{-1}$.$yr^{-1}$] & 142 $\pm$ 69 & -- & -- & -- \\ 
Quadratic drift $d_{2}$ [m.s$^{-1}$.$yr^{-2}$] & -279 $\pm$ 68 & -- & -- & -- \\
Orbital period $P_{out}$ [d] & -- & 3430 $\pm$ 1200 & 1686$^{_{+1000}}_{^{-450}}$ & 6154 $\pm$ 2100 \\
Periastron epoch $T_{p}$ [BJD$_\mathrm{TDB}$ - 2450000] & -- & $8045^{_{+3500}}_{^{-2300}}$ & 8643$^{_{+5200}}_{^{-2700}}$ & 7889.7$^{_{+4400}}_{^{-970}}$ \\
Orbital inclination $i_{out}$ [$^{\circ}$] & -- & 18.2$^{_{+18.0}}_{^{-5.5}}$ & 11.4$^{_{+9.3}}_{^{-4.6}}$ & 37.6 $\pm$ 11.0 \\
Orbital eccentricity $e_{out}$ & -- & 0.31$^{_{+0.37}}_{^{-0.21}}$ & 0.49 $\pm$ 0.30 & 0.21$^{_{+0.27}}_{^{-0.14}}$ \\
Argument of periastron $\omega_{out}$ [$^{\circ}$] & -- & 180 $\pm$ 110 & 175$^{_{+31}}_{^{-21}}$ & 140$^{_{+110}}_{^{-76}}$ \\
& & & & \\
\hline
\multicolumn{5}{l}{\it System}\\
& & & & \\
Distance from Earth $D$ [pc] & 836 $\pm$ 89 & 902 $\pm$ 110 & 1420 $\pm$ 160 & 1329 $\pm$ 240 \\
Interstellar absorption E(B$-$V) [mag] & 0.16 $\pm$ 0.04 & 0.10 $\pm$ 0.04 & 0.18 $\pm$ 0.04 & 0.17 $\pm$ 0.04 \\
Systemic velocity $\gamma$ [km.s$^{-1}$] & 10.104 $\pm$ 0.009 & 8.81 $\pm$ 0.42 & 9.776 $\pm$ 0.041 & 5.75$^{_{+1.3}}_{^{-0.56}}$ \\
& & & & \\
\hline
\multicolumn{5}{l}{\it Instrumental parameters}\\
& & & & \\
\multicolumn{5}{l}{\it \textit{Kepler} season 0}\\
Jitter [ppm] & 168 $\pm$ 21  & 166 $\pm$ 20 & 168 $\pm$ 22 & 167 $\pm$ 22 \\
Background contamination [\%] & 11.8 $\pm$ 1.5 & 11.9 $\pm$ 1.4 & 11.4 $\pm$ 1.4 & 11.3 $\pm$ 1.5 \\
& & & & \\
\multicolumn{5}{l}{\it \textit{Kepler} season 1}\\
Jitter [ppm] & 126 $\pm$ 28 & 129 $\pm$ 23 & 140 $\pm$ 22 & 141 $\pm$ 22 \\
Background contamination [\%] & 13.7 $\pm$ 1.5 & 13.9 $\pm$ 1.4 & 13.3 $\pm$ 1.4 & 13.2 $\pm$ 1.5 \\
& & & & \\
\multicolumn{5}{l}{\it \textit{Kepler} season 2}\\
Jitter [ppm] & 157 $\pm$ 27 & 155 $\pm$ 25 & 164 $\pm$ 23 & 163 $\pm$ 24 \\
Background contamination [\%] & 11.9 $\pm$ 1.5 & 11.9 $\pm$ 1.6 & 11.5 $\pm$ 1.4 & 11.3 $\pm$ 1.6 \\
& & & & \\
\multicolumn{5}{l}{\it \textit{Kepler} season 3}\\
Jitter [ppm] & 122 $\pm$ 46 & 118$^{_{+35}}_{^{-54}}$ & 110 $\pm$ 49 & 114 $\pm$ 47 \\
Background contamination [\%] & 11.4 $\pm$ 1.5 & 11.4 $\pm$ 1.5 & 11.0 $\pm$ 1.4 & 10.9 $\pm$ 1.6 \\
& & & & \\
\multicolumn{5}{l}{\it SOPHIE}\\
Radial velocity jitter [\ms] & 31 $\pm$ 9 & 32 $\pm$ 10 & 32 $\pm$ 10 & 35 $\pm$ 9 \\
V$_{span}$ offset [\ms] & 95 $\pm$ 13 & 68$^{_{+38}}_{^{-70}}$ & 59 $\pm$ 27 & -150 $\pm$ 130 \\
V$_{span}$ jitter [\ms] & 42 $\pm$ 16 & 39 $\pm$ 17 & 43 $\pm$ 19 & 54 $\pm$ 26 \\
FWHM jitter [\ms] & 111 $\pm$ 31 & 46 $\pm$ 43 & 48 $\pm$ 32 & 224 $\pm$ 54 \\
& & & & \\
\multicolumn{5}{l}{\it SED}\\
jitter [mags] & 0.11$^{_{+0.06}}_{^{-0.04}}$ & 0.10$^{_{+0.06}}_{^{-0.04}}$ & 0.10$^{_{+0.06}}_{^{-0.04}}$ & 0.10$^{_{+0.06}}_{^{-0.04}}$ \\
\label{ScenarioResult}
\end{longtable}%
\tablefoot{\tablefoottext{$\ast$}{This number also accounts for the out-of-transit flux for each light curve which are not listed in this table} ; \tablefoottext{$\star$}{This error does not account for the uncertainty of the stellar models.}}
\end{landscape}
} 

\onltab{
\begin{table*}[h]
\caption{Radial velocity measurements of KOI-1257}
\begin{center}
\begin{tabular}{cccccccccc}
\hline
\hline
Time & RV & $\sigma_{RV}$ & Vspan & $\sigma_{V_{span}}$ & FWHM & $\sigma_\mathrm{FWHM}$ & Texp & SNR\tablefootmark{$\dag$} & Moon\\
BJD & [\kms] & [\kms] & [\kms] & [\kms] & [\kms] & [\kms] & [s] & & flag\tablefootmark{\leftmoon} \\
\hline
\multicolumn{10}{l}{SOPHIE HE}\\
&&&&&&&&&\\
2456156.52163  &  9.5958  &  0.0137  &  -0.0356  &  0.0274  &  10.2559  &  0.0548  &  3600  &  17  & 0 \\
2456163.48998  &  9.5711  &  0.0251  &  -0.1068  &  0.0502  &  10.3591  &  0.1004  &  3604  &  14  & 0  \\
\vdots & \vdots & \vdots & \vdots & \vdots & \vdots & \vdots & \vdots & \vdots & \vdots \\
2456609.28745 &  10.1358  &  0.0226  &  -0.1002  &  0.0452  &  10.6818  &  0.0904  &  3600  &  14 & 1 \\
2456628.25761  &  10.1253  &  0.0189  &  -0.1099  &  0.0378  &  10.5455  &  0.0756  &  3600  &  15  & 0  \\
\hline
\multicolumn{10}{l}{HARPS-N}\\
&&&&&&&&&\\
2456566.34278  &  3.2224  &  0.0177  &  0.0684  &  0.0355  &  7.5882  &  0.0710  &  1800  &  10  & 0  \\
2456566.36392  &  3.2172  &  0.0165  &  0.2812  &  0.0330  &  7.4944  &  0.0660  &  1800  &  10  & 0  \\
\vdots & \vdots & \vdots & \vdots & \vdots & \vdots & \vdots & \vdots & \vdots & \vdots \\
2456576.36181  &  3.1944  &  0.0160  &  -0.0183  &  0.0319  &  7.5223  &  0.0639  &  1800  &  10  & 0  \\
2456576.38296  &  3.1918  &  0.0154  &  0.0129  &  0.0308  &  7.5161  &  0.0616  &  1800  &  11  & 0  \\
\hline
\hline
\end{tabular}
\tablefoot{\tablefoottext{$\dag$}: Signal-to-noise ratio per pixel measured at 550 nm;
\tablefoottext{\leftmoon}: Observations affected and corrected from the Moon background light. The amplitude of the radial velocity correction is (rv$_{\rm corr}$ - rv$_{\rm obs}$):
\begin{itemize}
\item 2456551.39898: 0.0886 \kms
\item 2456553.46555: 0.0693 \kms
\item 2456577.31138: -0.0215 \kms
\item 2456582.41201: 0.0200 \kms
\item 2456609.28745: 0.0192 \kms
\end{itemize}
}
\end{center}
\label{1257RVtable}
\end{table*}%
}

\onllongtab{
\begin{longtable}{cccccccc}
\caption{SOPHIE HE measurements of the constant star HD185144}\\
\hline
\hline
Time & RV & $\sigma_{RV}$ & Vspan & $\sigma_{V_{span}}$ & FWHM & $\sigma_\mathrm{FWHM}$ & SNR\\
BJD & [\kms] & [\kms] & [\kms] & [\kms] & [\kms] & [\kms] & \\
\hline
\endfirsthead
\caption{Continued.} \\
\hline
Time & RV & $\sigma_{RV}$ & Vspan & $\sigma_{V_{span}}$ & FWHM & $\sigma_\mathrm{FWHM}$ & SNR\\
\hline
\endhead
\hline
\endfoot
\hline
\hline
\endlastfoot
2455659.64178  &  26.7565  &  0.0014  &   0.0149  &  0.0028  &  8.7354  &  0.0056  &  158  \\
2455659.64495  &  26.7610  &  0.0007  &   0.0147  &  0.0014  &  8.7197  &  0.0028  &  295  \\
\vdots & \vdots & \vdots & \vdots & \vdots & \vdots & \vdots & \vdots \\
2456621.22269  &  26.7803  &  0.0007  &  -0.0022  &  0.0014  &  8.8162  &  0.0028  &  303  \\
2456628.21682  &  26.7798  &  0.0006  &   0.0014  &  0.0012  &  8.8298  &  0.0024  &  351  \\
\label{HD185144RVtable}
\end{longtable}%
}

\onllongtab{
\begin{longtable}{ccc}
\caption{Ground-based photometric data of KOI-1257. The reference for the time is BJD$_\mathrm{TDB}$ - 2456560.}\\
\hline
\hline
Time & Flux & $\sigma_\mathrm{flux}$ \\
\hline
\endfirsthead
\caption{Continued.} \\
\hline
Time & Flux & $\sigma_\mathrm{flux}$ \\
\hline
\endhead
\hline
\endfoot
\hline
\hline
\endlastfoot
\multicolumn{3}{l}{OHP-T120}\\
&&\\
6.30370 & 1.0012 & 0.0020\\
6.30630 & 1.0036 & 0.0020\\
\vdots & \vdots & \vdots\\
6.51099 & 0.9946 & 0.0020\\
6.51359 & 0.9922 & 0.0020\\
\hline
\multicolumn{3}{l}{IAC80}\\
&&\\
6.33684 & 1.0015 & 0.0015\\
6.33924 & 1.0001 & 0.0015\\
\vdots & \vdots & \vdots \\
6.54237 & 1.0005 & 0.0015\\
6.54476 & 0.9985 & 0.0015\\
\hline
\multicolumn{3}{l}{ROTAT}\\
&&\\
6.29281 & 1.0016 & 0.0061\\
6.29465 & 0.9899 & 0.0063\\
\vdots & \vdots & \\
6.53846 & 0.9886 & 0.0116\\
6.54032 & 1.0037 & 0.0108\\
\hline
\multicolumn{3}{l}{MOOS}\\
&&\\
6.30254 & 0.9968 & 0.0073\\
6.30475 & 0.9945 & 0.0062\\
\vdots & \vdots & \vdots \\
6.57980 & 0.9912 & 0.0070\\
6.58207 & 0.9947 & 0.0072\\
\hline
\multicolumn{3}{l}{Engarouines}\\
&&\\
6.29090 & 0.9963 & 0.0050\\
6.29436 & 1.0028 & 0.0050\\
\vdots & \vdots & \vdots\\
6.49472 & 0.9927 & 0.0050\\
6.49813 & 0.9881 & 0.0050\\
\label{PhotData}\\
\end{longtable}
}

\end{document}